\documentclass[12pt]{article}

\usepackage{graphicx,bbm}

\newcommand{\be}{\begin{equation}}
\newcommand{\ee}{\end{equation}}

\newcommand{\boldsymbol}[1]{\mbox{\boldmath ${#1}$}}

\newcommand{\he}{\tau}				
\newcommand{\hebar}{\overline{\tau}}		
\newcommand{\heprime}{\he'}			
\newcommand{\hebarprime}{\hebar{}'}		

\newcommand{\sind}{(\heprime \hebarprime)(\he \hebar)}	
\newcommand{\rind}{(\he \hebar)(\heprime \hebarprime)}	

\newcommand{\pr}{\mbox{\( e^-e^+ \rightarrow W^-W^+ \rightarrow
                        f_1\overline{f_2}f_3\overline{f_4} \)}}
\newcommand{\prpr}{\mbox{\( e^-e^+ \rightarrow W^-W^+ \)}}
\newcommand{\kvec}{\mbox{${\bf k}$}}		
\newcommand{\kbarvec}{\mbox{${\bf \overline{k}}$}}	
\newcommand{\qvec}{\mbox{${\bf q}$}}		
	
\newcommand{\s}{\sigma}

\newcommand{\lbar}{\overline{\lambda}}	 	
\newcommand{\lprime}{\lambda'}			
\newcommand{\lbarprime}{\overline{\lambda}{}'}	
\newcommand{\pind}{(\lambda \lbar)(\lprime \lbarprime)}
\newcommand{\tr}{\mbox{${\cal T}$}}		

\newcommand{\obi}{\mbox{${\cal O}_i$}}		
\newcommand{\obj}{\mbox{${\cal O}_j$}}		
\newcommand{\covo}{\mbox{$V(\cal{O})$}}		
\newcommand{\cpt}{\mbox{$C\!P\widetilde{T}$}}	
\newcommand{\cp}{\mbox{$C\!P$}}   		

\newcommand{\re}{\mbox{\rm Re}}
\newcommand{\im}{\mbox{\rm Im}}
\newcommand{\plh}{\hat{P}^L}
\newcommand{\prh}{\hat{P}^R}
\newcommand{\h}{h}		

\begin{document}


\begin{flushright}
  HD-THEP-02-34 \\
  hep-ph/0209229
\end{flushright}

\vspace{\baselineskip}

\begin{center}
\textbf{\Large Triple gauge couplings \\[0.3em]
	in polarised {\prpr} and their \\[0.3em]
        measurement using optimal observables \\}
\vspace{4\baselineskip}
{\sc M. Diehl}\footnote{email: mdiehl@physik.rwth-aachen.de} \\
\vspace{1\baselineskip}
\textit{Institut f\"ur Theoretische Physik E, RWTH Aachen, 52056
Aachen, Germany} \\
\vspace{2\baselineskip}
{\sc O. Nachtmann\footnote{email: O.Nachtmann@thphys.uni-heidelberg.de}
and F. Nagel\footnote{email: F.Nagel@thphys.uni-heidelberg.de}} \\
\vspace{1\baselineskip}
\textit{Institut f\"ur Theoretische Physik, Philosophenweg 16, 69120
Heidelberg, Germany} \\
\vspace{2\baselineskip}
\textbf{Abstract}\\
\vspace{1\baselineskip}
\parbox{0.9\textwidth}{The sensitivity of optimal integrated
observables to electroweak triple gauge couplings is investigated for
the process \mbox{\(\prpr \rightarrow 4\) fermions} at future linear
colliders.  By a suitable reparameterisation of the couplings we
achieve that all 28 coupling parameters have uncorrelated statistical
errors and are naturally normalised for this process.  Discrete
symmetry properties simplify the analysis and allow checks on the
stability of numerical results.  We investigate the sensitivity to the
couplings of the normalised event distribution and the additional
constraints that can be obtained from the total rate.  Particular
emphasis is put on the gain in sensitivity one can achieve with
longitudinal beam polarisation.  We also point out questions that may
best be settled with transversely polarised beams.  In particular we
find that with purely longitudinal polarisation one linear combination
of coupling parameters is hardly measurable by means of the normalised
event distribution. }
\end{center}
\vspace{\baselineskip}

\pagebreak

\tableofcontents

\pagebreak


\section{Introduction}
\label{sec-intr}

\suppressfloats

The Standard Model (SM) of electroweak interactions has been
thoroughly investigated both theoretically and experimentally and
turned out to be very successful.  After the discovery of the massive
gauge bosons $W$ and $Z$, the direct measurement of the triple gauge
couplings (TGCs), i.e.\ the couplings between two charged and one
neutral gauge boson, has been an important issue at the TEVATRON and
LEP as their values are determined by the non-Abelian electroweak
gauge group.  At $e^+e^-$ colliders the production of $W$ pairs,
single $W$s, single photons and single $Z$s is suitable for that.  In
this paper we consider the process \pr, where both the $\gamma WW$ and
the $ZWW$ couplings can be measured at the scale given by the
c.m.~energy.

Deviations from the SM at the $\gamma WW$ and $ZWW$ vertices can be
parameterised in a general framework.  Allowing for complex couplings
the most general vertex functions lead to altogether 28 real
parameters \cite{Hagiwara:1986vm}.  All four LEP collaborations have
investigated TGCs \cite{Heister:2001qt}, the tightest constraints
being of order $0.05$ for $\Delta g_1^Z$ and $\lambda_{\gamma}$, of
order $0.1$ for $\Delta \kappa_{\gamma}$ and of order $0.1$ to $0.6$
for the real and imaginary parts of $C$ and/or $P$ violating
couplings.  All these values correspond to single parameter fits.
Since only up to three-parameter fits have been performed, only a
small subset of couplings has been considered at a time, thereby
neglecting correlations between most of them.  Moreover many
couplings, notably the imaginary parts of $C$ and $P$ conserving
couplings, have been excluded from the analyses.

At a future linear $e^+e^-$ collider like TESLA \cite{Richard:2001qm}
or CLIC \cite{Ellis:1998wx}, respectively covering a c.m.\ energy
range from about 90~GeV to 800~GeV and from about 500~GeV to 5~TeV,
one will be able to study these couplings with unprecedented accuracy.
In this way one may begin to be sensitive to different extensions of
the SM, which predict deviations at the TGCs from the SM values,
typically through the effects of new particles and couplings in
radiative corrections.  Some examples, where effects of order
$10^{-3}$ may occur, are supersymmetric
models~\cite{Arhrib:1995dm,Argyres:1995ib}, models containing several
Higgs doublets~\cite{Couture:eu,He:qh}, $E_6$ vector
leptons~\cite{Couture:jn} or Majorana neutrinos~\cite{Katsuki:1994as}
and the minimal 3-3-1 model~\cite{Tavares-Velasco:2001vb}.  For
left-right symmetric models~\cite{Atwood:1990cm,Chang:1990fp,He:qh} and
mirror models~\cite{Chang:1990fp} the effects are predicted to be much
smaller, whereas models containing composite $W$
bosons~\cite{Suzuki:1985yh} or an additional gauge boson
$Z'$~\cite{Sharma:1996qv} may lead to larger effects.

Given the intricacies of a multi-dimensional parameter space, the full
covariance matrix for the errors on the couplings should best be
studied.  The high statistics needed for this will for instance be
available at TESLA, where the integrated luminosity is projected to be
about 500~${\rm fb}^{-1}$ or more per year at 500~GeV which for
unpolarised beams amounts to about 3.7~million produced $W$~pairs
(without cuts).  For a run at 800~GeV the luminosity is expected to be
twice as high, leading to 3.9~million $W$~pairs.  Moreover, polarised
beams will be particularly useful to disentangle different couplings.
In fact, certain directions in the parameter space of the couplings,
the so called right handed couplings, are difficult to measure in
$W$~pair production with unpolarised beams~\cite{Diehl:1993br}.  In
this case their effects are masked by the large contribution from the
neutrino exchange.  With polarised beams the strength of the neutrino
exchange contribution can in essence be varied freely.

In experimental analyses of TGCs and various other processes, optimal
observables~\cite{Atwood:1991ka,Diehl:1993br} have shown to be a
useful tool to extract physics parameters from the event
distributions.  These observables are constructed to have the smallest
possible statistical errors.  Due to this property they are also a
convenient means to determine the theoretically achievable sensitivity
in a given process.  In addition, they take advantage of the discrete
symmetries of the cross section.

The method proposed in~\cite{Diehl:1997ft} allows one to
simultaneously diagonalise the covariance matrix of the observables
and the part of the cross section which is quadratic in the couplings.
In this way one obtains a set of coupling constants that are naturally
normalised for the particular process and---in the limit of small
anomalous couplings---can be measured without statistical
correlations.  This allows one to see for which directions in
parameter space the sensitivity to the TGCs is high and for which it
is not, which is hardly possible by looking at covariance matrices of
large dimension without diagonalisation.  Moreover, the total cross
section acquires a particularly simple form and provides additional
constraints.

The basic purpose of this work is to use this extended optimal
observable method for a systematic investigation of the prospects to
measure the full set of TGCs in $W$ pair production at linear collider
energies, with special emphasis on initial-state polarisation.  The
usefulness of the method becomes particularly evident when considering
imaginary parts of $\cp$ conserving TGCs.  In this subspace of
couplings we find one direction to which---in the linear approximation
and for longitudinally polarised beams---there is no sensitivity in
the process we consider.  In~\cite{Diehl:1997ft} this was not taken
into account and led to numerical instabilities.  It is therefore
essential to disentangle the measurable TGCs from the hardly
measurable one. As a historical motivation one may think of the
electromagnetic nucleon form factors $F_1$ and $F_2$, where the choice
of linear combinations $G_E$ and $G_M$ leads to a simplification of
the Rosenbluth formula for the differential cross section of
electron-nucleon scattering (see e.g.~\cite{bib:kall}).  Since we deal
with 28 couplings here, an appropriate choice of parameters is
essential.

We restrict ourselves to the semileptonic decays of the $W$ pair,
where one $W$~boson decays into a quark-antiquark pair and the other
decays leptonically, but leave aside the decay into \( \tau \nu_{\tau}
\) since these events have a completely different experimental
signature.  The selected channels have a branching ratio of altogether
$8/27$, which is six times larger than that of both $W$s decaying into
\( e \nu_e \) or \( \mu \nu_{\mu} \).  On the other hand, the channels
where both $W$s decay hadronically are difficult to
reconstruct~\cite{bib:dham}.  The semileptonic channels have only one
ambiguity in the kinematical reconstruction if the charges of the two
jets from the hadronically decaying $W$ are not tagged.  Then one
cannot associate the jets to the up- and down-type (anti)quark of the
$W$~decay, and therefore has access only to the sum of the
distributions corresponding to the two final states.

This work is organised as follows: In Sect.~\ref{sec-ampl} we recall
the helicity amplitudes and cross sections of the process using a spin
density matrix formalism.  In Sect.~\ref{sec-opti} the
optimal-observable method is presented in the form as it is used in
our numerical calculations.  We explain our technique to implement the
simultaneous diagonalisation in a numerically stable way.  The role of
discrete symmetries in the framework of optimal observables is
described.  Many of the symmetry relations are well known, notably the
classification of the TGCs into four symmetry
classes~\cite{Hagiwara:1986vm} and its applicability to the
optimal-observable method~\cite{Diehl:1993br,Diehl:1997ft}.  Other
properties are used for a check on the numerics.  In
Sect.~\ref{sec-pola} the dependence of the sensitivity on longitudinal
beam polarisation is illustrated by a simple model.  In
Sect.~\ref{sec-hmc} we show analytically that one is insensitive to
one of the imaginary $\cp$ conserving couplings in the case of
longitudinal beam polarisation.  However, this particular coupling
becomes accessible with transverse beam polarisation.  In
Sect.~\ref{sec-resu} we present our numerical results, in
Sect.~\ref{sec-concl} our conclusions.


\section{Cross Section}
\label{sec-ampl}

\begin{figure}[h]
\vspace{0.6cm}
\centering
\includegraphics[totalheight=5.1cm]{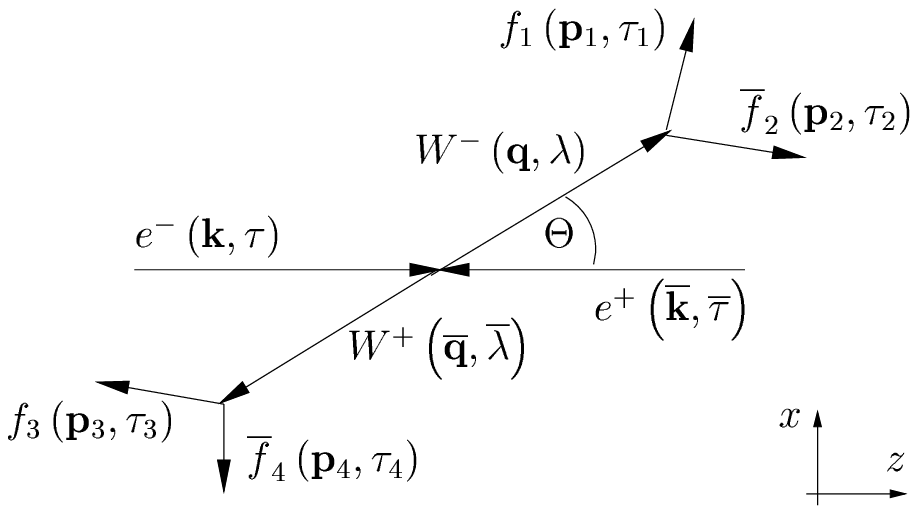}
\caption{\label{fig:pro}Momenta and helicities of the
	particles in the $e^+e^-$ c.m.~frame.}
\vspace{1.0cm}
\end{figure}

First we briefly recall the differential cross section of the process

\be
\label{eq:proc}
\begin{array}{llll}
e^- + e^+ \to \ & W^- & + \ & W^+ \\
		& \ \hookrightarrow f_1 + \overline{f}_2 &
		& \ \hookrightarrow f_3 + \overline{f}_4
\end{array}
\ee

\noindent
for arbitrary initial beam polarisations, where the final state
fermions are leptons or quarks.  Our notation for particle momenta and
helicities is shown in Fig.~\ref{fig:pro}.  Our coordinate axes are
chosen such that the $e^-$~momentum points in the positive
$z$-direction and the $y$~unit vector is given by \mbox{\( \hat{e}_y =
(\kvec \times \qvec) / |\kvec \times \qvec | \)}.

\begin{figure}
\centering
\includegraphics[totalheight=4.3cm]{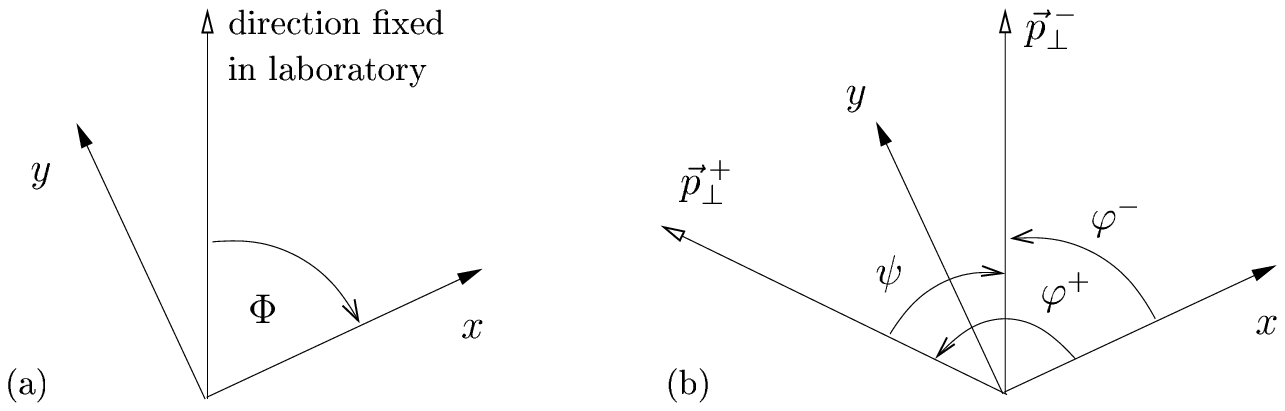}
\caption{\label{fig:angle}Definition of azimuthal angles.}
\end{figure}

In the $e^+e^-$ c.m.~frame and at a given c.m.~energy $\sqrt{s}$, a pure
initial state of longitudinally polarised $e^-$ and $e^+$ is uniquely
specified by the $e^-$ and $e^+$ helicities:
\be
|\he \hebar \rangle = |e^-(\kvec ,\he ) e^+(\kbarvec ,\hebar )
\rangle\;\;\;\;\;\; (\he,\hebar = \pm 1).
\ee
Note that fermion helicity indices are normalised to 1 throughout this
work.  A mixed initial state of arbitrary polarisation is given by a spin
density matrix
\be
\boldsymbol{\rho} = \sum_{(\he)} |\he \hebar \rangle \rho_{\rind}
\langle \heprime \hebarprime | ,
\ee
where $(\he)$ denotes summation over $\he$, $\hebar$, $\heprime$ and
$\hebarprime$, and the matrix elements satisfy \(\rho^{\ast}_{\rind} =
\rho_{\sind} \) and \(\sum_{\he,\hebar} \rho_{(\he \hebar )(\he \hebar
)} = 1 \).  We define the cross section operator
\be
\boldsymbol{d\s} = \sum_{(\he )} |\heprime \hebarprime \rangle
d\s_{\sind} \langle \he \hebar |
\ee
by requiring the differential cross section for arbitrary \(
\boldsymbol{\rho} \) to be
\be
d\s |_{\rho} = {\textstyle \rm tr}(\boldsymbol{d\s \rho}) =
\sum_{(\he )} d\s_{\sind}\, \rho_{\rind}. \label{eq:trace}
\ee
The matrix \( d\s_{\sind} \) is given by
\be
d\s_{\sind} = \frac{1}{2s} \int d\Gamma \, \langle f | \tr | \he
\hebar \rangle \, \langle f | \tr | \heprime \hebarprime \rangle
^{\ast}, \label{eq:dcr}
\ee
where we neglect the electron mass in the flux factor.  Here
$\tr$ is the transition operator, \( |f\rangle = | f_1\left( {\bf
p}_1, \he_1 \right)\overline{f_2}\left( {\bf p}_2, \he_2
\right)f_3\left( {\bf p}_3, \he_3 \right)\overline{f_4}\left( {\bf
p}_4, \he_4 \right) \rangle \) the final state and
\begin{equation}
d\Gamma = \left(
\prod_{i = 1}^4 \frac{d^3p_i}{(2\pi)^3 2p_i^0} \right) (2\pi)^4
\delta^{(4)} (k + \overline{k} - \sum_i p_i)
\end{equation}
the usual phase space measure for final states.  Using the
narrow-width approximation for the $W$s the result is
\begin{eqnarray}
d\s_{\sind} & = & R \sum_{(\lambda)} d{\cal P}_{\sind}^{\pind} \,
d{\cal D}^{}_{\lprime\lambda} \, d\overline{\cal D}_{\lbarprime\lbar}\;\,, \\
R & = & \frac{\beta}{2^{18}\pi^6 (m_W\Gamma_W )^2 s}.
\end{eqnarray}
Here $m_W$ is the $W$~boson mass, $\Gamma_W$ its width and \( \beta =
\sqrt{1 - 4 m_W^2 / s} \) its velocity in the $e^+e^-$ c.m.~frame.
The sum $(\lambda )$ runs over $\lambda$, $\lprime$, $\lbar$ and
$\lbarprime$.  Furthermore
\be
d{\cal P}_{\sind}^{\pind} = d(\cos\!\Theta)\:d\Phi\;\langle \lambda
\lbar, \Theta | \tr | \he \hebar \rangle \, \langle \lprime
\lbarprime, \Theta | \tr | \heprime \hebarprime \rangle ^{\ast}
\label{eq:prod}
\ee
is the differential production tensor for the $W$~pair and
\begin{eqnarray}
d{\cal D}_{\lprime\lambda} & = & d(\cos\!\vartheta)\,d\varphi\;\langle
f_1 \overline{f_2} | \tr | \lambda \rangle \, \langle
f_1\overline{f_2} | \tr | \lprime \rangle ^{\ast}, 
\nonumber \\
d\overline{\cal D}_{\lbarprime\lbar} & = &
d(\cos\!\overline{\vartheta})\,d\overline{\varphi}\;\langle f_3
\overline{f_4} | \tr | \lbar \rangle \, \langle f_3\overline{f_4} |
\tr | \lbarprime \rangle ^{\ast} \label{eq:deca}
\end{eqnarray}
are the tensors of the $W^-$ and $W^+$~decays in their respective
c.m.~frames.  Note that in the narrow-width approximation the
intermediate $W$s are treated as on-shell.  We define the $W$ helicity
states which occur on the right hand side of (\ref{eq:prod}) in the
frame of Fig.~\ref{fig:pro}, i.e.\ we choose the \prpr\ scattering
plane as the $x$-$z$-plane and define $\Theta$ to be the polar angle
between the $W^-$ and $e^-$ momenta.  We choose a fixed direction
transverse to the beams in the laboratory.  By $\Phi$ we denote the
azimuthal angle between this fixed direction and the \prpr\ scattering
plane (see Fig.~\ref{fig:angle}(a)).
The respective frames for the decay tensors (\ref{eq:deca}) are
defined by a rotation by $\Theta$ about the $y$-axis of the frame in
Fig.~\ref{fig:pro} (such that the $W^-$~momentum points in the
positive $z$-direction) and a subsequent rotation-free boost into the
c.m.~system of the corresponding $W$.  The spherical coordinates are
those of the $f_1$ and $\overline{f_4}$ momentum directions,
respectively.  In its rest frame, the quantum state of a $W$~boson is
determined by its polarisation.  For real $W$s we take as basis the
eigenstates of the spin operator $S_z$ with the three eigenvalues \(
\lambda = \pm 1,0 \).  For off-shell $W$s a fourth, scalar
polarisation occurs but is suppressed by \(m_f/m_W\) in the decay
amplitude.  

Neglecting the electron mass we have in the SM
\be
d\s_{\sind} = 0 \qquad \qquad \mbox{for~} 
     \he = \hebar \mbox{~or~} \heprime = \hebarprime ,	
  \label{eq:zero}
\ee
which we will use in the following (this point is further discussed in
Sect.~\ref{ssec-disc}).  At a linear collider the initial beams are
uncorrelated so that their spin density matrix factorises, i.e.\
\be
\rho_{\rind} = \rho_{\he \heprime}
\overline{\rho}{}_{\, \hebar\,\hebarprime}\;\,, \label{eq:fact}
\ee
where \( \rho_{\he \heprime} \) and
\( \overline{\rho}_{\hebar\,\hebarprime} \) are the two Hermitian and
normalised spin density matrices of $e^-$ and $e^+$ respectively.  We
parameterise these matrices by
\be
\rho_{\he \heprime} = \frac{1}{2} \left( \mathbbm1 +  \vec{p}\,^-
\cdot \vec{\s}  \right)_{\he \heprime}\;\,,\;\;\;
\overline{\rho}_{\,\hebar\,\hebarprime} = \frac{1}{2} 
\left( \mathbbm1 - \vec{p}\,^+ \cdot \vec{\s}^{\,\ast}
                         \right)_{\hebar\,\hebarprime}\;\,,
\label{eq:dens}
\ee
with
\be
\vec{p}\,^{\pm} = P_t^{\pm} \left( \begin{array}{c} 
					\cos{\varphi^{\pm}} \\
					\sin{\varphi^{\pm}} \\
					0
					\end{array} \right) 
		+ P_l^{\pm} \left( \begin{array}{c}
					0 \\
					0 \\
					\mp 1
					\end{array} \right), 
\label{eq:polvec}
\ee
where \(0 \le \varphi^{\pm} < 2 \pi\), and the vector components of
$\vec{\s}$ are the Pauli matrices.  The degrees $P_t^{\pm}$ of
transverse and $P_l^{\pm}$ of longitudinal polarisation obey the
relations \( (P_t^{\pm})^2 + (P_l^{\pm})^2 \le 1\) and \( P_t^{\pm}
\ge 0 \).  The components of \(\vec{p}\,^{\pm}\) in~(\ref{eq:polvec})
refer to the frame of \mbox{Fig.~\ref{fig:pro}}.  Note that choosing
the same reference frame for $\vec{p}\,^-$ and $\vec{p}\,^+$ while
specifying each spinor in its respective helicity basis results in
different forms of the density matrices in~(\ref{eq:dens}).  The
relative azimuthal angle \( \psi = \varphi^- - \varphi^+ \) between
\(\vec{p}\,^-\) and \(\vec{p}\,^+\) is fixed by the experimental
conditions, whereas the azimuthal angle $\varphi^-$ of  $\vec{p}\,^-$
in the system of Fig.~\ref{fig:pro} depends on the final state (see
Fig.~\ref{fig:angle}(b)).  For the case where \( P_t^- \ne 0 \) we
choose the transverse part of the vector \(\vec{p}\,^-\) as fixed
direction in the laboratory.  Then we have \( \Phi = - \varphi^- \).
Using~(\ref{eq:trace}) and~(\ref{eq:zero}) to (\ref{eq:polvec}), we
obtain the differential cross section for arbitrary polarisation:
\begin{eqnarray}
d\s |_{\rho} & = & \frac{1}{4}\, \bigg\{ (1 + P_l^-)(1 - P_l^+)\,
 d\s_{(+-)(+-)}
\nonumber \\
 & & \hspace{0.3em} {}+ (1 - P_l^-)(1 + P_l^+)\, d\s_{(-+)(-+)} 
\phantom{\bigg\{ \bigg\}}
\nonumber \\
 & & \hspace{0.3em} {}- 2 P_t^- P_t^+ 
	\Big[ \, {\rm Re}\, d\s_{(+-)(-+)}\, \cos{(\psi + 2 \Phi)} 
\phantom{\bigg\{ \bigg\}}
\nonumber \\
 & & \hspace{4.2em}
             {}+ {\rm Im}\, d\s_{(+-)(-+)}\, \sin{(\psi + 2 \Phi)}
 \Big] \bigg\}.	
\label{eq:genpol} 
\end{eqnarray}
In the absence of transverse polarisation, (\ref{eq:genpol}) is
independent of $\Phi$ due to rotational invariance.

\begin{figure}
\centering
\includegraphics[totalheight=3cm]{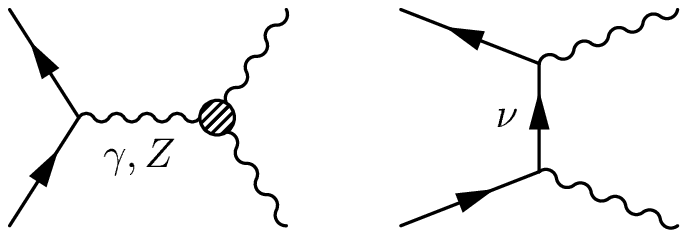}
\caption{\label{fig:feynm} Feynman diagrams for the process
	\prpr\ with anomalous TGCs.}
\end{figure}

In our analysis we evaluate the matrix elements in~(\ref{eq:prod}) at
tree level in the SM, replacing the $\gamma WW$ and $ZWW$ vertices by
their most general forms allowed by Lorentz invariance.  The
corresponding Feynman diagrams are shown in Fig.~\ref{fig:feynm}.
New physics may also lead to deviations from the SM values at the
fermion-boson vertices~\cite{Casalbuoni:1999hd}.  For these vertices
we however retain the SM expressions, following the rationale that at
present they are confirmed experimentally to a much higher precision
than the triple gauge couplings.  We remark that there are scenarios
of physics beyond the SM where such a treatment is not adequate, since
the process \prpr\ can receive non-standard contributions that cannot
be expressed in terms of anomalous fermion-boson or three-boson
couplings (an example are box graph contributions in supersymmetric
theories~\cite{Arhrib:1995dm,Hahn:2000bf}).  Such effects can still be
parameterised within a more general form factor
ansatz~\cite{Hagiwara:1996kf}.  If they are important, an analysis in
terms of only TGCs will not give a correct picture of the underlying
physics, but it will still correctly signal a deviation from the SM.
We finally remark that radiative corrections in the SM itself can be
included in the analysis procedure we develop here (see
Sect.~\ref{sec-opti}).  The purpose of the present study is however to
investigate the pattern of sensitivity to TGCs and its dependence on
beam polarisation, and for this purpose it should be sufficient to
take the SM prediction at tree level.
 
For the three-boson vertices we use the parameterisation~(2.4) in
Ref.~\cite{Hagiwara:1986vm}, which we express in terms of the coupling
parameters $g_1^V$, $\kappa_V$, $\lambda_V$, $g_4^V$, $g_5^V$,
$\tilde{\kappa}_V$ and $\tilde{\lambda}_V$ (\(V = \gamma, Z\)) by
means of the transformation~(2.5) in the same reference.  In other
words, we understand these parameters as form factors of the
three-boson vertices, which depend on the boson virtualities and can
take complex values, and not as coupling constants in an effective
Lagrangian, which by definition are energy independent and real-valued.

For a given $e^-$ beam helicity $\he$ the process~(\ref{eq:proc}) is not
sensitive to all couplings, but only to the linear combinations
$g_1^L$, $\kappa_L$, etc.\ for left handed ($\he = -1)$ and $g_1^R$,
$\kappa_R$, etc.\ for right handed ($\he = 1)$ electrons, where
\begin{eqnarray}
g_1^L & = & 4\sin^2\theta_W\, g_1^{\gamma} + (2 - 4\sin^2\!\theta_W )\,
\xi\, g_1^Z,	\nonumber \\
g_1^R & = & 4\sin^2\theta_W\, g_1^{\gamma} - 4\sin^2\!\theta_W\, \xi\,
g_1^Z,	\label{eq:lrgz}
\end{eqnarray}
and similarly for the other
couplings~\cite{Hagiwara:1986vm,Diehl:1993br}.  Here $e$ denotes the
positron charge, $\theta_W$ the weak mixing angle, and \( \xi =
s/(s-m_Z^2) \) the ratio of the $Z$ and photon propagators.  The
parameterisation~(\ref{eq:lrgz}) will in the following be called the
L(eft)-R(ight)-basis.

The expressions of the amplitudes can be found
in~\cite{Hagiwara:1986vm}.  For convenience of the reader we rewrite
the $W^-W^+$ production part in terms of the LR-basis.  The matrix
element of~(\ref{eq:prod}) is given by
\be
\langle \lambda \lbar, \Theta | \tr | \he \hebar \rangle = - \sqrt{2} e^2
{\cal M} (\he; \lambda, \lbar; \Theta ) \, \eta \, d_{\Delta \he,
\Delta \lambda}^{J_0} (\Theta), \label{eq:pram}
\ee
where \( \eta = \Delta \he (-1)^{\lbar} \), \( \Delta \he =
\frac{1}{2} (\he - \hebar) \), \( \Delta \lambda = \lambda - \lbar \),
and \( J_0 = \max(|\Delta \he |, |\Delta \lambda |) \).  The definition
of the $d$-functions and our spinor conventions, as well as the SM
matrix elements for the $W$ decays in (\ref{eq:deca}) are listed in
the Appendix.  The production amplitude \( {\cal M} = {\cal M}^{\rm
TGC} + {\cal M}^{\nu} \) consists of two terms given by
\begin{eqnarray}
{\cal M}^{\rm TGC} (\he; \lambda, \lbar; \Theta) & = & -
\frac{\beta}{4 \sin^2 \theta_W} A_{\lambda \lbar}^{\he}\;\,,
\label{eq:mtgcs} \\
{\cal M}^{\nu} (\he; \lambda, \lbar; \Theta) & = & \frac{1}{2 \sin^2
\theta_W \beta} \delta_{\he, -1} \left( B_{\lambda \lbar} - \frac{1}{1
+ \beta^2 - 2 \beta \cos \Theta} C_{\lambda \lbar} \right).
\label{eq:mnu}
\end{eqnarray}
The expressions for $A_{\lambda \lbar}^{\he}$ (which contains the left
handed couplings for \( \he = -1 \) and the right handed ones for \(
\he = +1 \)) and for $B_{\lambda \lbar}$ and $C_{\lambda \lbar}$ are
listed in Table~\ref{tab:amp2}.  Since the vector bosons carry spin 1,
the TGCs do not contribute to the $WW$ helicity combinations \( (+-)
\) and \( (-+) \).  For the other helicity amplitudes, the largest
power of~$\gamma$ in the coefficients $A_{\lambda \lbar}^{\he}$
coincides with the number of longitudinal $W$s.  An exception are the
couplings $\lambda_a$ and $\smash{\tilde{\lambda}_a}$, which
correspond to dimension-six operators in the effective Lagrangian
(cf.~\cite{Hagiwara:1986vm}) and occur with an additional factor
of~$\gamma^2$.  Note that the largest kinematical factors in
$A_{\lambda \lbar}^{\he}$ behave like $\gamma^2$ at high energies, in
contrast to the basis of form factors $f_i$ used in Table~4
of~\cite{Hagiwara:1986vm}, where huge factors of $\gamma^4$ appear.
In the SM at tree level one has
\be
g_1^V = 1,\;\;\; \kappa_V = 1 \;\;\;\;\;\;\;(V = \gamma, Z)
\label{eq:coupsm}
\ee
and all other couplings equal to zero. For the anomalous parts of the
couplings we write $\Delta g_1^V = g_1^V - 1$ and $\Delta \kappa_V =
\kappa_V - 1$ as usual.

\begin{table}
\begin{center}
\leavevmode
\caption{\label{tab:amp2} Coefficients $A_{\lambda \lbar}^{\he}$,
$B_{\lambda \lbar}$ and $C_{\lambda \lbar}$ of the matrix elements
(\protect\ref{eq:mtgcs}) and (\protect\ref{eq:mnu}).  The indices of
the couplings are $a=L$ for $\tau=-1$ and $a=R$ for $\tau=+1$.  The
relativistic factors are defined by \( \gamma = \sqrt{s} / 2 m_W \)
and \( \beta = (1 - 4 m_W^2 / s)^{1/2} \).}
\[
\begin{array}{cccc}
\hline
&&& \\[-.45cm]
(\lambda \lbar) & A_{\lambda \lbar}^{\he} & ~~~ B_{\lambda \lbar} ~~~ &
C_{\lambda \lbar} \\[.1cm]
\hline
& \\[-.45cm]
(+-),(-+) & 0 & 0 & 2 \sqrt{2}\,\beta \\[.1ex]
(+0)	  & \gamma\, [g_1^a + \kappa_a + \lambda_a - i g_4^a + \beta g_5^a + i
\beta^{-1} (\tilde{\kappa}_a - \tilde{\lambda}_a)] & 2\gamma & 2(1
+\beta)/\gamma \\[.15ex]
(0-) 	  & \gamma\, [g_1^a + \kappa_a + \lambda_a + i g_4^a + \beta g_5^a - i
\beta^{-1} (\tilde{\kappa}_a - \tilde{\lambda}_a)] & 2\gamma & 2(1
+\beta)/\gamma \\[.15ex]	
(0+)	  & \gamma\, [g_1^a + \kappa_a + \lambda_a + i g_4^a - \beta g_5^a + i
\beta^{-1} (\tilde{\kappa}_a - \tilde{\lambda}_a)] & 2\gamma & 2(1
-\beta)/\gamma \\[.15ex]
(-0)	  & \gamma\, [g_1^a + \kappa_a + \lambda_a - i g_4^a - \beta g_5^a - i
\beta^{-1} (\tilde{\kappa}_a - \tilde{\lambda}_a)] & 2\gamma & 2(1
-\beta)/\gamma \\[.15ex]
(++)	  & g_1^a + 2\gamma^2\lambda_a + i \beta^{-1} \tilde{\kappa}_a -
i(\beta^{-1} + 2\gamma^2\beta) \tilde{\lambda}_a & 1 & 1/\gamma^2 \\[.15ex]
(--)	  & g_1^a + 2\gamma^2\lambda_a - i \beta^{-1} \tilde{\kappa}_a +
i(\beta^{-1} + 2\gamma^2\beta) \tilde{\lambda}_a & 1 & 1/\gamma^2 \\[.15ex]
(00) 	  & g_1^a + 2\gamma^2\kappa_a & 2\gamma^2 & 2/\gamma^2 \\[.15ex]
\hline
\end{array}
\]
\end{center}
\end{table}

\begin{figure}
\begin{center}
\leavevmode
\includegraphics[totalheight=11.5cm]{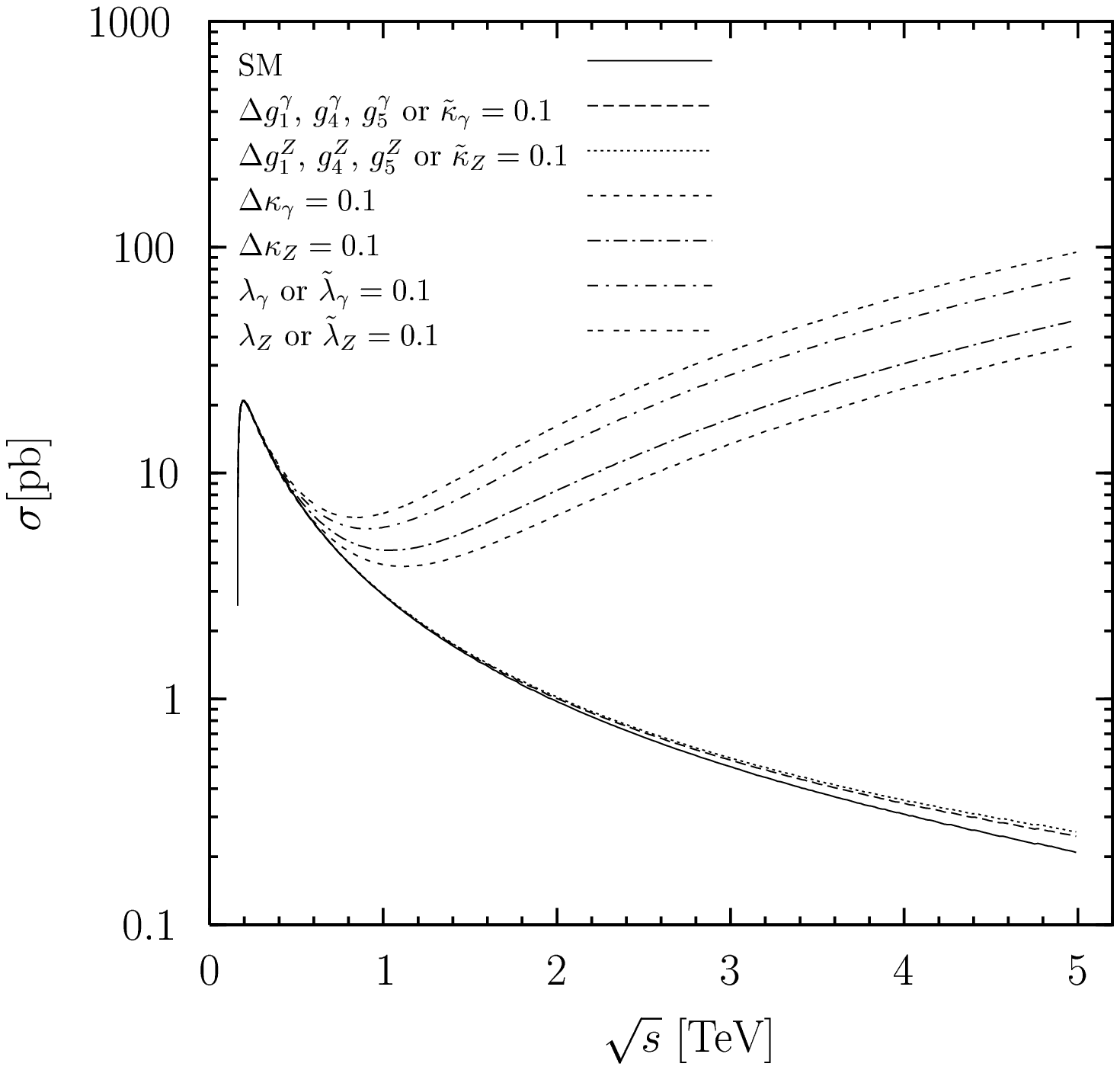}
\end{center}
\caption{\label{fig:norm} Total cross section with unpolarised beams
as a function of the c.m.~energy in the SM and for {\em one} anomalous
coupling differing from zero.  Some curves coincide as explained in
the text.}
\end{figure}

A detailed discussion of the differential cross section is given
in~\cite{Hagiwara:1986vm}.  Here we only point out some salient
features of the high-energy limit.  In~Fig.~\ref{fig:norm} we show the
total cross section for unpolarised beams as a function of~$\sqrt{s}$.
It rises rapidly from threshold up to a maximum of about 20~pb at
$\sqrt{s}\approx$~200~GeV, and in the SM decreases for higher
c.m.~energies.  In the SM each $Z$-coupling is equal to the
corresponding photon coupling.  Since \( \xi = 1 + O(\gamma^{-2}) \)
the \mbox{L-couplings} are then of order 1 and the \mbox{R-couplings}
of order \( \gamma^{-2} \).  For \( \he = + 1\) this leads to a high
energy behaviour of at most \( {\cal M} \sim O(1) \).  For \( \he = -
1\) we have \( g_1^L = \kappa_L = 2 \xi \) in the SM, and the
coefficients $A_{\lambda \lbar}$ and $B_{\lambda \lbar}$ only differ
by a factor $2 + O(\gamma^{-2})$ according to Table~\ref{tab:amp2}.
As they occur with different sign in (\ref{eq:mtgcs}) and
(\ref{eq:mnu}) this again results in a high-energy behaviour ${\cal M}
\sim O(1)$, except for very forward $W^-$ momentum where there is an
enhancement by the propagator factor $(1 + \beta^2 - 2 \beta \cos
\Theta )^{-1}$.  Altogether, these gauge cancellations preserve the
unitarity of the SM.  We also plot in Fig.~\ref{fig:norm} the total
cross section for {\em one} anomalous coupling differing from zero.
At high energies each coupling mainly contributes via the $W$~helicity
amplitude where it occurs with the highest power of~$\gamma$, i.e.\
either linearly or quadratically according to Table~\ref{tab:amp2}.
At sufficiently high energy, the square of an anomalous term dominates
over its interference term with the SM amplitude.  In the limit \(
\beta \rightarrow 1 \) the couplings $g_1$, $g_4$, $g_5$ and
$\tilde{\kappa}$ enter with a factor $\gamma$, whereas $\kappa$,
$\lambda$, $\tilde{\lambda}$ enter with a factor $\gamma^2$, which
explains their different behaviour in Fig.~\ref{fig:norm}.  Some
couplings have equal coefficients in this limit, which leads to a
degeneracy of the curves.  We also remark that even if more than one
anomalous coupling differs from zero, anomalous amplitudes belonging
to couplings of different $C$ or $P$ eigenvalue do not interfere in
the total cross section with unpolarised beams
(cf.~Sect.~\ref{ssec-disc}).

\section{Optimal Observables}
\label{sec-opti}

In this section we discuss how the method of optimal integrated
observables is applied to our case.  In an experiment one measures the
differential cross section 
\begin{equation}
 S = d\s |_\rho / d\phi ,
\label{eq:diffpol}
\end{equation}
where $\phi$ denotes the set of all measured phase space variables.
We distinguish between the information from the total cross section \(
\sigma = \int d\sigma \) and from the normalised distribution $S/\s$
of the events. We first investigate how well TGCs can be extracted
from the latter, and then use $\sigma$ to get constraints on those
directions in the space of couplings to which the normalised
distribution is not sensitive.

Since $S$ is a polynomial of second order in the anomalous couplings
(defined as the couplings minus their tree-level values in the SM) one
can write
\be
S = S_0 + \sum_i S_{1i} \h_i + \sum_{i,j} S_{2ij} \h_i \h_j\;\,.
\label{eq:distri}
\ee
Here $S_0$ is the tree-level cross section in the SM, and each $\h_i$
is the real or imaginary part of an anomalous TGC.  Thus there are 28
real parameters $h_i$ altogether.  In the remainder of this section
$i$ and $j$ each run from 1 to 28.

One way to extract the coupling constants from the measured
distribution~(\ref{eq:distri}) is to look for a suitable set of
observables $\obi (\phi)$ whose expectation values
\be
	E[\obi ] = \frac{1}{\s} \int \!d\phi S \obi	\label{eq:exdef}
\ee
are sensitive to the dependence of~$S$ on the couplings~$\h_i$.  Given
the present experimental constraints, we expect these variations to be
small.  Thus we expand the expectation values to first order in~$\h_i$:
\begin{eqnarray}
E[\obi ]   & = & E_0[\obi ] + \sum_j c_{ij} \h_j + O(\h^2),
\label{eq:exexp}
\end{eqnarray}
with
\begin{eqnarray}
E_0[\obi ] & = & \frac{1}{\s_0} \int \!d\phi\, S_0 \obi\;, \\[.5ex]
c_{ij}     & = & \frac{1}{\s_0} \int \!d\phi\, \obi S_{1j} 
               - \frac{\sigma_{1j}}{\s_0^{2}} \int \!d\phi\, S_0 \obi\;,	
\label{eq:cij} \\[.5ex]
\sigma_{1j}   & = & \int \!d\phi\, S_{1j}\;.	
 \label{eq:tcf}
\end{eqnarray}
Here $E_0[\obi ]$ is the expectation value for zero anomalous
couplings, whereas $c_{ij}$ gives the sensitivity of $E[\obi ]$ to
$\h_j$.  Solving (\ref{eq:exexp}) for the set of the $\h_j$ we get
estimators for the anomalous couplings, whose covariance matrix is
given by
\be
V(\h) = \frac{1}{N}\, c^{-1} \covo\, (c^{-1})^T ,	\label{eq:vh}
\ee
where we use matrix notation.  Here $N$ is the number of events, and
\be
\covo_{ij} = \frac{1}{\s_0}
             \int \!d\phi S_0 \obi \obj - E_0[\obi ] E_0[\obj
]\;\;+\;\;O(h)	\label{eq:covo}
\ee
is the covariance matrix of the observables, which we have Taylor
expanded around its value in the SM.  As observables we choose
\be
\obi = \frac{S_{1i}(\phi)|_\rho}{S_0(\phi)|_\rho}. 
\label{eq:defob}
\ee
{}From (\ref{eq:cij}) and (\ref{eq:covo}) one obtains for this
specific choice
\be
\covo = c + O(h),	\label{eq:ccov}
\ee
and therefore
\be
V(\h) = \frac{1}{N }\, c^{-1} + O(h).	\label{eq:geo}
\ee
The observables~(\ref{eq:defob}) are ``optimal'' in the sense that for
\( \h_i \rightarrow 0 \) the errors~(\ref{eq:geo}) on the couplings
are as small as they can be for a given probability
distribution.\footnote{For details on this so called
Rao-Cram\'er-Fr\'echet bound see e.g.\
\protect\cite{bib:cram,Groom:in}.}
Apart from being useful for actual experimental analyses, the
observables~(\ref{eq:defob}) thus provide insight into the sensitivity
that is at best attainable by {\em any} method, given a certain
process and specified experimental conditions.  In the case of one
parameter this type of observable was first proposed
in~\cite{Atwood:1991ka}, the generalisation to several parameters was
made in~\cite{Diehl:1993br}.  Moreover, it has been shown that optimal
observables are unique up to a linear
reparameterisation~\cite{Diehl:1997ft}.  We further note that phase
space cuts, as well as detector efficiency and acceptance have no
influence on the observables being ``optimal'' in the above sense,
since their effects drop out in the ratio~(\ref{eq:defob}).  This is
not the case for detector resolution effects, but the
observables~(\ref{eq:defob}) are still close to optimal if such
effects do not significantly distort the differential
distributions~$S_{1i}$ and~$S_0$ (or tend to cancel in their ratio).
To the extent that they are taken into account in the data analysis,
none of these experimental effects will bias the estimators.

In the present work we use the method of optimal observables in the
linear approximation valid for small anomalous couplings.  But we
emphasise that the method has been extended to the fully non-linear
case where one makes no a priori assumptions on the size of anomalous
couplings in~\cite{Diehl:1997ft}.  Such a non-linear analysis for real
data was presented in~\cite{bib:dham} and turned out to be very
convenient and highly efficient.

Given the projected accuracy at linear colliders, it will in general
be necessary to take into account radiative corrections to the process
$e^-e^+\to f_1\overline{f_2}f_3\overline{f_4}$ within the SM, which
have been worked out in detail in the
literature~\cite{Grunewald:2000ju}.  One possibility to include them
in searches for non-standard TGCs would be to ``deconvolute'' these
corrections.  For this write the one-loop corrected differential cross
section in the SM as
\begin{equation}
S_{\rm SM,\, corr}(\phi) = \int \!d\phi'\, S_0(\phi')  F(\phi',\phi) ,
\end{equation}
where $S_0$ is the tree-level expression, and the integral kernel can
e.g.\ be obtained from an event generator by generating events
according to both $S_{\rm SM,\, corr}$ and $S_0$.  Approximating the
true physical cross section as
\begin{equation}
S_{\rm phys}(\phi) = \int \!d\phi'\, S(\phi')  F(\phi',\phi),
  \label{conv-approx}
\end{equation}
with $S$ given as in~(\ref{eq:distri}), one could invert this
convolution bin by bin, and then extract the couplings~$h_i$ from the
deconvoluted Born level cross section $S$ as described before.  The
error made in~(\ref{conv-approx}) is that the SM radiative corrections
encoded in $F$ do of course not apply to the anomalous part $S - S_0$
of the cross section, but this error is of order $h_i$ times the weak
coupling constant $\alpha_{\mathrm{w}}$.  Should effects beyond the SM
be found in such an analysis, one would in a second step have to
consider more sophisticated methods to quantitatively disentangle them
from SM radiative corrections.


\subsection{Simultaneous Diagonalisation}
\label{sec-simul}

Although discrete symmetry properties (see Sect.~\ref{ssec-disc})
reduce the 28$\times$28 matrices $V(\h)$, $c$, etc.\ to blocks of
8$\times$8 and 6$\times$6~matrices, these blocks still contain many
non-negligible off-diagonal entries.  To make explicit how sensitive
the process is to each direction in the space of couplings and to
identify the role of polarisation we need to know directions and
lengths of the principal axes of the error ellipsoid defined
by~(\ref{eq:vh}), i.e.\ we have to know the eigenvalues and
eigenvectors of $V(\h)$.  Using optimal observables, to leading order
in the $h_i$, the three matrices $V(\h)$, $c$ and \covo\ are
automatically diagonalised simultaneously due to~(\ref{eq:ccov})
and~(\ref{eq:geo}).  This means that after such a transformation each
observable is sensitive to exactly one coupling, and the observables
as well as the estimators of the couplings are statistically
independent.  Since $V(\h)$ is symmetric the diagonalisation could be
achieved by an orthogonal transformation.  Following the proposal
of~\cite{Diehl:1997ft} we take however a different choice and
transform simultaneously $V(\h)$ into diagonal form and the normalised
second-order part of the total cross section into the unit matrix:
\be
\hat{\s}_{2ij} \equiv \frac{1}{\s_0} \int \!d\phi\, S_{2ij} 
\;\rightarrow\; \delta_{ij}\;.
\label{eq:trsg2}
\ee
This can always be done since $\hat{\s}_2$ is symmetric and positive
definite.  We therefore arrive at the following prescription for the
transformation of the couplings (using vector and matrix notation):
\begin{eqnarray}
\boldsymbol{\h} & \rightarrow & \boldsymbol{\h'} 
= A^{-1} \boldsymbol{\h},
\label{eq:gtrafo} \\[0.1em]
V(\h)^{-1} 	& \rightarrow & V(\h')^{-1} = A^T V(\h)^{-1} A = {\rm
diag}\left( (\delta \h_1')^{-2}, (\delta \h_2')^{-2}, \ldots, (\delta \h_{28}')^{-2} \right), 
\label{eq:vgtrafo} \\
\hat{\sigma}_2	& \rightarrow & \hat{\sigma}_2' = A^T \hat{\sigma}_2\, A
= \mathbbm1 ,	
\label{eq:strafo}
\end{eqnarray}
where $\delta h'_i$ are the one-sigma errors on the new couplings.
This transformation exists and is unique up to permutations and a sign
ambiguity for each $\h'_i$.  Note that the matrix $A$ is in general
not orthogonal.  {}From (\ref{eq:distri}), (\ref{eq:defob}) and
(\ref{eq:gtrafo}) the transformation of all other quantities follows as
\begin{eqnarray}
\boldsymbol{S_1}	& \rightarrow & \boldsymbol{S'_1} = A^T
\boldsymbol{S_1},	
\nonumber \\
\boldsymbol{\cal O}	& \rightarrow & \boldsymbol{\cal O'} = A^T
\boldsymbol{\cal O},	
\nonumber \\
c         & \rightarrow & c'= A^T c\, A,
\nonumber \\
\covo     & \rightarrow & V({\cal O'}) = A^T \covo\, A.
\label{eq:alltrafo}
\end{eqnarray}
The meaning of~(\ref{eq:strafo}) is that all quadratic terms
contribute to the total cross section with equal strength:
\be
\sigma = \sigma_0 \left( 1+ \sum_{i = 1}^8 \hat{\sigma}'_{1i} \h'_i +
\sum_{i = 1}^{28} (\h'_i)^2 \right),	\label{eq:stotnorm}
\ee
where \( \hat{\s}_{1i}' = \sigma_0^{-1} \int \!d\phi\, S_{1i}' \).
Thus the anomalous couplings do not mix in $\sigma$ and are
``naturally'' normalised for the process which we consider.  This is
not true in the conventional basis, where changing different anomalous
couplings by the same amount has completely different effects on the
total cross section (see Fig.~\ref{fig:norm} in Sect.~\ref{sec-ampl}).

Moreover, the particularly simple form~(\ref{eq:stotnorm}) of $\sigma$
easily allows one to use the information from the total rate: it
constrains the couplings to lie between two hyperspheres in the space
of the $h_i'$, whose difference in radius depends on the measurement
error on $\sigma$.  Making in addition use of the sign ambiguity in
(\ref{eq:gtrafo}), one can for all $\hat{\sigma}'_{1i} \neq 0$ choose
the sign of $\h'_i$ such that $\hat{\sigma}'_{1i} > 0$.  This choice
is however not relevant for the analysis.

We finally note that the presented method of simultaneous
diagonalisation is quite similar to the way one analyses the normal
modes of a multi-dimensional harmonic oscillator in classical
mechanics~\cite{bib:gold}.  There the harmonic potential
(corresponding to $V$) is diagonalised with respect to the scalar
product defined by the kinetic energy (corresponding
to~$\hat{\sigma}_2$).


\subsection{Numerical Realisation}
\label{sec-real}

We now give some details of how the simultaneous diagonalisation can
be carried out numerically.  Although the procedure finally aims at
the disentanglement of the couplings as achieved
in~(\ref{eq:vgtrafo}), the numerical computation of~$V(\h)$
or~$V(\h)^{-1}$ from~(\ref{eq:vh}) needs the inverse of the matrix~$c$
or~$\covo$ and might therefore be unstable, because before the
diagonalisation one cannot single out those directions in parameter
space where the errors are large.  However, as $V(\h')$ and $V({\cal
O}')$ are simultaneously diagonal for our observables, we can
according to~(\ref{eq:ccov}) equally well compute the diagonal entries
of
\be
V({\cal O}') = c' = {\rm diag}(c_1', c_2', \ldots, c_{28}'),	
\label{eq:vprime}
\ee
and extract the errors on the couplings using~(\ref{eq:geo}):
\be
\delta \h_i' = \frac{1}{\sqrt{N c_i'}}.	\label{eq:delg1p}
\ee
Hence, using the shorthand notation \( V = \covo \) and \( V' =
V({\cal O}') \), we have to solve the \( n^2 + n \) equations
\begin{eqnarray}
A^T \hat{\sigma}_2 A	& = & \mathbbm1 ,	\nonumber \\
A^T V A			& = & V'	\label{eq:shortdiag}
\end{eqnarray}
for the $n^2$~entries of~$A$ and the $n$~diagonal elements of $V'$.
Since the multiplication of $S$ by a constant changes neither the
observables~(\ref{eq:defob}) nor~$V$ nor~$\hat{\s}_2$, the
matrices~$A$ and~$V'$ only depend on the normalised distribution of
the events but not on the total rate~$N$.  The latter enters the
errors on the transformed couplings only through the statistical
factor~$N^{-1/2}$ in~(\ref{eq:delg1p}).  {}From~(\ref{eq:shortdiag})
it follows that
\be
V A = \hat{\sigma}_2\, A\, V' .	\label{eq:gep}
\ee
This is a generalised eigenvalue problem, with the $c_i'$ being the
generalised eigenvalues and the columns of~$A$ being the generalised
eigenvectors.  The pair~$(V',A)$ is called the ``eigensystem''
of~(\ref{eq:shortdiag}).  A standard method for solving (\ref{eq:gep})
is to first perform a \emph{Cholesky}
decomposition~\cite{bib:goll,bib:numr}
\be 
\hat{\sigma}_2 = M M^T ,
\label{eq:chol} 
\ee
where $M$ is a lower triangular matrix, i.e.\ \( M_{ij} = 0 \) for \(
j > i \).  Algorithms for the computation of $M$ can be found in the
same references.  Then (\ref{eq:gep}) is equivalent to
\be
C X = X V',	\label{eq:cx}
\ee
where
\begin{eqnarray}
C & = & M^{-1} V ( M^{-1} ) ^T,	\label{eq:c} \\
X & = & M^T A.			\label{eq:x}
\end{eqnarray}
Equation~(\ref{eq:cx}) denotes a (usual) eigenvalue problem for the
matrix~$C$, whose eigenvalues are the same as the original ones and
whose eigenvectors are the columns of~$X$.  Since $C$ is symmetric,
(\ref{eq:cx}) can be solved and the eigenvectors are orthogonal with
respect to the standard scalar product (assuming non-degeneracy of the
eigenvalues).  Requiring the eigenvectors to be normalised to 1, we
have $n$~conditions
\be
X^T X = \mathbbm1 ,\label{eq:normone}
\ee
which together with the $n^2$~equations~(\ref{eq:cx}) are equivalent
to (\ref{eq:shortdiag}).  The generalised eigenvectors are obtained by
solving (\ref{eq:x}) for $A$.  The procedure has to be followed for
each initial state polarisation and c.m.~energy separately, leading in
general to different eigenvalues and transformation matrices; the
dependence of those quantities on polarisation is investigated in
Sect.~\ref{sec-pola}.  In each case we use the procedure iteratively,
i.e.\ once $A$ is obtained, we compute $V$ and $\hat{\s}_2$ for the
transformed observables and couplings, and then diagonalise
these---already approximately diagonal---matrices again.  We found
this to be essential to assure the numerical stability of the results.
A stable value was reached in most cases by the second evaluation and
at the latest by the fourth.  In all cases at least five evaluation
steps were carried out.  The numbers presented in Sect.~\ref{sec-resu}
were obtained by averaging over the results of several subsequent
steps where the stable value had already been reached.

We note that for situations where $\hat{\sigma}_2$ and $V$ have the
same block diagonal structure, the diagonalisation can of course be
carried out for each block separately.  This is relevant in the
presence of discrete symmetries.


\subsection{Discrete Symmetries}
\label{ssec-disc}

Let us now discuss the special role of the combined symmetry
operations $\cp$ and $\cpt$ in the context of our
reaction~\cite{Hagiwara:1986vm,Diehl:1993br,Diehl:1997ft}.  Here $C$
denotes charge conjugation, $P$ parity reversal, and $\widetilde{T}$
``na\"{\i}ve time reversal'', i.e.\ the reversal of all particle
momenta and spins without the interchange of initial and final state.
Under the condition that the initial state, as well as phase space
cuts and detector acceptance are invariant under a $\cp$
transformation, a $\cp$ odd observable gets a nonzero expectation
value only if $\cp$ is violated in the interaction.  Similarly, if the
initial state, phase space cuts and acceptance are invariant under
$\cpt$ followed by a rotation by 180$^\circ$ around an axis
perpendicular to the beam momenta, a nonzero expectation value of a
$\cpt$ odd observable implies the interference between absorptive and
nonabsorptive amplitudes in the cross section.  In terms of the
three-boson couplings one finds that to $O(h)$ the expectation values
of $\cp$ even (odd) observables only involve the $\cp$ conserving
(violating) couplings $g_1$, $\kappa$, $\lambda$, $g_5$ ($g_4$,
$\tilde{\kappa}$, $\smash{\tilde{\lambda}}$).  Similarly, $\cpt$ even
(odd) observables are to first order only sensitive to the real
(imaginary) parts of the coupling para\-meters.  The coefficient
matrix $c$ is thus block diagonal in the following groups of
observables:
\begin{quote}
(a):~~$\cp$ and \cpt\ even, \\
(b):~~$\cp$ even and \cpt\ odd, \\
(c):~~$\cp$ odd and \cpt\ even, \\
(d):~~$\cp$ and \cpt\ odd.
\end{quote}
One further finds that the first-order terms $\sigma_{1i}$ in
the integrated cross section can only be non-zero for couplings of
class (a).

\begin{figure}
\begin{center}
\leavevmode
\includegraphics[totalheight=2.8cm]{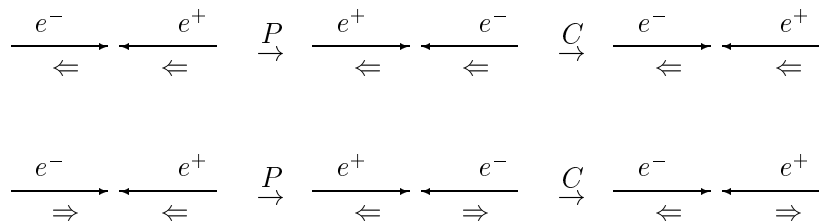}
\end{center}
\caption{\label{fig:CP} Effect of a $\cp$ transformation on an
$e^+e^-$ state with helicities aligned (top) or antialigned (bottom).}
\end{figure}

The above requirements on the initial $e^+e^-$ state are nontrivial in
the case of polarised beams.  Since charge conjugation exchanges $e^+$
and $e^-$, a $\cp$ invariant spin density matrix requires $\vec{p}\,^-
= \vec{p}\,^+$, in particular $P^-_t = P^+_t$ and $P_l^- = -P_l^+$.
Under the same conditions the density matrix is also invariant under
$\cpt$ times a rotation by 180$^\circ$ around $\vec{k}\times
\vec{p}\,^-$.  Let us investigate the situation where these
requirements are not satisfied.  {}From Fig.~\ref{fig:CP} we see that
the $e^+e^-$ states where the beam helicities are aligned are $\cp$
eigenstates.  But the antialigned ($\Delta\tau = 0$) states are
interchanged under $\cp$ and therefore are not eigenstates.  On the
other hand, the amplitudes for these helicity combinations are
proportional to the electron mass in the SM (supplemented by the most
general TGCs).  They are thus generically suppressed by $m_e / M_W$
compared with the amplitudes for aligned beam helicities ($\Delta\tau
= \pm 1$).\footnote{To be precise, one must exclude final states $e^+
\nu_e\, e^- \bar{\nu}_e$, where nonresonant graphs contribute in which
the initial $e^+$ and $e^-$ do not annihilate.}
With transverse beam polarisation, the two types of amplitudes can
interfere, giving small $m_e /M_W$ effects in the cross section.  With
purely longitudinally polarised beams they do not interfere, so that
effects due to the $\Delta\tau=0$ combinations of $e^+e^-$ are of
order $(m_e /M_W)^2$ and thus beyond experimental accuracy.  We remark
that the same holds e.g.\ in the minimal supersymmetric extension of
the SM, where left- and right handed leptons as well as their
superpartners only mix with a strength proportional to the lepton
mass.  To investigate what can happen in more generic models is beyond
the scope of this work.

For general beam polarisation, a nonzero mean value of a $\cp$ odd
observable can be generated by genuine $\cp$ violation in the
reaction, or by the $\cp$ odd part of the spin density matrix in the
initial state.  According to the above estimate, the latter would
require nonstandard physics to be experimentally visible, especially
for longitudinal beam polarisation, and thus be interesting in its own
right.  Similarly, a nonzero result for a $\cpt$ odd observable can
originate from absorptive parts in the process, or from effects of the
amplitudes with zero total helicity of the initial beams.  If any such
effects were observed, one could in a next step investigate their
dynamical origin, using their different dependences on the beam
polarisation.  One possibility is to use that in the absence of
$\Delta \tau = 0$ amplitudes the cross section depends on transverse
beam polarisation only via the product $P_t^- P_t^+$, as seen in
(\ref{eq:genpol}).  If $P_t^+ = 0$, the cross section is then
independent of $P_t^-$ and $\varphi^-$, whereas the interference
between $\Delta \tau = 0$ and $\Delta \tau = \pm 1$ amplitudes leads
to terms with $P_t^- \, \cos \varphi^-$ and $P_t^- \, \sin \varphi^-$
to the cross section, which can experimentally be identified via their
angular dependence on $\varphi^-$.  A possibility to search for the
presence of $\Delta \tau = 0$ amplitudes with only longitudinal beam
polarisation will be discussed at the end of Sect.~\ref{sec-pola}.

Returning to purely longitudinal beam polarisation and the assumption
that \mbox{$\Delta \tau=0$} amplitudes are negligible, we remark that
constraints similar to the ones discussed above cannot be derived for
$C$ and $P$ separately, since neutrino exchange maximally violates
both symmetries.  One can however classify the TGCs according to their
$C$ and $P$ behaviour as shown in Table~\ref{tab:disc}.  As can be
seen from the amplitudes in Sect.~\ref{sec-ampl}, terms of distinct
$P$ or $C$ do not mix in the quadratic part of the total cross section
\be
\s_{2ij} = \int \!d\phi\, S_{2ij}\;,
\ee
provided that phase space cuts are separately invariant under $P$ and
$C$.  Under the same conditions the linear terms $\sigma_{1i}$ in the
integrated cross section for the couplings $g_4^R$, $g_5^R$,
$\smash{\tilde{\kappa}_R}$ and $\smash{\tilde{\lambda}_R}$ vanish.
This is due to the absence of the neutrino exchange graph for right
handed electrons and the fact that those couplings differ in their $P$
or $C$ eigenvalues from the TGCs in the SM.  Finally, real and
imaginary parts of couplings do not mix in $\sigma_{2ij}$.

\begin{table}
	\begin{minipage}[l]{.45\textwidth} \caption{\label{tab:disc}
		Properties of TGCs under parity and charge
		conjugation.}
	\end{minipage}
	\hspace{1cm}
	\begin{minipage}[r]{.55\textwidth}
		\(
		\begin{array}{c|c|c|c|c}
		\hline \hline
		 &&&& \\[-0.45cm]
		 & g_1, \kappa, \lambda & g_4 & g_5 & \tilde{\kappa},
		\tilde{\lambda} \\[.1ex]
		\hline \hline
		 &&&& \\[-0.45cm]
		C & + & - & - & + \\[.1ex]
		\hline
		 &&&& \\[-0.45cm]
		P & + & + & - & - \\[.1ex]
		\hline
		 &&&& \\[-0.45cm]
		CP & + & - & + & - \\[.1ex]
		\hline \hline
		\end{array}
		\)
	\end{minipage}\
\end{table}

In the LR-basis, $\sigma_{2ij}$ has an additional block diagonality,
with two separate blocks for the R- and the L-couplings, which cannot
mix in the total cross section.  Any block diagonality of
$\sigma_{2ij}$ means that already before the simultaneous
diagonalisation (\ref{eq:alltrafo}) the subspaces corresponding to
these blocks are perpendicular to each other with respect to the
scalar product $(\boldsymbol{a}, \boldsymbol{b}) = \sum_{ij} a_i\,
\hat{\s}_{2ij}\, b_j$.  As a consequence, two row vectors of $A$ or
two column vectors of $A^{-1}$ which correspond to different blocks
are perpendicular with respect to the standard scalar product.  In the
case of $A^{-1}$ this follows from the first equation
of~(\ref{eq:shortdiag}) by solving for $\hat{\s}_2$, and in the case
of $A$ by solving for $\hat{\s}_2^{-1}$.  The comparison of the matrix
products $AA^T$ and $(A^{-1})^T A^{-1}$ with this expected
orthogonality thus provides a good way to test the numerical results
for $A$ and $A^{-1}$.  Note however that the transformation described
in Sect.~\ref{sec-simul} cannot be carried out on smaller blocks than
those given by the four classes (a) to~(d) because the left handed
couplings mix with the right handed ones in \covo{}, $c$ and $V(h)$.


\section{Polarisation}
\label{sec-pola}

In this section we explore how the sensitivity of our process to
anomalous TGCs depends on the longitudinal polarisation of the initial
beams.  To this end we will introduce an appropriate polarisation
parameter $P$ and analyse how the eigensystem determined
by~(\ref{eq:shortdiag}) depends on it.  This may be seen as a
preparation for interpreting the numerical results in
Sect.~\ref{sec-resu}, which will be given in terms of the same
parameter.

To start with, we introduce a convenient notation to make explicit the
polarisation dependence of the matrices $\hat{\s}_2$ and $V$, which
have to be diagonalised simultaneously according to
(\ref{eq:shortdiag}).  In the following we restrict ourselves to the
case of longitudinally polarised beams.  Since in the limit \( m_e
\rightarrow 0 \) only two beam helicity combinations contribute to the
amplitudes, we can use (\ref{eq:genpol}) to write the differential
cross section \( S = d\s |_{\rho} / d\phi \) in terms of the cross
sections $S^L$ and $S^R$ for purely left and right handed $e^-$ beams
(and opposite $e^+$ helicities),
\be
S = P^L S^L + P^R S^R,	\label{eq:sdef}
\ee
where
\begin{eqnarray}
P^L & = & \frac{1}{4}(1 - P^-)(1 + P^+),	\qquad
P^R \;=\; \frac{1}{4}(1 + P^-)(1 - P^+),	
\label{eq:pdef}
\\[0.2em]
S^L & = & d\s_{(-+)(-+)} / d\phi,               \qquad\qquad\;\;
S^R \;=\; d\s_{(+-)(+-)} / d\phi	
\label{eq:slsr}
\end{eqnarray}
and \( d\phi = d(\cos\!\Theta)\, d(\cos\!\vartheta)\, d\varphi\,
d(\cos\!\overline{\vartheta})\, d\overline{\varphi} \).  Since we only
deal with longitudinal polarisation in this section we drop the
subscript '$l$' in the polarisation parameters.  Note that \( 0 \leq
P^L, P^R \leq 1 \).  Integrating over $d\phi$ one obtains the total
cross section as
\be
\s = P^L \s^L + P^R \s^R. 	\label{eq:sigpol}
\ee
In the LR-basis we have
\begin{eqnarray}
S^a & = & S_0^a + \sum_i S_{1i}^a \h_i^a + \sum_{i,j} S_{2ij}^a \h_i^a
\h_j^a\;, \\
\s^a & = & \s_0^a + \sum_i \s_{1i}^a \h_i^a + \sum_{i,j} \s_{2ij}^a
\h_i^a \h_j^a \;.	\label{eq:siga}
\end{eqnarray}
We denote again the real or imaginary parts of the anomalous TGCs by
$\h_i^a$, but in contrast to~(\ref{eq:distri}) we now explicitly write
an index \( a = L, R \), so that $i$ and $j$ only run from 1 to 14.
Using vector and matrix notation we can rewrite the total cross
section as
\be
\s = \s_0 (1 + \boldsymbol{\h}^T \boldsymbol{\hat{\s}_1} +
\boldsymbol{\h}^T \hat{\s}_2 \boldsymbol{\h}) ,	\label{eq:sigmat}
\ee
where
\begin{eqnarray}
\s_0 & = & P^L \s_0^L + P^R \s_0^R, 
\nonumber \\[0.4em]
\boldsymbol{\hat{\s}_1} & = & \frac{1}{\s_0} 	
	\left( \begin{array}{c}
	  P^L \boldsymbol{\s_1^L} \\[.3ex]
	  P^R \boldsymbol{\s_1^R}
	\end{array} \right),\;\;\;\;\;\;
\boldsymbol{h} = \left( 	\begin{array}{c}
					\boldsymbol{h^L} \\
					\boldsymbol{h^R}
				\end{array} \right), 	
\nonumber \\[0.4em]
\hat{\s}_2 & = & \frac{1}{\s_0}	\left( \begin{array}{cc}
					P^L \overline{\s}_2 & 0 \\
					0 & P^R \overline{\s}_2
				\end{array} \right),	
\hspace{0.8cm}	\label{eq:stwoh}
\end{eqnarray}
with vectors \( (\boldsymbol{h^a})_i = h_i^a \) and \(
(\boldsymbol{\s_1^a})_i^{} = \s_{1i}^a \) and the matrix \(
(\overline{\s}_2)_{ij} = \s_{2ij}^L = \s_{2ij}^R \).  In the LR-basis
and for longitudinal polarisation we thus obtain the following
expression for the covariance matrix~(\ref{eq:covo}):
\be
V_{ij}^{ab} = P^a P^b \left( \frac{1}{\s_0} \int \!d\phi \frac{S_{1i}^a
S_{1j}^b}{S_0} - \frac{\s_{1i}^a \s_{1j}^b}{\s_0^2} \right),
\label{eq:covopol}
\ee
with \( a, b = L, R \).  Let us now investigate in detail how the
eigensystem of $\hat{\sigma}_2$ and $V$ depends on $P^L$ and $P^R$.
It is useful to express $P^L$ and $P^R$ by new variables $r$ and $P$
(to be specified below), so that $\hat{\s}_2$, $V$ and hence also
their eigensystem only depend on $P$.  For this purpose we introduce
$\plh$ and $\prh$ through
\be
P^{L,R} = r \hat{P}^{L,R}(P)	\label{eq:defphut}
\ee
for some well-behaved rescaling function \( r(P^L, P^R) \), and define
the ``$r$-normalised'' quantities
\begin{eqnarray}
\s_0^r & = & \plh \s_0^L + \prh \s_0^R,	\\
S_0^r & = & \plh S_0^L + \prh S_0^R.
\end{eqnarray}
We then have from (\ref{eq:stwoh}) and (\ref{eq:covopol}):
\begin{eqnarray}
\hat{\s}_2 & = & \frac{1}{\s_0^r(P)}	
			\left( \begin{array}{cc}
				\plh(P)\,
				\overline{\s}_2 & 0 \\
				0 & \prh(P)\, \overline{\s}_2
			\end{array} \right),
\label{eq:sthpol} \\[0.2em]
V_{ij}^{ab} & = & \hat{P}^a(P)\, \hat{P}^b(P) 
				\left( \frac{1}{\s_0^r(P)}
				\int \!d\phi\,
				\frac{S_{1i}^a\,
				S_{1j}^b}{S_0^r(P)} -
				\frac{\s_{1i}^a\,
				\s_{1j}^b}{[\, \s_0^r(P)]^2}
				\right) ,
\label{eq:covohpol}
\end{eqnarray}
where we have made the dependence on $P$ explicit.  Note that the left
hand sides of~(\ref{eq:sthpol}) and~(\ref{eq:covohpol}) depend on $P$
but not on $r$, since $\hat{\s}_2$ and $V$ do not change when $S$ is
multiplied by a constant.  For the matrix $M$ in~(\ref{eq:chol}) we get
\be
M = \frac{1}{\sqrt{\s_0^r(P)}}	\left( \begin{array}{cc}
					\sqrt{\plh}\, \overline{M} & 0 \\
					0 & \sqrt{\prh}\, \overline{M}
				\end{array} \right),
\ee
where \( \overline{\s}_2 = \overline{M}\,\overline{M}^T \) is the
Cholesky decomposition of the $P$ independent submatrix
of~$\hat{\s}_2$.  Then the result for the transformation
(\ref{eq:gtrafo}) is
\begin{eqnarray}
\boldsymbol{\h'} & = & A^{-1}\boldsymbol{\h} = X^{-1} M^T
 			\boldsymbol{\h} 
\nonumber \\[0.4em]
 & = & X^{-1}(P) \frac{1}{\sqrt{\s_0^r(P)}}	
	\left( \begin{array}{cc}
		\sqrt{\plh}\, \overline{M} & 0 \\
		0 & \sqrt{\prh}\, \overline{M}
	\end{array} \right) 
		\boldsymbol{\h}.	\label{eq:restrafo}
\end{eqnarray}
The factors~$\sqrt{\plh}$ and $\sqrt{\prh}$ in the rightmost matrix
determine the mutual normalisation of the blocks of left and right
handed couplings.  They let $A^{-1}$ become singular in the limits
$\plh$ or \(\prh \rightarrow 0\).  This is not surprising because with
beams of purely longitudinal polarisation one is sensitive to only
half of the couplings.  The coefficient $(\s_0^r(P))^{-1/2}$
in~(\ref{eq:restrafo}) leads to an overall normalisation which
strongly depends on the polarisation.  At \( \sqrt{s} = 500\)~GeV we
have for instance
\be
t_{LR} \;\equiv\; \sqrt{\s_0^L / \s_0^R} \;\approx\; 17 , \;\;\;
\sigma_0^r \approx \sigma_0^L (\plh + \prh / 17^2),
\label{eq:slbysr}
\ee
whereas at \( \sqrt{s} = 3\) TeV the ratio $t_{LR}$ is about~30.
{}From~(\ref{eq:normone}) we see that the matrix $X^{-1}$ is
orthogonal for any $P$ .  In the case of pure polarisation it is block
diagonal in the left and right handed couplings.  This is however not
the case for general (longitudinal) polarisation since the
diagonalisation cannot be reduced to smaller blocks than those given
by the four discrete symmetry classes introduced in
Sect.~\ref{ssec-disc}.

We now specify the transformation~(\ref{eq:defphut}) by choosing
\be
r = \frac{1}{4} \left( \sqrt{P^R} + \sqrt{P^L} \right)^2,
\label{eq:defh}
\ee
and defining a polarisation parameter
\be
P = \frac{\sqrt{P^R} - \sqrt{P^L}}{\sqrt{P^R} + \sqrt{P^L}},
\label{eq:defp}
\ee
with values \( -1 \leq P \leq +1 \).  We then have
\be
\hat{P}^{R, L} = \left( 1 \pm P \right) ^2.	\label{eq:phut}
\ee
In terms of the individual beam polarisations $P^-$ and $P^+$ the
parameters $r$ and $P$ are given as
\begin{eqnarray}
r & = & \frac{1}{8} \left( 1 - P^+ P^- + \sqrt{\left(1 - P^+ P^-
\right)^2 - \left(P^+ - P^- \right)^2} \,\right), \\
P & = & \frac{P^- - P^+}{1 - P^+ P^- + \sqrt{\left(1 - P^+ P^-
\right)^2 - \left(P^+ - P^-\right)^2}}.
\label{eq:rp}
\end{eqnarray}
The reason for this particular choice is as follows.  For electron
polarisation~$P_0^-$ and positron polarisation~\mbox{\(P_0^+ =
-P_0^-\)} one simply has \mbox{\( P = P_0^- \)}.  For general
polarisations $P$ is between $P^-$ and $-P^+$, and the differential
cross section $S$ for \mbox{\((P^-,P^+)\)} equals the one for
\mbox{\((P_0^- \!\! = \! P,\, P_0^+ \!\! = \!  -P)\)} up to a
constant.  The eigenvalues $c_i'$ (cf.\ (\ref{eq:vprime})) are hence
the same for \mbox{\((P^-, P^+)\)} and for \mbox{\((P_0^-, P_0^+)\)}.

\begin{figure}
\begin{center}
\leavevmode
\includegraphics[totalheight=5.38cm]{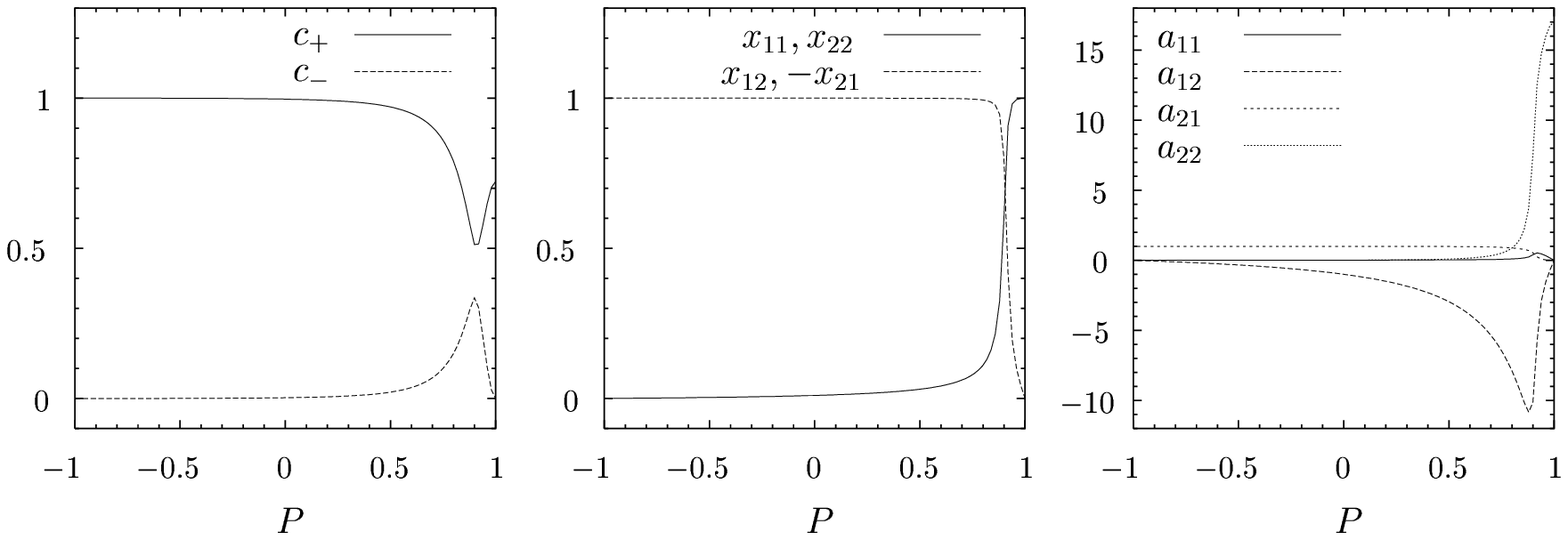}
\end{center}
\caption{\label{fig:par1} Eigenvalues (\protect\ref{eq:eigval}) and
entries of the matrices (\protect\ref{eq:xtil}) and
(\protect\ref{eq:notatil}) for $\s_0^L=1$, $\s_0^R=(1/17)^2$,
$\tilde{v}^{LL}=1$, $\tilde{v}^{LR}=0.01$, $\tilde{v}^{RR}=0.0025$ and
$\left(\overline{\s}_2\right)_{11}=1$.}
\vspace{2\baselineskip}
\begin{center}
\leavevmode
\includegraphics[totalheight=5.38cm]{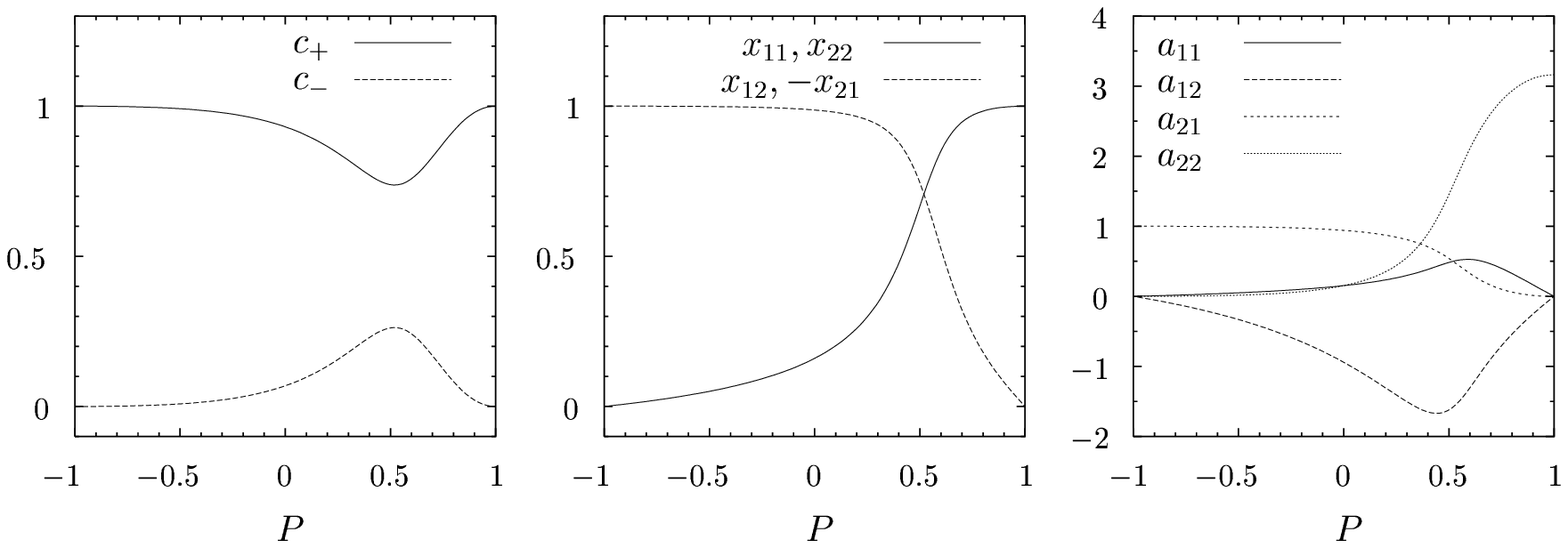}
\end{center}
\caption{\label{fig:par2} Same as Fig.~\protect\ref{fig:par1} but
using $\s_0^L=1$, $\s_0^R=1/10$, $\tilde{v}^{LL}=1$,
$\tilde{v}^{LR}=0.15$, $\tilde{v}^{RR}=0.1$ and
$\left(\overline{\s}_2\right)_{11}=1$.}
\end{figure}

To develop some intuition of how the generalised eigenvalues of
(\ref{eq:sthpol}) and (\ref{eq:covohpol}) depend on $P$, we consider
the case of only one left and one right handed coupling.  Moreover, we
neglect the second term in~(\ref{eq:covohpol}), which appears only in
symmetry class~(a).  The matrix $\overline{\s}_2$ in (\ref{eq:stwoh})
then reduces to a single number \( \left(\overline{\s}_2\right)_{11}
\), the vector $(S_1)_i^a$ has only one component \( s^a \equiv
(S_1)_1^a \), and the 2$\times$2 matrices which have to be
diagonalised according to (\ref{eq:shortdiag}) can be written as
\begin{eqnarray}
\hat{\s}_2 & = & \frac{\left(\overline{\s}_2\right)_{11}}{\s_0^r(P)}
					\left( \begin{array}{cc}
					\plh	&	0 \\
					0	& 	\prh
					\end{array} \right),	
\label{eq:s2r2d} \\[0.2em]
V & = & \frac{1}{\s_0^r(P)}	\left(\begin{array}{cc}
					\plh 	& 	0 \\
					0	&	\prh
				\end{array} \right)
				\left(\begin{array}{cc}
					v^{LL}	&	v^{LR} \\
					v^{LR}	&	v^{RR}
				\end{array} \right)
				\left(\begin{array}{cc}
					\plh 	& 	0 \\
					0	&	\prh
				\end{array} \right) ,
\end{eqnarray}
where
\be
v^{LL} = \int \!d\phi\, \frac{(s^L)^2}{S_0^r(P)} , \qquad
v^{LR} = \int \!d\phi\, \frac{s^L s^R}{S_0^r(P)} , \qquad
v^{RR} = \int \!d\phi\, \frac{(s^R)^2}{S_0^r(P)} .	
\label{eq:abddef} 
\ee
As in~(\ref{eq:c}) we construct a symmetric matrix
\be
C =  \frac{1}{\left(\overline{\s}_2\right)_{11}}  
\left(\begin{array}{cc}
	\sqrt{\plh} 	& 	0 \\
	0		&	\sqrt{\prh}
\end{array} \right)
\left(\begin{array}{cc}
	v^{LL} 	&	v^{LR} \rule{0pt}{1.2em} \\
	v^{LR}	&	v^{RR} \rule{0pt}{1.2em}
\end{array} \right)
\left(\begin{array}{cc}
	\sqrt{\plh} 	& 	0 \\
	0		&	\sqrt{\prh}
\end{array} \right) ,	\label{eq:cmat}
\ee
whose usual eigenvalues are equal to the generalised eigenvalues of
$V$, i.e.\ to the diagonal entries of the transformed matrix $V'$
in~(\ref{eq:shortdiag}).  They are given by
\be
c_{\pm} = \frac{1}{2 \left(\overline{\s}_2\right)_{11}} 
\left( \plh v^{LL} + \prh\, v^{RR} 
       \pm \sqrt{\Big( \plh v^{LL} - \prh\, v^{RR} \Big)^2 
                 + 4 \plh \prh\, \Big(v^{LR}\Big)^2} 
\;\right).	\label{eq:eigval}
\ee
We approximate the matrix entries (\ref{eq:abddef}) by
\be
v^{ab}(P) = \frac{\tilde{v}^{ab}}{\plh \s_0^L + \prh \s_0^R} ,
\qquad a,b = L,R, 
\label{eq:chi}
\ee
with constants $\tilde{v}^{ab}$, which should take into account their
$P$ dependence sufficiently well for a qualitative model.  In
Fig.~\ref{fig:par1} we plot the eigenvalues $c_{\pm}$ in arbitrary
units for \( \s_0^L = 1 \), and \( \s_0^R = \s_0^L / t_{LR}^2 =
(1/17)^2\), with the last number taken from~(\ref{eq:slbysr}).  The
ratios \( \tilde{v}^{LL} / \tilde{v}^{LR} \approx \tilde{v}^{LR} /
\tilde{v}^{RR} \approx t_{LR} \) in our choice of parameters are
motivated by the power of $s^L$ in~(\ref{eq:abddef}), which
corresponds to the power of the neutrino exchange amplitude in the
cross section.  One can show analytically that the slopes of the
curves for $c_\pm(P)$ tend to zero for \( P \rightarrow \pm 1 \).  For
\( P = 1 \) this cannot be seen in the plot, since for large $t_{LR}$
(i.e.\ $\sigma_0^L \gg \sigma_0^R$) the eigenvalues change rapidly as
the $e^-$ beam becomes purely right-handed.  To see the horizontal
tangent we plot a second example in Fig.~\ref{fig:par2} with a more
moderate value of $t_{LR}$.  Notice that for nonzero $v^{LR}$ the two
curves for $c_+$ and $c_-$ do not touch.  If $\prh$ or $\plh$ is zero
the matrix $\hat{\s}_2$ in~(\ref{eq:s2r2d}) and hence $C$
in~(\ref{eq:cmat}) is singular, which leads to a zero eigenvalue $c_-$
at \( P = \pm 1 \).

As in Sect.~\ref{sec-real} let
\be
X = \left( \boldsymbol{x_-}, \boldsymbol{x_+} \right)
= \left(\begin{array}{cc}
		x_{11} & x_{12} \\
		x_{21} & x_{22}
	\end{array} \right)	\label{eq:xtil}
\ee
be the matrix whose columns are the normalised eigenvectors of $C$.
We can see from Figs.~\ref{fig:par1} and~\ref{fig:par2} that for
\mbox{\( P = -1 \)} the vector $\boldsymbol{x_+}$ with large
eigenvalue has only an upper component (corresponding to $h^L$),
whereas the vector $\boldsymbol{x_-}$ with zero eigenvalue has only a
lower component (corresponding to $h^R$).  For \( P = +1 \) the
situation is reversed.  This reflects the fact that one is only
sensitive to the left handed couplings for \( P = -1 \) and to the
right handed ones for \( P = +1 \).  We finally plot the elements of
the transformation matrix $A^{-1}$ in~(\ref{eq:restrafo}) using the
notation
\be
A^{-1} =	\left( \begin{array}{cc}
			a_{11} & a_{12} \\
			a_{21} & a_{22}
		\end{array} \right).	\label{eq:notatil}
\ee
Writing the transformed couplings as
\be
\boldsymbol{h'} = \left( \begin{array}{c}
				h_-' \\
				h_+'
 			 \end{array} \right)
 = A^{-1} \left( \begin{array}{c}
				h^L \\
				h^R
 			 \end{array} \right),
\ee
we can see that for \( P = -1 \) the right handed contributions to
both $h_-'$ and $h_+'$ vanish, whereas for \( P = +1 \) the same
happens to the left handed ones.  This behaviour of $A^{-1}$ has to be
taken into account when carrying out the simultaneous diagonalisation
for high degrees of longitudinal polarisation, because it leads to a
singularity of its inverse in the limit \( P \rightarrow \pm 1 \).

We have seen that under the condition that only beam helicity
combinations with \( \Delta \he = \pm 1 \) contribute to the cross
section, the expectation values of \emph{normalised} observables only
depend on the polarisation parameter $P$.  Such a statement no longer
holds if beams with helicities coupled to $\Delta\tau=0$ contribute as
well.  This provides a possibility to disentangle effects from genuine
$\cp$ violation or absorptive parts from those due to nonzero \(
\Delta \he = 0 \) amplitudes.  In particular, the $\cp$ odd part of
the spin density matrix for longitudinally polarised $e^+e^-$ beams is
proportional to $P^+ + P^-$ and will give different contributions to
normalised observables if $P^+ + P^-$ is varied for fixed $P$.


\section{Hardly Measurable Couplings}
\label{sec-hmc}

The particular form of the SM amplitudes for the process \prpr\ has
consequences on its sensitivity to the couplings in the $\cp$
conserving sector, which we shall now discuss.  To this end we write
the TGC part of the transition operator as
\be
\tr^{\rm TGC} = \sum_{a=L,R} \; \sum_{i=1 \rule{0pt}{0.45em}}^7 \;
	(\h_{0,i}^a + H_i^a)\, \tr_i^a\;,
\label{eq:traop}
\ee
where for simplicity of notation we label the respective couplings
$g_1$, $\kappa$, $\lambda$, $g_4$, $g_5$, $\tilde{\kappa}$,
$\tilde{\lambda}$ by an index $i=1,\ldots,7$.  Here $\h_{0,i}^a$ are
the SM couplings in the LR-basis and \( H_i^a = {\rm Re}H_i^a + i\,
{\rm Im} H_i^a \) are the complex anomalous couplings (which we write
in uppercase to distinguish them from the real-valued
parameters~$h^a_i$).  The nonzero SM couplings are
(see~(\ref{eq:coupsm}),~(\ref{eq:lrgz}))
\be
\h_{0,1}^{L} = \h_{0,2}^{L} = 2 \xi + 4 \sin^2 \theta_W (1 - \xi)\;, \qquad
\h_{0,1}^{R} = \h_{0,2}^{R} = 4 \sin^2 \theta_W (1 - \xi)\;.
\label{eq:direction}
\ee

We first consider longitudinal polarisation where the differential
cross section~$S$ (cf.~\ref{eq:diffpol})) is given by~(\ref{eq:sdef}).
For $S^R$ there is no neutrino exchange contribution, so that we have
from~(\ref{eq:dcr}), (\ref{eq:slsr}) and (\ref{eq:traop})
\begin{eqnarray}
S^R & \propto & 
\sum_{i,j} \,\langle f | \tr_i^R |\, {+-} \rangle \,
      \langle f | \tr_j^R |\, {+-} \rangle^{\ast} 
\nonumber \\
      &   & \hspace{1em} {}\times (\h_{0,i}^R + \re H_i^R + i\,\im
      H_i^R) (h_{0,j}^R + \re H_j^R - i\,\im H_j^R) .
\label{eq:reim}
\end{eqnarray}
Consider now the following direction in the space of right handed
anomalous couplings:
\be
\left( 	\begin{array}{c}
		\re\, \boldsymbol{H^R} \\
		\im\, \boldsymbol{H^R} 
		\end{array} \right) =  
\left( 	\begin{array}{c}
		0 \\
		\omega \boldsymbol{h_{0}^R} 
		\end{array} \right) ,
\label{eq:insdir}	
\ee
where we assume \( \omega \ll 1 \) and use the vector notation
\be
\boldsymbol{H^R} = \left(	\begin{array}{c}
			H_1^R \\
			\vdots \\[0.3em]
			H_7^R
			\end{array} \right),\hspace{0.8cm}
\boldsymbol{\h_{0}^R} = \left(	\begin{array}{c}
			\h_{0,1}^R \\
			\vdots \\[0.3em]
			\h_{0,7}^R
			\end{array} \right) .
\ee
With~(\ref{eq:insdir}) the second line of~(\ref{eq:reim}) equals
\be
(h_{0,i}^R + i\omega \h_{0,i}^R )\, (h_{0,j}^R -
i\omega \h_{0,j}^R) = (1 + \omega^2)\ h_{0,i}^R \,\h_{0,j}^R\;.	
\label{eq:canc}
\ee
In the space of the {\em imaginary} parts of right handed couplings
there is hence a direction in which the differential cross section for
unpolarised or longitudinally polarised beams has no linear term in
$\omega$, but is only sensitive to order $\omega^2$.  This direction
is determined by the \emph{real} SM couplings as given
by~(\ref{eq:direction}).  Therefore, one of the functions $S_{1i}'$
and the corresponding observable ${\cal O}_i^{\,\prime}$
in~(\ref{eq:alltrafo}) are identically zero, and \covo\ contains one
(usual as well as generalised) eigenvalue in symmetry~class~(b) that
is zero for all values of $P$.  This is confirmed by our numerical
results shown in Figs.~\ref{fig:eigval500b}, \ref{fig:eigval800b} and
\ref{fig:eigval3b}.  In the tables in Sect.~\ref{sec-resu} the
eigenvalues of symmetry~class~(b), \(c_9',\ldots,c_{16}' \), are given
in decreasing order.  Using this notation we have
\be
S_{1,16}'(\phi) \equiv 0,	\hspace{1.2cm}	c_{16}' = 0.
\label{eq:c16pr}
\ee
{}From the total rate one can derive constraints on this coupling as
explained in Sect.~\ref{sec-opti}.

Now consider
\be
\left( 	\begin{array}{c}
		\re\, \boldsymbol{H^R} \\
		\im\, \boldsymbol{H^R} 
		\end{array} \right) =  
\left( 	\begin{array}{c}
		\omega \boldsymbol{\h_{0}^R} \\
		0  
		\end{array} \right),	\label{eq:coupre}
\ee
which merely ``stretches'' the right handed SM couplings by a factor
\( (1 + \omega ) \).  Then the last line of~(\ref{eq:reim}) becomes
\be
(1 + \omega )^2\, \h_{0,i}^R \, \h_{0,j}^R\;.	\label{eq:enha}
\ee
In the case of purely right handed electrons or left handed positrons,
i.e.\ for \( P^L = 0 \), we have $S \propto S^R$ from~(\ref{eq:sdef}),
so that the anomalous coupling~(\ref{eq:coupre}) only increases the
total rate but does not affect the normalised distribution~\( \s^{-1}S
\).  This holds both to order $\omega$ and to order $\omega^2$.
Symmetry~class~(a) therefore contains a fifth zero eigenvalue for \( P
= +1 \), in addition to the four zero eigenvalues from the left handed
couplings Re$\,g_1^L$, Re$\,\kappa_L$, Re$\,\lambda_L$ and
Re$\,g_5^L$, which cannot be measured for \( P^L = 0 \).  This is
again confirmed by our numerical results
(cf.~Figs.~\ref{fig:eigval500a}, \ref{fig:eigval800a},
\ref{fig:eigval3a}).  For \( P^L \neq 0 \), however, $S^L$ also
contributes to~$S$.  Since the functions \( \s^{-1} S^a \) are not
identical for \( a = L \) and \( a = R \), the enhancement of $S^R$
due to~(\ref{eq:enha}) will not just change $S$ by an overall factor,
but also modify the normalised distribution \( \s^{-1} S \).  The
latter is sensitive to the anomalous TGC in~(\ref{eq:coupre}) in the
linear approximation, since ({\ref{eq:enha}) contains a term linear in
$\omega$.  In contrast to symmetry~(b), there is thus no eigenvalue
which is identical to zero for all values of $P$.  Note that $S^L$
contains interference terms of the left handed anomalous amplitudes
and the SM neutrino exchange, so that the arguments above do not apply
to the subspace of the left handed couplings.  Also for the
symmetries~(c) and~(d) there is no similar argument because $\cp$
violating TGCs are absent in the SM at tree level.

The direction~(\ref{eq:insdir}) in coupling space becomes measurable
in the linear approximation with $e^+e^-$ beams of transverse
polarisation.  In fact, abbreviating
\begin{eqnarray}
{\cal A}_0^{\tau=+1} & = & 
	\langle f | {\cal T}_0 |\, {+-} \rangle , \qquad\;\,
{\cal A}_0^{\tau=-1} \;=\; 
	\langle f | {\cal T}_0 |\, {-+} \rangle ,
\nonumber \\[0.2em]
{\cal A}_i^{\tau=+1} & = & 
	\langle f | {\cal T}_i^R |\, {+-} \rangle , \qquad
{\cal A}_i^{\tau=-1} \;=\;
	\langle f | {\cal T}_i^L |\, {-+} \rangle ,
\end{eqnarray}
where ${\cal T}_0$ is the transition operator in the SM at tree level
and ${\cal T}_i^a$ is defined in~(\ref{eq:traop}), we can write the
part of the differential cross section~(\ref{eq:genpol}) that is
linear in the anomalous TGCs as
\begin{eqnarray}
d\s |_{\rho}^{\rm lin} & \propto & \sum_{i,a} \bigg( P^a \Big[
  \re \left( {\cal A}_0^{\tau \ast} {\cal A}_i^\tau \right) \re H_i^a 
- \im \left( {\cal A}_0^{\tau \ast} {\cal A}_i^\tau \right) \im H_i^a
\Big]
\label{eq:slin} \\
 & & {} - \frac{P_t^- P_t^+}{4}\, \bigg\{\cos(\psi + 2\Phi) \left[
  \re \left( {\cal A}_0^{-\tau \ast} {\cal A}_i^\tau \right) \re H_i^a
- \im \left( {\cal A}_0^{-\tau \ast} {\cal A}_i^\tau \right) \im H_i^a
\right] 
\nonumber \\[0.2em]
 & & \hspace{3.4em} {} - \tau \sin(\psi + 2\Phi) \left[
  \im \left( {\cal A}_0^{-\tau \ast} {\cal A}_i^\tau \right) \re H_i^a
+ \re \left( {\cal A}_0^{-\tau \ast} {\cal A}_i^\tau \right) \im H_i^a
\right]\bigg\} \bigg) ,
\nonumber
\end{eqnarray}
where of course $a=R$ implies $\tau=1$ and $a=L$ implies $\tau=-1$.
As in Sect.~\ref{sec-ampl} we denote the degrees longitudinal and
transverse polarisation of the $e^+$ and $e^-$ beams by $P_l^{\pm}$
and $P_t^{\pm}$, and as in Sect.~\ref{sec-pola} we abbreviate $P^L =
(1 - P^-_l)(1 + P^+_l) / 4$ and $P^R = (1 + P^-_l)(1 - P^+_l) / 4$.
Note that $d\s |_{\rho}^{\rm lin}$ depends on the $W^-$ azimuthal
angle $\Phi$ only via the explicit trigonometric functions
in~(\ref{eq:slin}).  One thus only has $d\s |_{\rho}^{\rm lin} \equiv
0$ if the three lines of~(\ref{eq:slin}) vanish separately.  The first
line is however the same as what we had for purely longitudinal
polarisation, so that it vanishes for generic polarisations $P^a$ only
if condition~(\ref{eq:insdir}) is fulfilled and $\boldsymbol{H^L}=0$.
Then relation~(\ref{eq:slin}) becomes:
\begin{eqnarray}
d\s |_{\rho}^{\rm lin} & \propto & \omega \sum_i h_{0,i}^R \bigg( 
- P^R\, \im \left( {\cal A}_0^{+1 \ast} {\cal A}_i^{+1} \right) \\
 & & {} + \frac{P_t^- P_t^+}{4}
\bigg\{ \cos (\psi + 2\Phi)\, \im \left(
	{\cal A}_0^{-1 \ast} {\cal A}_i^{+1} \right) 
      + \sin (\psi + 2\Phi)\, \re \left( 
 	{\cal A}_0^{-1 \ast} {\cal A}_i^{+1} \right) \bigg\}\bigg) \nonumber \\[.5em]
 & & = \omega\, \frac{P_t^- P_t^+}{4}
\bigg\{ \cos (\psi + 2\Phi)\, \im \left(
	{\cal A}_0^{-1 \ast} {\cal A}_0^{+1} \right) 
      + \sin (\psi + 2\Phi)\, \re \left( 
 	{\cal A}_0^{-1 \ast} {\cal A}_0^{+1} \right) \bigg\},  \nonumber
\end{eqnarray}
where for the equality we have used the fact that  \( {\cal A}_0^{+1}
= \sum_i h_{0,i}^R \, {\cal A}_i^{+1} \).  Since ${\cal A}_0^{-1}$
contains the neutrino exchange graph, $d\s |_{\rho}^{\rm lin}$ no
longer vanishes.  Transverse beam polarisation thus allows for the
measurement of the anomalous coupling~(\ref{eq:insdir}), which is
hardly possible using only longitudinal polarisation.

Before presenting our numerical results for unpolarised and
longitudinally polarised beams, we must explain how to take into
account the zero eigenvector (\ref{eq:insdir}) of $V({\cal O})$ in the
analysis.  {}From~(\ref{eq:geo}) and~(\ref{eq:alltrafo}) we obtain
for the inverse covariance matrix of the couplings $h$
\be V(h)^{-1} = N (A^{-1})^T c' A^{-1}, \ee
where the $h_i$ are again 28 real parameters, viz.\ the real and
imaginary parts of $g_1^R$, $g_1^L$, $\kappa_R$, $\kappa_L$, etc.  We
number the couplings in the order of their symmetry class (a) to (d),
and within each symmetry class take the L-couplings first.  We then
have $h_{13} = \im\,g_1^R$ and $h_{14} = \im\,\kappa_R$.  Note that
$V(h)^{-1}$ always exists, even in our case where one parameter is
unmeasurable.  In this case $V(h)^{-1}$ is a singular matrix with a
one-dimensional zero eigenspace coming from $c_{16}' = 0$.
Geometrically speaking, the error ellipsoid defined by $V(h)^{-1}$ is
degenerate in such a way that the length of one principal axis is
infinite.  Instead of an ellipsoid we have a cylinder whose axis
corresponds to the direction of the unmeasurable coupling and whose
cross-section (orthogonal to the axis) is an ellipsoid giving the
errors on the couplings in the orthogonal space.  We know from
(\ref{eq:direction}) and (\ref{eq:insdir}) that the unmeasurable
direction is given by $\im\,g_1^R = \im\,\kappa_R \ne 0$ with all
other couplings being zero.  Therefore the projection of the cylinder
onto the $(\im\,g_1^R)$-$(\im\,\kappa_R)$-plane is a band in the $\im
(g_1^R + \kappa_R)$-direction, see Fig.~\ref{fig:band}(a).
\begin{figure}
\begin{center}
\includegraphics[totalheight=5.5cm]{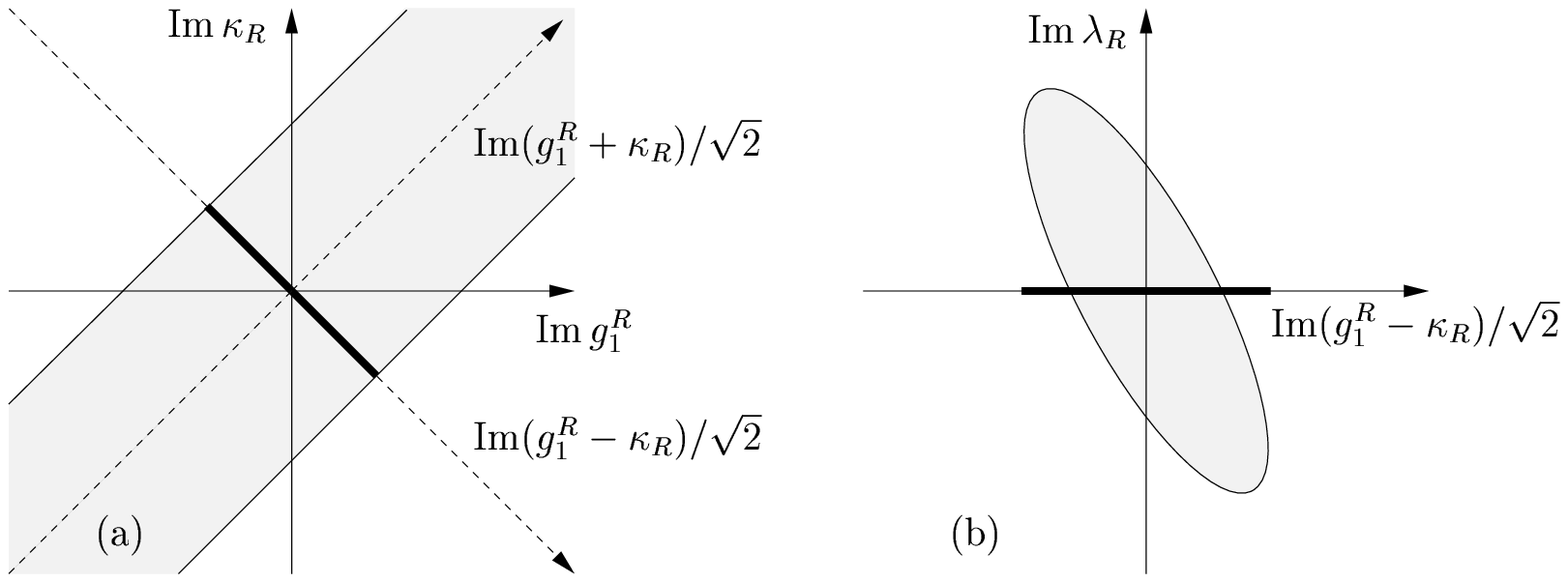}
\end{center}
\caption{\label{fig:band}Schematic view of the constraints in the case
of symmetry class (b).}
\end{figure}
This shows that we cannot obtain any constraint on $\im\,g_1^R$ {\em
or} $\im\,\kappa_R$ unless one of them is known.  We can however
choose coordinate axes parallel and orthogonal to the band shown in
Fig.~\ref{fig:band}(a).  In other words, we perform a rotation by
$-45^{\circ}$ in the $(\im\,g_1^R)$-$(\im\,\kappa_R)$-plane,
\be
\boldsymbol{\tilde{h}} = R\, \boldsymbol{h},
\ee
where $R$ is the identity matrix except for the
$(\im\,g_1^R)$-$(\im\,\kappa_R)$-block, which reads
\be
\frac{1}{\sqrt{2}}
\left(	\begin{array}{rr}
		1 & -1 \\
		1 &  1 
	\end{array}	\right).
\ee
The new couplings $\tilde{h}_i$ are the same as the $h_i$, except for
$\tilde{h}_{13} = \im (g_1^R - \kappa_R) / \sqrt{2}$ and
$\tilde{h}_{14} = \im (g_1^R + \kappa_R) / \sqrt{2}$, which replace
$\im\,g_1^R$ and $\im\,\kappa_R$.  The inverse covariance matrix of
the new couplings is
\be
V(\tilde{h})^{-1} = R\, V(h)^{-1} R^T.
\ee
All entries in the 14th row and in the 14th column of
$\smash{V(\tilde{h})^{-1}}$ are equal to zero: there is no correlation
between the unmeasurable $\im (g_1^R + \kappa_R) / \sqrt{2}$-direction
and the couplings $\tilde{h}_i$ with $i\neq 14$.  These couplings are
hence constrained by a 27-dimensional ellipsoid, which is drawn
schematically in Fig.~\ref{fig:band}(b) for one further coupling,
taken to be $\im\,\lambda_R$.  This ellipsoid is determined by the
``reduced'' $27 \times 27$ matrix $\smash{V_{red}^{-1}(\tilde{h})}$
obtained from $\smash{V(\tilde{h})^{-1}}$ by deleting the 14th row and
the 14th column.  Its inverse $\smash{V_{red}(\tilde{h})}$ is the
covariance matrix of $\im (g_1^R - \kappa_R) / \sqrt{2}$ and the other
26 measurable couplings.  In particular, the width of the band in
Fig.~\ref{fig:band}(a) gives the error on $\im (g_1^R - \kappa_R) /
\sqrt{2}$ in the presence of all other 27 couplings $\tilde{h}_i$,
cf.\ Fig.~\ref{fig:band}(b).  We finally mention that because of the
discrete symmetries explained in Sect.~\ref{ssec-disc}, the matrix
$\smash{V(h)^{-1}}$ is block diagonal with one block for each symmetry
class (a) to (d), so that the errors on the couplings of class (a),
(c) and (d) are entirely unaffected by the previous discussion.


\section{Numerical Results}
\label{sec-resu}

In this section we present the results for the generalised eigenvalues
$c_i'$ of the covariance matrix $V({\cal O})$ and the corresponding
errors $\delta h_i' = (N c_i')^{-1/2}$ on the transformed couplings.
The covariance matrix for the couplings in any other parameterisation
is then obtained by conventional error propagation.  We discuss its
most important properties in the LR-basis for \( \sqrt{s} = \mbox{500
GeV} \) and unpolarised beams in Sect.~\ref{ssec-unpol}.  In
Sect.~\ref{ssec-pola} we investigate the gain in sensitivity by
longitudinal $e^-$ as well as additional $e^+$ polarisation.  The
results for higher c.m.~energies are reported in
Sect.~\ref{ssec-ener}.  In Sect.~\ref{ssec-totr} we finally give the
constraints which can be obtained from the total rate according
to~(\ref{eq:stotnorm}).  Numerical rounding errors on the results
presented in this section are typically of order~1\%.

We use $m_W = 80.42 {\rm \ GeV}$ and $m_Z = 91.19 {\rm \ GeV}$
from~\cite{Groom:in}, and the definition $\sin^2\theta_W = 1 - m_W^2
/m_Z^2$ for the weak mixing angle.  For the total event rate $N$ of
the semileptonic channels with $e$ and $\mu$ summed over we use the
values listed in Table~\ref{tab:events}.  They correspond to an
effective electromagnetic coupling constant \( \alpha = 1/128 \) and
integrated luminosities of \mbox{500 ${\rm fb}^{-1}$}, \mbox{1 ${\rm
ab}^{-1}$} and \mbox{3 ${\rm ab}^{-1}$} at \( \sqrt{s} = \mbox{500
GeV} \), \mbox{800 GeV} and \mbox{3 TeV}, respectively.  We assume
full kinematical reconstruction of the final state, except that the
jet charges are not tagged.  Due to this twofold ambiguity we cannot
take the zeroth and first order parts of the differential cross
section $S$ as the denominator and the numerator of the optimal
observables ${\cal O}$, but use their respective sums over the two
experimentally undistinguished final states~(cf.~\cite{Diehl:1993br}).

\begin{table}
\caption{\label{tab:events} Total event rate $N$ in units of $10^3$
for the semileptonic channels with $e$ and $\mu$ summed over.
Corresponding luminosities are given in the text.  $P^-$ and $P^+$
respectively denote the degrees of longitudinal polarisation of the
$e^-$ and $e^+$ beams, and $P$ is given in~(\protect\ref{eq:rp}).}
\begin{center}
\leavevmode
\begin{tabular}{c|c|c|rrr}
\hline \hline
\multicolumn{3}{c|}{polarisation} & 
\multicolumn{3}{c}{$\sqrt{s}$~[GeV]} \\
$P^-$ & $P^+$ & $P$ & \mbox{500} & \mbox{800} & \mbox{3000} \\
\hline \hline
&&&& \\[-2.5ex]
$-80\%$ & $+60\%$ & $-71\%$ & 3280 & 3410 & 1280 \\
$-80\%$ & $0$ & $-50\%$ & 2050 & 2130 & 799 \\
$0$ & $0$ & $0$ & 1140 & 1190 & 444 \\
$+80\%$ & $0$ & $+50\%$ & 235 & 242 & 89.7 \\
$+80\%$ & $-60\%$ & $+71\%$ & 103 & 103 & 37 \\
\hline \hline
\end{tabular}
\end{center}
\end{table}


\subsection{Unpolarised Beams at 500 GeV}
\label{ssec-unpol}

We first consider the sensitivity at $\sqrt{s}=$~\mbox{500 GeV} with
unpolarised beams.  In Tables~\ref{tab:sensa}--\ref{tab:sensd} we list
the standard-deviation \( \delta h_i = [V(h)_{ii}]^{1/2} \) for each
coupling $h_i$, which gives its error in the presence of all other
couplings.  Notice the difference of this to $( N c_{ii} )^{-1/2}$,
which corresponds to the error on $h_i$ when all other couplings are
assumed to be zero.  We also give the correlation matrix
\begin{equation}
 W(h)_{ij} =  \frac{V(h)_{ij}}{ \sqrt{V(h)_{ii}\, V(h)_{jj}} }
\end{equation}
of the couplings for each symmetry~class~(a) to~(d).  In the case of
symmetry~(b) we use the reduced matrix $V_{red}(\tilde{h})$ introduced
in Sect.~\ref{sec-hmc}.  Since $W(h)$ is symmetric we only list its
upper triangular part.

The values of $\delta h$ range from about \( 5 \cdot 10^{-4} \) to \(
10^{-2} \) within each symmetry class.  The smallest are those for
$\lambda_L$, $\kappa_L$ and $\tilde{\lambda}_L$ since at high energies
the corresponding terms in the helicity amplitudes contain a factor $2
\gamma^2$ (cf.~Table~\ref{tab:amp2}).  In all cases the errors on the
R-couplings are larger than those on the respective L-couplings, viz.\
by a factor 1.5 to 3.4 for Im$\,\kappa$, Re$\,\Delta \kappa$ and
Re$\,g_5$, and by a factor between 4 and 7 for the other couplings.
This is because (for unpolarised or longitudinally polarised beams)
the $\nu$-exchange interferes with the amplitudes containing the
$h^L$, but not with those containing the $h^R$.  In general, the
sensitivity to the real part of a specific coupling is roughly of the
same size as the sensitivity to its imaginary part, the errors on the
latter being rather larger for $\cp$ conserving couplings and smaller
for the $\cp$ violating ones.  To get an accurate picture of the
sensitivities, correlations have to be taken into account.  Looking at
the 2$\times$2 blocks corresponding to $h^L_i$ and $h^R_i$ for a given
index~$i$, i.e.\ at the diagonal entries of the right upper block in
the correlation matrices, we see that the absolute values of the
correlations are smaller than about 0.2, except for Re$\,\Delta
\kappa$, Im$\,\kappa$ and Im$\,g_5$, where they are still smaller than
0.6.  The corresponding correlations would be substantial in the basis
of the $\gamma$- and $Z$-form factors, which is hence not a very
suitable parameterisation for the present reaction, compared with the
LR-basis (cf.~\cite{Diehl:1993br}).  Considering the matrix blocks of
correlations among different L-couplings or among different
R-couplings, we find that about half of them have an absolute value
larger than 0.4.  Note that there are correlations of order 0.5
between couplings with different $C$ or $P$ eigenvalues.

By simultaneous diagonalisation (see Sect.~\ref{sec-simul}) we
determine the generalised eigenvalues $c_i'$ and corresponding errors
$\delta h_i'$ given in Tables~\ref{tab:eveven} and~\ref{tab:evodd}.
For symmetry~class~(b) we give the transformation matrix~$A$ and its
inverse in Tables~\ref{tab:amat} and~\ref{tab:ainvmat}.\footnote{The
matrices $A^{-1}$ for symmetry~classes~(a), (c) and (d) can be
obtained from the authors.  Further numerical results for unpolarised
and longitudinally polarised beams at various c.m.~energies are given
in \protect\cite{nagel}.}
In our numerical calculation we find the smallest eigenvalue in
symmetry~class~(b) to be \( c_{16}' \sim 10^{-14} \).  We can however
use the result of our analytical considerations in Sect.~\ref{sec-hmc}
and set $c_{16}^{\prime}$ to zero.  The same holds for the last column
of~$A$ except for its Im$\,g_1^R$- and Im$\,\kappa_R$-components,
which determine the corresponding ``blind'' direction in the LR-basis.
As explained in Sect.~\ref{ssec-disc} the $P$ odd coupling $g_5$ does
not mix with the other couplings in~$\hat{\s}_2$, and the same is true
for the left and right handed couplings.  {}From this block structure
of $\hat{\s}_2$ and the relation $\hat{\sigma}_2 A = (A^{-1})^T$ it
follows that in the last row of $A^{-1}$ we can set all entries to
zero, except for the Im$\,g_1^R$-, Im$\,\kappa_R$- and
Im$\,\lambda_R$-components.  Numerically we find that the absolute
values of those matrix entries which we set to zero are smaller than
$10^{-8}$ for $A$ and smaller than $10^{-4}$ for $A^{-1}$.  We remark
that we have computed the matrix $A^{-1}$ by inverting $A$ using
singular value decomposition~\cite{bib:numr}.  As mentioned at the end
of Sect.~\ref{ssec-disc} we have as a further check evaluated the
products $AA^T$ and \( (A^{-1})^T A^{-1} \).


\subsection{Polarised Beams}
\label{ssec-pola}

At future $e^+e^-$-colliders longitudinal polarisation of both initial
beams is
envisaged~\cite{Moortgat-Pick:2002gb,Menges:2001gg,Assmann:2001uq}.
An electron polarisation of \( P^- = \pm 80\% \) and a positron
polarisation of \( P^+ = \pm 60\% \) is considered to be achievable.

In Tables~\ref{tab:pol1} and~\ref{tab:pol2} we give the errors $\delta
h$ on the real couplings (in the presence of all couplings) for
$\sqrt{s}=$~\mbox{500 GeV} and various combinations of beam
polarisations.  For all couplings $h^L$ and all couplings $h^R$ we
find roughly the following gain or loss in sensitivity using always
the event rates of Table~\ref{tab:events}.  Turning on
$e^-$~polarisation of $-80\%$ we gain a factor of 1.4 for $h^L$ and
loose a factor of 6 for $h^R$.  If in addition \( P^+ = +60\% \) we
gain a factor of 1.8 for $h^L$ and loose a factor of 17 for $h^R$
compared to unpolarised beams.  For \( P^- = +80\% \) we loose a
factor of 2.6 for $h^L$ and gain a factor of 3.0 for $h^R$.  If
furthermore \( P^+ = -60\% \) we loose a factor of 5 for $h^L$ and
gain a factor of 5.5 for~$h^R$ compared to unpolarised beams.
Especially for the right handed couplings the gain from having both
beams polarised is thus appreciable.

The behaviour of the generalised eigenvalues as a function of the
parameter $P$ introduced in Sect.~\ref{sec-pola} is shown in
Figs.~\ref{fig:eigval500a}--\ref{fig:eigval3d}.  Although the four
largest eigenvalues are more or less constant for \( P < 0 \), the
transformation matrix $A^{-1}$ is not.  This can be seen from
Tables~\ref{tab:vec1}, \ref{tab:vec2} and~\ref{tab:vec3}.  For the
largest eigenvalue $c_1'$ of symmetry~(a) we find that the smaller $P$
is, the more the R-components are suppressed, i.e.\ the more one
purely measures the $h^L$.  Going from \( P = 0 \) to \( P = 1 \) we
become more and more sensitive to the $h^R$.  For the fourth lowest
curve in Fig.~\ref{fig:eigval500a}, corresponding to $c_5'$, as well
as for the smallest eigenvalue $c_8'$ of symmetry~(a) we find the
opposite tendency.  Note that in the case of $\pm 100\%$ electron or
positron polarisation we can only be sensitive to at most half of the
couplings.  This is seen for symmetries~(c) and~(d) in
Figs.~\ref{fig:eigval500c}, \ref{fig:eigval500d},
\ref{fig:eigval800c}, \ref{fig:eigval800d}, \ref{fig:eigval3c}
and~\ref{fig:eigval3d}: half of the curves go to zero at \( P = \pm 1
\).  For class~(a) (cf.~Figs.~\ref{fig:eigval500a},
\ref{fig:eigval800a} and~\ref{fig:eigval3a}) we find one additional
eigenvalue going to zero at \( P = +1 \) and for class~(b)
(cf.~Figs.~\ref{fig:eigval500b}, \ref{fig:eigval800b}
and~\ref{fig:eigval3b}) there is a zero eigenvalue for all $P$, as
explained in Sect.~\ref{sec-hmc}.  Comparing with Fig.~\ref{fig:par1}
we see that for symmetries~(b) to~(d) the shape of the curves is
qualitatively well described by the simple model of
Sect.~\ref{sec-pola}.  Although the lower and upper curves for
$c_{\pm}$ do not intersect in our examples there, an intersection like
in Fig.~\ref{fig:eigval500c} is not excluded.  In general, it is
however not possible to associate a certain pair of couplings to a
pair of curves in Figs.~\ref{fig:eigval500a}--\ref{fig:eigval3d} for
the full range of~$P$.  This is particularly obvious from the
eigenvalues of symmetry~(c) at \( \sqrt{s} = 3 \) TeV
(Fig.~\ref{fig:eigval3c}), where some curves alternately play the role
of the lower-type and upper-type curves in the simplified model.
Moreover, for symmetry~class~(a) the description of the shape of the
eigenvalue curves is less obvious due to the second term in the
brackets of~(\ref{eq:covopol}).


\subsection{Energy Dependence}
\label{ssec-ener}

The gain in sensitivity when going up from 500~GeV to 800~GeV---using
always the event rates of Table~\ref{tab:events}---lies between 1.4
and 2.7 for all couplings except for Im$\,\kappa_R$, where it is 3.6.
At 3~TeV we gain a factor of about 25 compared to 800~GeV for this
coupling, and of 1.5 to 8 for all others.  For symmetries~(a) and~(c)
we give $\delta h$ in Tables~\ref{tab:energ1} and~\ref{tab:energ2}.

Note that this gain is not due to the the total rate, which actually
decreases with energy (cf.~Table~\ref{tab:events}).  The largest gains
are achieved for $\kappa$, $\lambda$ and $\tilde{\lambda}$, which have
a prefactor $2\gamma^2$ in the amplitude.  We remark that both for
real and imaginary parts the gains in sensitivity for an L-coupling
and the corresponding R-coupling are of the same size, except for
Im$\,\kappa_R$.  Furthermore, except for $\Delta \kappa_L$, $g_4^L$,
$g_4^R$ and $\tilde{\kappa}_R$, the gain is slightly larger for the
imaginary than for the real parts.  For the real parts of the
couplings we also give the errors on the transformed couplings $\delta
h_i'$ in Tables~\ref{tab:energ3} and~\ref{tab:energ4}.  Note that the
transformations~(\ref{eq:gtrafo}) are not identical at the various
c.m.~energies, and neither are the couplings~$h_i'$.  Due to the
different normalisation of the $h_i'$ achieved by~(\ref{eq:gtrafo})
their errors $\delta h_i'$ may well increase with rising energy
although the errors in a fixed basis as in Tables~\ref{tab:energ1}
and~\ref{tab:energ2} decrease.


\subsection{Constraints from the Total Rate}
\label{ssec-totr}

As explained in Sect.~\ref{sec-opti} the measurement of the total
cross section restricts the anomalous TGCs in the $h_i'$-basis to a
shell between two hyperspheres in the multi-dimensional parameter
space.  For the couplings given in a basis before the transformation
we have hyperellipsoids instead of hyperspheres.  With \( \sqrt{s} =
\mbox{500 GeV} \) and unpolarised beams the expansion of the total
cross section~(\ref{eq:stotnorm}) is
\begin{eqnarray}
\lefteqn{
\sigma / \sigma_0  =  1 - 0.026  
}
\label{eq:trate500}
\\[0.6em]
 & & \mbox{} + (h_1' + 0.16)^2 + (h_2' + 0.026)^2 + (h_3' + 0.0042)^2
 + (h_4' + 0.0061)^2 
\nonumber \\
 & & \mbox{} + (h_5' - 0.013)^2 + (h_6' - 0.022)^2 + (h_7' - 0.0093)^2
 + (h_8' + 0.00013)^2
 + \sum_{i = 9}^{28} (h_i')^2.
\nonumber 
\end{eqnarray}
At 800~GeV we obtain
\begin{eqnarray}
\lefteqn{
\sigma / \sigma_0  =  1 - 0.016  
}
\\[0.6em]
 & & \mbox{} + (h_1' + 0.13)^2 + (h_2' + 0.0078)^2 + (h_3' + 0.0025)^2
 + (h_4' + 0.0027)^2 \nonumber \\
 & & \mbox{} + (h_5' - 0.0066)^2 + (h_6' - 0.013)^2 + (h_7' -
     0.0062)^2 + (h_8' - 0.00023)^2 
 + \sum_{i = 9}^{28} (h_i')^2,
\nonumber
\end{eqnarray}
and, finally, at 3~TeV
\begin{eqnarray}
\lefteqn{
\sigma / \sigma_0  =  1 - 0.0071  
}
\\[0.6em]
 & & \mbox{} + (h_1' + 0.084)^2 + (h_2' + 0.00083)^2 + (h_3' + 4.3
 \cdot 10^{-6})^2 + (h_4' + 0.00060)^2 \nonumber \\
 & & \mbox{} + (h_5' + 0.0030)^2 + (h_6' - 0.0052)^2 + (h_7' -
     0.0017)^2 + (h_8' + 3.2 \cdot 10^{-5})^2
 + \sum_{i = 9}^{28} (h_i')^2 .
\nonumber
\end{eqnarray}
We remark again that the couplings $h_i'$ are not the same at
different energies.  For a measurement of the rate~$N$ with a (purely
statistical) error \( \sqrt{N} \) the thickness of the shell is
$5.8\cdot 10^{-3}$ at 500~GeV, $7.2\cdot 10^{-3}$ at 800~GeV and
$18\cdot 10^{-3}$ at 3~TeV (cf.~Table~\ref{tab:events}).  Systematic
errors could be more important.  The results~(\ref{eq:trate500}) agree
quite well with those of~\cite{Diehl:1997ft} for all couplings except
for the smallest term $h_8'$ and for~$h_3'$.  Note that these results
strongly depend on a reliable transformation matrix $A$.
In~\cite{Diehl:1997ft} numerical instabilities occurred in the
diagonalisation procedure, whereas here $A$ is obtained iteratively as
explained in Sect.~\ref{sec-opti} and was found to be stable.

It has been pointed out~\cite{Diehl:1997ft} that the constraints from
the total rate are in general of the same size as the largest error on
the couplings determined from the normalised distribution, which we
confirm.


\section{Conclusions}
\label{sec-concl}

\suppressfloats

In this paper we have investigated the sensitivity of optimal
observables to anomalous TGCs in \pr\ at future linear colliders.  We
have treated all 28 couplings at a time and disentangled them by a
simultaneous diagonalisation of the covariance matrix of the
observables and the term of the integrated cross section which is
quadratic in the anomalous couplings.  Several relations arising from
discrete symmetry properties of the differential and the total cross
section, based on the classification of the TGCs according to their
$\cp$ and $\cpt$ eigenvalues, have been used to simplify the analysis
and to check the numerical stability of the results.  We have
re-examined the conditions for investigating $\cp$ violation in the
presence of polarised $e^+e^-$ beams.  Suitably defined $\cp$~odd
observables receive contributions only from genuine $\cp$~violation in
the interaction and from the $\cp$~odd part of the spin density matrix
of the initial $e^+e^-$~state.  The two kinds of effects can be
separated using their different dependence on the beam polarisations,
either with transversely polarised beams (cf.\ Sect.~\ref{ssec-disc})
or with suitable combinations of longitudinal polarisations (cf.\
Sect.~\ref{sec-pola}).  An analogous statement holds for the
investigation of absorptive parts in the scattering amplitude by
$\cpt$~odd observables.

We find that already at a 500~GeV collider with unpolarised beams and
the event rates of Table~\ref{tab:events}, the statistical errors
obtained with the optimal-observable method, treating {\em all}
couplings at a time, are considerably smaller than the constraints
obtained from {\em single} parameter fits of LEP2 data.  Our errors
for $\sqrt{s}=$~500~GeV and 800~GeV treating all couplings are of the
same size or smaller than those obtained at generator level for TESLA
by a spin density matrix method with a restricted number of
couplings~\cite{Richard:2001qm,Menges:2001gg}.

We have performed a detailed study of the sensitivity to anomalous
TGCs for different longitudinal polarisations and c.m.~energies of the
lepton beams.  The polarisation dependence is conveniently expressed
through a parameter~$P$.  A simple model has been analysed
(see~Figs.~\ref{fig:par1}, \ref{fig:par2}), which provides an
understanding of the dependence of sensitivities on~$P$.  We find that
beam polarisation can provide an appreciable gain in sensitivity,
especially for right handed couplings (cf.\ Tables~\ref{tab:pol1}
and~\ref{tab:pol2}).  At $\sqrt{s} = 500$~GeV the sensitivity to these
couplings increases by a factor of about 3 when going from unpolarised
beams to $+80\%$ electron polarisation.  With additional $-60\%$
positron polarisation this factor is about 5 or larger.

In the sector of $\cp$ conserving imaginary TGCs we have found one
linear combination of couplings which appears only quadratically in
the differential cross section for longitudinally polarised beams
(cf.~Sect.~\ref{sec-hmc}).  Therefore the normalised event
distribution of $W$~pair production for unpolarised or longitudinally
polarised beams does not provide a good way to measure this
coupling---regardless of whether the analysis uses optimal observables
or any other method.  Information on this coupling can however be
obtained either from the total event rate, or from the normalised
event distribution with {\em transversely} polarised beams.

Since our numerical results are at the ``theory level'', a study using
Monte Carlo event samples and including a detector simulation will at
some point be necessary.  As noted in Sect.~\ref{sec-opti} inclusion
of detector resolution and phase space cuts is easy in the framework
of the method presented here.

In this work we have analysed the normalised event distributions in
the linear approximation for anomalous TGCs.  We emphasise that the
extension of our method to the fully non-linear case is
available~\cite{Diehl:1997ft}.  Such an analysis for the non-linear
case has indeed been performed successfully with LEP2
data~\cite{bib:dham}.

To summarise, we have performed a detailed study of the sensitivities
achievable in the measurements of anomalous $\gamma W^+ W^-$ and  $Z
W^+ W^-$ couplings at future linear $e^+e^-$~colliders.  We advocate
the use of integrated optimal observables for this purpose.  We have
shown that our method allows for a very good overview of the
sensitivities and of their dependence on the beam polarisations in the
space of the 28 anomalous couplings.  Such measurements will be a
crucial check wether the structure imposed by the fundamental gauge
group \( SU(2)_L \times U(1) \) of the electroweak interactions in the
Standard Model for the couplings of three gauge bosons is realised in
nature.


\section*{Acknowledgements}

The authors are grateful to W.~Buchm\"uller, A.~Denner, W.~Menges,
K.~M\"onig, G.~Moortgat-Pick and P.~Zerwas for useful discussions.
This work was supported by the German Bundesminesterium f\"ur Bildung
und Forschung, project~no.~05HT9HVA3, and the Graduiertenkolleg
``Physikalische Systeme mit vielen Freiheitsgraden'' in Heidelberg.

\clearpage



\renewcommand{\arraystretch}{1.1}

\begin{table}[p]
\caption{\label{tab:sensa} Errors $\delta h$ in the presence of all
other couplings and correlation matrix $W(h)$ at $\sqrt{s}=$~500~GeV
with unpolarised beams for the couplings of symmetry (a) (see
Sect.~\ref{ssec-disc}), i.e.\ for the real parts of the $\cp$ even
couplings.}
\begin{center}
\leavevmode
\footnotesize
\begin{tabular}{l|r|rrrr|rrrr}
$\;\;\;h$ & \( \delta h\!\times\!10^3 \) & Re$\,\Delta g_1^L$ & 
Re$\,\Delta \kappa_L$ & Re$\,\lambda_L$ & Re$\,g_5^L$ & Re$\,\Delta
g_1^R$ & Re$\,\Delta \kappa_R$ & Re$\,\lambda_R$ & Re$\,g_5^R$ \\
\hline
&&&&&&&&&\\[-2.5ex]
Re$\,\Delta g_1^L$		& $2.6$   	& 1 		&
$-0.60$	& $-0.35$	& $0.21$	& $-0.070$	& $0.25$
& $-0.054$	& $0.15$ \\
Re$\,\Delta \kappa_L$	& $0.85$   	& 		& 1
& $0.096$	& $-0.054$	& $0.20$	& $-0.59$	&
$0.13$	& $0.019$ \\
Re$\,\lambda_L$	& $0.59$   	& 		& 		& 1
& $-0.034$	& $0.099$	& $-0.080$	& $0.030$	&
$0.10$ \\
Re$\,g_5^L$		& $2.0$   	& 		&
& 		& 1 		& $-0.084$	& $0.11$	&
$-0.13$	& $0.075$ \\
\hline
&&&&&&&&&\\[-2.5ex]
Re$\,\Delta g_1^R$		& $10$   	& 		&
& 		& 		& 1 		& $-0.70$	&
$-0.56$	& $0.65$ \\
Re$\,\Delta \kappa_R$	& $2.4$   	& 		&
& 		& 		& 		& 1 		&
$0.023$	& $-0.34$ \\
Re$\,\lambda_R$	& $3.6$   	& 		& 		&
& 		& 		& 		& 1 		&
$-0.25$ \\
Re$\,g_5^R$		& $6.7$   	& 		&
& 		& 		& 		& 		&
& 1 
\end{tabular}\\
\normalsize
\end{center}
\end{table}

\begin{table}[p]
\caption{\label{tab:sensb} Same as Table \protect\ref{tab:sensa}, but
for symmetry (b), i.e.\ the imaginary parts of the $\cp$ even
couplings.  As explained in Sect.~\protect\ref{sec-hmc} we have
$\delta\,\im (g_1^R + \kappa_R) = \infty$ and no correlation
of this coupling with the others.  Thus we only give the reduced $7
\times 7$ matrix here.}
\begin{center}
\leavevmode
\footnotesize
\begin{tabular}{l|r|rrrr|rrr}
$\;\;\;h$ & \( \delta h\!\times\!10^3 \) & Im$\,g_1^L$ & Im$\,\kappa_L$ & 
Im$\,\lambda_L$ & Im$\,g_5^L$ & $\frac{1}{\sqrt{2}} \im(g_1^R - \kappa_R)$ & 
Im$\,\lambda_R$ & Im$\,g_5^R$ \\
\hline
&&&&&&&&\\[-2.5ex]
Im$\,g_1^L$     & $2.7$  	& 1 		& $-0.47$	&
$-0.50$	& $-0.12$	& $0.028$	& $0.16$	& $0.038$ \\
Im$\,\kappa_L$ 	& $1.7$   	& 		& 1 		&
$0.0070$	& $0.41$	& $0.33$	&	$-0.10$	& $0.68$ \\
Im$\,\lambda_L$	& $0.48$   	& 		& 		& 1
& $-0.15$	& $-0.00069$  & $-0.21$	&	$-0.22$ \\
Im$\,g_5^L$     & $2.5$   	& 		& 		&
& 1 		& $0.081$	& $0.22$	&$0.50$ \\
\hline
&&&&&&&&\\[-2.5ex]
$\frac{1}{\sqrt{2}} \im(g_1^R - \kappa_R)$     	& $11$  & & & & & 1 	& $-0.53$ & $0.60$ \\
Im$\,\lambda_R$				& $3.1$ & & & & & & 1 	& $-0.11$ \\
Im$\,g_5^R$     			& $17$  & & & & & & & 1 
\end{tabular}\\
\normalsize
\end{center}
\end{table}

\begin{table}[p]
\caption{\label{tab:sensc} Same as Table \protect\ref{tab:sensa}, but
for symmetry (c), i.e.\ the real parts of the $\cp$ violating
couplings.}
\begin{center}
\leavevmode
\footnotesize
\begin{tabular}{l|r|rrr|rrr}
$\;\;\;h$ & \( \delta h\!\times\!10^3 \) & Re$\,g_4^L$ & 
Re$\,\tilde{\lambda}_L$ &
Re$\,\tilde{\kappa}_L$ & Re$\,g_4^R$ & Re$\,\tilde{\lambda}_R$ &
Re$\,\tilde{\kappa}_R$ \\
\hline
&&&&&&&\\[-2.5ex]
Re$\,g_4^L$		& $2.5$   	& 1 		& $-0.055$
& $-0.49$	& $-0.091$	& $-0.18$	& $0.073$ \\
Re$\,\tilde{\lambda}_L$	& $0.60$   	& 		& 1
& $0.27$	& $0.073$	& $0.0088$	& $-0.16$ \\
Re$\,\tilde{\kappa}_L$	& $2.7$   	& 		&
& 1 		& $0.036$	& $0.11$	& $0.14$ \\
\hline
&&&&&&&\\[-2.5ex]
Re$\,g_4^R$		& $10$   	& 		&
& 		& 1 		& $-0.24$	& $-0.47$ \\
Re$\,\tilde{\lambda}_R$	& $3.8$   	& 		&
& 		& 		& 1 		& $0.65$ \\
Re$\,\tilde{\kappa}_R$	& $11$   	& 		&
& 		& 		& 		& 1 
\end{tabular}\\
\normalsize
\end{center}
\end{table}

\begin{table}[p]
\caption{\label{tab:sensd} Same as Table \protect\ref{tab:sensa}, but
for symmetry (d), i.e.\ the imaginary parts of the $\cp$ violating
couplings.}
\begin{center}
\leavevmode
\footnotesize
\begin{tabular}{l|r|rrr|rrr}
$\;\;\;h$ & \( \delta h\!\times\!10^3 \) & Im$\,g_4^L$ & 
Im$\,\tilde{\lambda}_L$ &
Im$\,\tilde{\kappa}_L$ & Im$\,g_4^R$ & Im$\,\tilde{\lambda}_R$ &
Im$\,\tilde{\kappa}_R$ \\
\hline
&&&&&&&\\[-2.5ex]
Im$\,g_4^L$		& $1.9$   	& 1 		& $-0.059$
& $0.092$	& $0.20$	& $0.22$	& $-0.017$ \\
Im$\,\tilde{\lambda}_L$	& $0.46$   	& 		& 1
& $0.53$	& $-0.15$	& $-0.18$	& $-0.015$ \\
Im$\,\tilde{\kappa}_L$	& $2.0$   	& 		&
& 1 		& $-0.33$	& $-0.099$	& $0.14$ \\
\hline
&&&&&&&\\[-2.5ex]
Im$\,g_4^R$		& $7.7$   	& 		&
& 		& 1 		& $-0.12$	& $-0.68$ \\
Im$\,\tilde{\lambda}_R$	& $2.9$   	& 		&
& 		& 		& 1 		& $0.56$ \\
Im$\,\tilde{\kappa}_R$	& $8.6$   	& 		&
& 		& 		& 		& 1 
\end{tabular}\\
\normalsize
\end{center}
\end{table}


\begin{table}[p]
\caption{\label{tab:eveven} Generalised eigenvalues $c_i'$ of the
covariance matrix $V({\cal O})$ and the corresponding errors $\delta
h_i'$ on the transformed couplings, obtained from
(\protect\ref{eq:delg1p}) and Table \protect\ref{tab:events} at
\mbox{500 GeV} with unpolarised beams for symmetries (a) and (b).}
\begin{center}
\leavevmode
\footnotesize
\begin{tabular}{r|r|r||r|r|r}
$i$ & $c_i'$ & $\delta h_i'\!\times\!10^3$ & $i$ & $c_i'$ & $\delta
h_i'\!\times\!10^3$ \\ 
\hline
&&&&& \\[-2.5ex]
1 & 1.44 & 0.780 & 9 & 1.27 & 0.831 \\
2 & 1.17 & 0.866 & 10 & 1.01 & 0.931 \\
3 & 0.751 & 1.08 & 11 & 0.791 & 1.05 \\
4 & 0.557 & 1.25 & 12 & 0.287 & 1.75 \\
5 & 0.318 & 1.66 & 13 & 0.0584 & 3.88 \\
6 & 0.108 & 2.85 & 14 & 0.0221 & 6.30 \\
7 & 0.0366 & 4.90 & 15 & 0.0102 & 9.29 \\
8 & 0.0147 & 7.72 & 16 & $0$ & $\infty$
\end{tabular}
\normalsize
\end{center}
\end{table}

\begin{table}[p]
\caption{\label{tab:evodd} Same as Table \protect\ref{tab:eveven} but
for symmetries (c) and (d).}
\begin{center}
\leavevmode
\footnotesize
\begin{tabular}{r|r|r||r|r|r}
$i$ & $c_i'$ & $\delta h_i'\!\times\!10^3$ & $i$ & $c_i'$ & $\delta
h_i'\!\times\!10^3$ \\ 
\hline
&&&&& \\[-2.5ex]
17 & 1.17 & 0.868 & 23 & 1.40 & 0.792 \\
18 & 0.585 & 1.23 & 24 & 1.02 & 0.929 \\
19 & 0.320 & 1.66 & 25 & 0.829 & 1.03 \\
20 & 0.0645 & 3.69 & 26 & 0.219 & 2.00 \\
21 & 0.0262 & 5.78 & 27 & 0.0316 & 5.27 \\
22 & 0.0131 & 8.18 & 28 & 0.0241 & 6.04
\end{tabular}
\normalsize
\end{center}
\end{table}


\begin{table}[p]
\caption{\label{tab:amat} Coefficient matrix \( A \times 10^2 \) for
symmetry (b)}
\begin{center}
\leavevmode
\footnotesize
\begin{tabular}{l|rrrrrrrr}
& $h_9'$ & $h_{10}'$ & $h_{11}'$ & $h_{12}'$ & $h_{13}'$ & $h_{14}'$
& $h_{15}'$ & $h_{16}'$  \\
\hline
&&&&&&&\\[-2.5ex]
Im$\,g_1^L$    	& $140$  	& $-85$  	& $-130$  	&
$97$  	& $12$  	& $7.1$  	& $0.86$  	& $0$ \\
Im$\,\kappa_L$ 	& $-4.3$  	& $3.4$  	& $-1.3$  	&
$-64$  	& $-15$  	& $0.80$  	& $12$  	& $0$ \\
Im$\,\lambda_L$	& $6.6$  	& $46$  	& $10$  	&
$-6.1$  	& $0.87$  	& $-1.9$  	& $-0.94$  	& $0$
\\
Im$\,g_5^L$    	& $120$  	& $-52$  	& $150$  	&
$-7.9$  	& $-16$  	& $13$  	& $12$  	& $0$
\\ \hline
Im$\,g_1^R$    	& $6.3$  	& $-11$  	& $8.2$  	&
$-23$  	& $170$  	& $-91$  	& $110$  	& $-50$ \\
Im$\,\kappa_R$ 	& $-0.89$  	& $1.7$  	& $-1.4$  	&
$5.1$  	& $-37$  	& $7.6$  	& $-21$  	&
$-50$ \\
Im$\,\lambda_R$	& $-1.8$  	& $2.3$  	& $-1.2$  	&
$-0.50$  	& $0.62$  	& $47$  	& $-6.5$  	& $0$
\\
Im$\,g_5^R$    	& $-22$  	& $16$  	& $1.4$  	&
$55$  	& $-86$  	& $24$  	& $170$  	& $0$ 
\end{tabular}\\
\normalsize
\end{center}
\end{table}

\begin{table}[p]
\caption{\label{tab:ainvmat} Coefficient matrix \( A^{-1} \times 10^2
\) for symmetry (b)}
\begin{center}
\leavevmode
\footnotesize
\begin{tabular}{r|rrrr|rrrr}
$i$ & Im$\,g_1^L$ & Im$\,\kappa_L$ & Im$\,\lambda_L$ & Im$\,g_5^L$ & 
Im$\,g_1^R$ & Im$\,\kappa_R$ & Im$\,\lambda_R$ & Im$\,g_5^R$ \\
\hline
&&&&&&&\\[-2.5ex]
$9$	& $42$  & $45$  & $110$	& $29$
& $0.49$  	& $-0.49$  	& $-6.7$ & $-5.3$ \\
$10$	& $1.5$  	& $-12$	& $210$ & $-12$ 	
& $-1.2$	& $1.2$  	& $7.3$ & $3.9$ \\
$11$	& $-33$  	& $-54$ & $-15$ & $37$
& $1.1$  	& $-1.1$  	& $-2.6$ & $0.35$ \\
$12$	& $-0.73$  	& $-140$  	& $2.0$  	& $-1.9$
& $-5.0$  	& $5.0$  	& $-14$ & $13$ \\
$13$	& $-2.1$  	& $-38$  	& $7.2$  	& $-4.0$
& $36$  	& $-36$  	& $87$  & $-21$ \\
$14$	& $1.3$  	& $4.6$  	& $-6.6$  	& $3.1$
& $2.2$  	& $-2.2$  	& $210$ & $5.7$ \\
$15$	& $4.5$  	& $34$  	& $-1.5$  	& $2.8$
& $19$  	& $-19$  	& $17$  & $41$ \\
$16$	& $0$   	& $0$   	& $0$   	& $0$
& $-35$  	& $-160$  	& $-40$ & $0$  
\end{tabular}
\normalsize
\end{center}
\end{table}


\begin{table}[p]
\caption{\label{tab:pol1} Errors \( \delta h\!\times\!10^3 \) on the
couplings of symmetry (a) at \mbox{500 GeV} for different initial beam
polarisations.}
\begin{center}
\leavevmode
\footnotesize
\begin{tabular} {c|c|rrrr|rrrr}
$P^-$ & $P^+$ & Re$\,\Delta g_1^L$ & Re$\,\Delta
\kappa_L$ & Re$\,\lambda_L$ & Re$\,g_5^L$ & Re$\,\Delta g_1^R$ & 
Re$\,\Delta \kappa_R$ & Re$\,\lambda_R$ & Re$\,g_5^R$ \\
\hline 
&&&&&&&&& \\[-2.5ex]
$-80\%$ & $+60\%$ & 1.5 & 0.47 & 0.34 & 1.1 & 169 & 40 & 57 & 112 \\
$-80\%$ & $0$     & 1.9 & 0.60 & 0.43 & 1.5 & 62 & 14 & 21 & 41 \\
$0$     & $0$     & 2.6 & 0.85 & 0.59 & 2.0 & 10 & 2.4 & 3.6 & 6.7 \\
$+80\%$ & $0$     & 6.9 & 2.3 & 1.5 & 5.3 & 3.5 & 0.83 & 1.2 & 2.3 \\
$+80\%$ & $-60\%$ & 13 & 4.5 & 2.8 & 10 & 2.0 & 0.47 & 0.67 & 1.3
\end{tabular}
\normalsize
\end{center}
\end{table}

\begin{table}[p]
\caption{\label{tab:pol2} Same as Table \protect\ref{tab:pol1}, but
for symmetry (c).}
\begin{center}
\leavevmode
\footnotesize
\begin{tabular} {c|c|rrr|rrr}
$P^-$ & $P^+$ & Re$\,g_4^L$ & Re$\,\tilde{\lambda}_L$ &
Re$\,\tilde{\kappa}_L$ & Re$\,g_4^R$ & Re$\,\tilde{\lambda}_R$ &
Re$\,\tilde{\kappa}_R$ \\
\hline
&&&&&&& \\[-2.5ex]
$-80\%$ & $+60\%$ & 1.4 & 0.34 & 1.5 & 174 & 61 & 193 \\
$-80\%$ & $0$     & 1.8 & 0.43 & 1.9 & 62 & 22 & 69 \\
$0$     & $0$     & 2.5 & 0.60 & 2.7 & 10 & 3.8 & 11\\
$+80\%$ & $0$     & 6.5 & 1.5 & 6.9 & 3.2 & 1.3 & 3.7 \\
$+80\%$ & $-60\%$ & 13 & 2.9 & 13 & 1.8 & 0.70 & 2.0
\end{tabular}
\normalsize
\end{center}
\end{table}


\begin{table}[p]
\caption{\label{tab:vec1} Vector components belonging to the largest
eigenvalue $c_1'$ of \mbox{symmetry (a)} at \mbox{500 GeV} for
different longitudinal polarisations (two digits precision). Each line
is the first row of \( A^{-1} \times 10^2 \) in the LR-basis.}
\begin{center}
\leavevmode
\footnotesize
\begin{tabular} {c|c|rrrr|rrrr}
$P^-$ & $P^+$ & Re$\,\Delta g_1^L$ & Re$\,\Delta \kappa_L$ &
Re$\,\lambda_L$ & Re$\,g_5^L$ & Re$\,\Delta g_1^R$ & Re$\,\Delta
\kappa_R$ & Re$\,\lambda_R$ & Re$\,g_5^R$ \\ 
\hline
&&&&&&&&& \\[-2.5ex]
$-80\%$ & $+60\%$ & $-$34 & $-$150 & $-$33 & 18 & $-$0.19 & $-$1.2 &
$-$0.089 & 0.11 \\
$-80\%$ & $0$	  & $-$34 & $-$150 & $-$33 & 17 & $-$0.72 & $-$4.5 &
$-$0.36 & 0.41 \\
$0$ 	& $0$ 	  & $-$32 & $-$150 & $-$29 & 11 & $-$9.0 & $-$48 &
$-$7.1 & 4.8 \\ 
$+80\%$ & $0$	  & $-$13 & $-$84 & $-$10 & $-$3.9 & $-$91 & $-$390 &
$-$110 & 41 \\ 
$+80\%$ & $-60\%$ & $-$5.5 & $-$42 & $-$4.9 & $-$2.6 & $-$190 & $-$840 &
$-$240 & 79 
\end{tabular}
\normalsize
\end{center}
\end{table}

\begin{table}[p]
\caption{\label{tab:vec2} Same as Table \protect\ref{tab:vec1}, but
for the eigenvalue $c_5'$ of symmetry (a).}
\begin{center}
\leavevmode
\footnotesize
\begin{tabular} {c|c|rrrr|rrrr}
$P^-$ & $P^+$ & Re$\,\Delta g_1^L$ & Re$\,\Delta \kappa_L$ & 
Re$\,\lambda_L$ & Re$\,g_5^L$ & Re$\,\Delta g_1^R$ & Re$\,\Delta
\kappa_R$ & Re$\,\lambda_R$ & Re$\,g_5^R$ \\
\hline
&&&&&&&&& \\[-2.5ex]
$-80\%$ & $+60\%$ & $-$1.5 & $-$7.9 & 1.1 & $-$2.1 & 7.0 & 14 &
8.3 & $-$5.6 \\
$-80\%$ & $0$	  & $-$2.9 & $-$16 & 2.1 & $-$4.1 & 13 & 28 &
16 & $-$10 \\ 
$0$ 	& $0$ 	  & $-$9.5 & $-$50 & 6.9 & $-$14 & 37 & 82 &
44 & $-$27 \\
$+80\%$ & $0$	  & $-$13 & $-$41 & 150 & $-$20 & 14 & 42 & $-$9.6
& $-$39 \\ 
$+80\%$ & $-60\%$ & $-$20 & $-$72 & 100 & $-$24 & 26 & 120 & $-$28 &
$-$18 
\end{tabular}
\normalsize
\end{center}
\end{table}

\begin{table}[p]
\caption{\label{tab:vec3} Same as Table \protect\ref{tab:vec1}, but
for the smallest eigenvalue $c_8'$ of symmetry (a).}
\begin{center}
\leavevmode
\footnotesize
\begin{tabular} {c|c|rrrr|rrrr}
$P^-$ & $P^+$ & Re$\,\Delta g_1^L$ & Re$\,\Delta \kappa_L$ & 
Re$\,\lambda_L$ & Re$\,g_5^L$ & Re$\,\Delta g_1^R$ & Re$\,\Delta
\kappa_R$ & Re$\,\lambda_R$ & Re$\,g_5^R$ \\
\hline
&&&&&&&&& \\[-2.5ex]
$-80\%$ & $+60\%$ & 0.029 & 1.0 & 0.071 & $-$0.16 & 0.81 & $-$
4.1 & 38 & $-$0.91 \\ 
$-80\%$ & $0$	  & 0.054 & 1.8 & 0.17 & $-$0.31 & 1.6 &
$-$7.5 & 73 & $-$1.7 \\
$0$ 	& $0$ 	  & 0.15 & 4.8 & 0.64 & $-$1.0 & 5.4 & $-$21 &
220 & $-$4.8 \\
$+80\%$ & $0$	  & 0.26 & 14 & 1.2 & $-$3.2 & 17 & $-$61 &
650 & $-$13 \\
$+80\%$ & $-60\%$ & $-$3.2 & 47 & $-$5.1 & $-$8.5 & 38 & $-$170 & 1100
& $-$24 \\
\end{tabular}
\normalsize
\end{center}
\end{table}


\begin{table}[p]
\caption{\label{tab:energ1} Errors \( \delta h\!\times\!10^3 \) on the
couplings of symmetry (a) for different c.m.\ energies.}
\begin{center}
\leavevmode
\footnotesize
\begin{tabular} {r|rrrr|rrrr}
$\sqrt{s}$ [GeV] & Re$\,\Delta g_1^L$ & Re$\,\Delta \kappa_L$ & 
Re$\,\lambda_L$ & Re$\,g_5^L$ & Re$\,\Delta g_1^R$ & Re$\,\Delta
\kappa_R$ & Re$\,\lambda_R$ & Re$\,g_5^R$ \\
\hline
&&&&&&&& \\[-2.5ex]
500  & 2.6 & 0.85 & 0.59 & 2.0 & 10 & 2.4 & 3.6 & 6.7 \\
800  & 1.6 & 0.35 & 0.24 & 1.4 & 6.2 & 0.92 & 1.8 & 4.8 \\
3000 & 0.93 & 0.051 & 0.036 & 0.88 & 3.1 & 0.12 & 0.36 & 3.2
\end{tabular}
\end{center}
\end{table}

\begin{table}[p]
\caption{\label{tab:energ2} Same as Table \protect\ref{tab:energ1} but
for symmetry (c).}
\begin{center}
\leavevmode
\footnotesize
\begin{tabular} {r|rrr|rrr}
$\sqrt{s}$ [GeV] & Re$\,g_4^L$ & Re$\,\tilde{\lambda}_L$ &
Re$\,\tilde{\kappa}_L$ & Re$\,g_4^R$ & Re$\,\tilde{\lambda}_R$ &
Re$\,\tilde{\kappa}_R$ \\
\hline
&&&&&& \\[-2.5ex]
500  & 2.5 & 0.60 & 2.7 & 10 & 3.8 & 11 \\
800  & 1.7 & 0.24 & 1.8 & 6.5 & 1.8 & 6.8 \\
3000 & 0.90 & 0.036 & 0.97 & 3.4 & 0.36 & 3.2
\end{tabular}
\normalsize
\end{center}
\end{table}


\begin{table}[p]
\caption{\label{tab:energ3} Errors \( \delta h_i'\!\times\!10^3 \) on
the transformed couplings of symmetry (a) at different c.m.\
energies.}
\begin{center}
\leavevmode
\footnotesize
\begin{tabular}{r|rrr}
$i$ & 500 GeV & 800 GeV & 3 TeV \\
\hline
&&& \\[-2.5ex]
1 & 0.780 & 0.765 & 1.26 \\
2 & 0.866 & 0.841 & 1.35 \\
3 & 1.08 & 1.16 & 2.02 \\
4 & 1.25 & 1.26 & 2.39 \\
5 & 1.66 & 1.83 & 4.18 \\
6 & 2.85 & 3.07 & 5.29 \\
7 & 4.90 & 4.96 & 8.54 \\
8 & 7.72 & 9.27 & 20.8
\end{tabular}
\normalsize
\end{center}
\end{table}

\begin{table}[p]
\caption{\label{tab:energ4} Same as Table \protect\ref{tab:energ3} but
for symmetry (c).}
\begin{center}
\leavevmode
\footnotesize
\begin{tabular}{r|rrr}
$i$ & 500 GeV & 800 GeV & 3 TeV \\
\hline
&&& \\[-2.5ex]
17 & 0.868 & 0.832 & 1.35 \\
18 & 1.23 & 1.22 & 2.03 \\
19 & 1.66 & 1.58 & 2.52 \\
20 & 3.69 & 3.39 & 5.12 \\
21 & 5.78 & 5.54 & 8.74 \\
22 & 8.18 & 9.53 & 20.9
\end{tabular}
\normalsize
\end{center}
\end{table}


\clearpage


\begin{figure}
\begin{center}
\leavevmode
\includegraphics[totalheight=15cm]{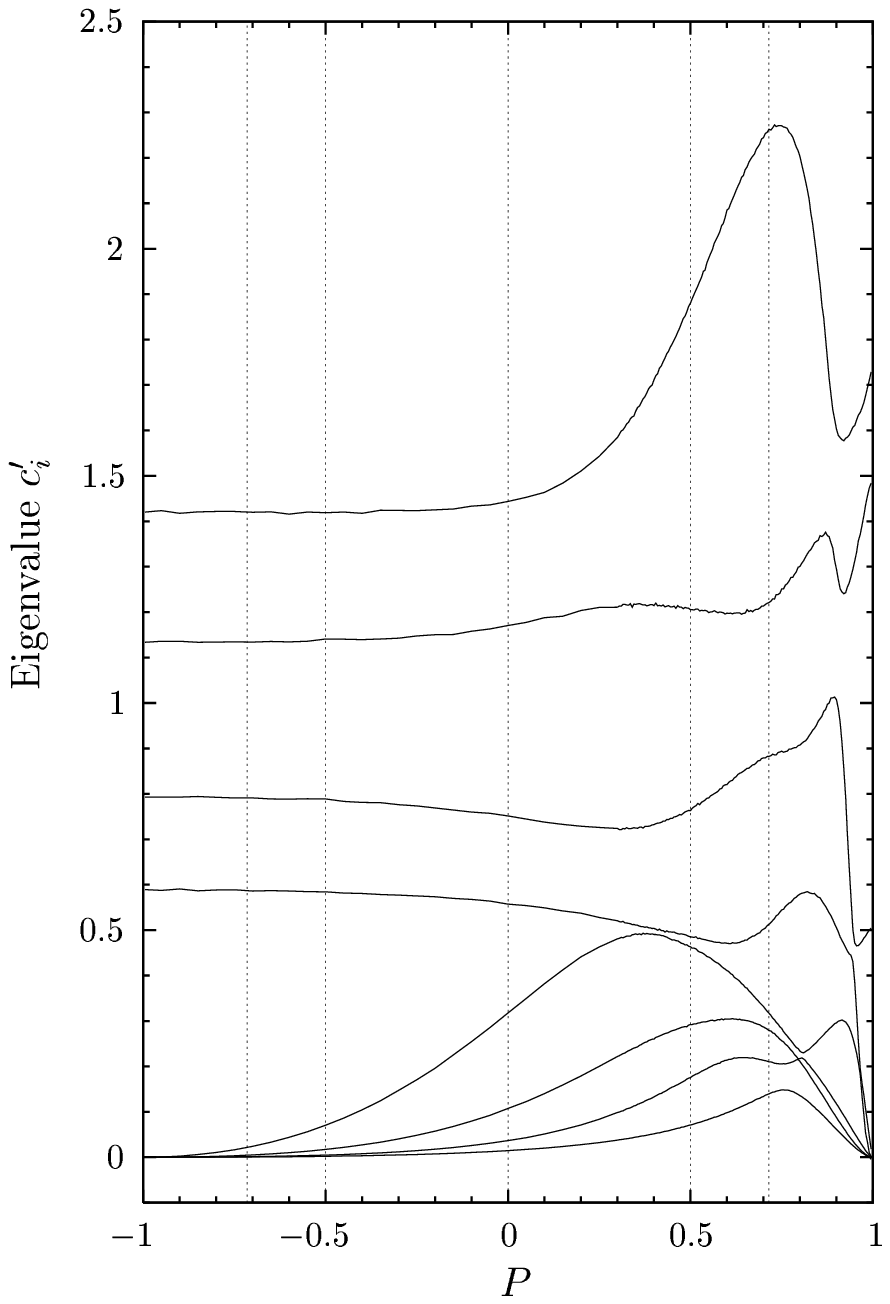}
\end{center}
\caption{\label{fig:eigval500a} Generalised eigenvalues $c_i'$ of the
correlation matrix $V({\cal O})$ for the couplings of symmetry class
(a) (cf.~Sect.~\protect\ref{ssec-disc}) at \mbox{\( \sqrt{s} = 500
{\rm \ GeV} \)}.  The $c_i'$ do not depend on the total rate $N$.
Errors on the transformed couplings $h_i'$ (\protect\ref{eq:gtrafo})
are obtained as $\delta h_i' = (N c'_i)^{-1/2}$.  Vertical lines mark
the five cases investigated in detail in
Sect.~\protect\ref{ssec-pola}, i.e.\ from left to right $(P^-, P^+) =
(-80\%,+60\%)$, $(-80\%,0)$, $(0,0)$, $(+80\%,0)$,
$(+80\%,-60\%)$. $P$ is given in terms of $P^-$ and $P^+$
by~(\protect\ref{eq:rp}).}
\end{figure}

\begin{figure}
\begin{center}
\leavevmode
\includegraphics[totalheight=15cm]{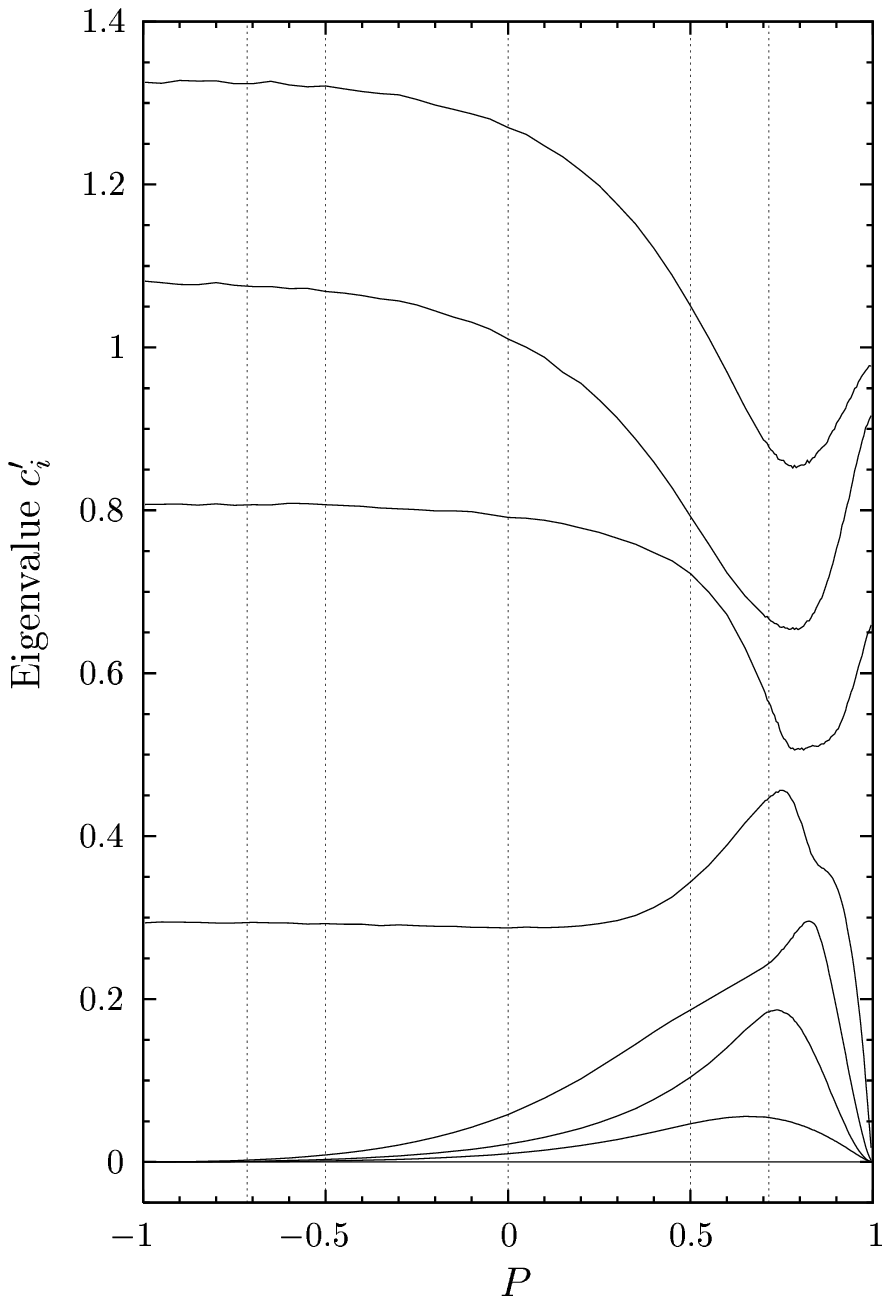}
\end{center}
\caption{\label{fig:eigval500b} Same as Fig.\
\protect\ref{fig:eigval500a} for symmetry class (b).}
\end{figure}

\begin{figure}
\begin{center}
\leavevmode
\includegraphics[totalheight=15cm]{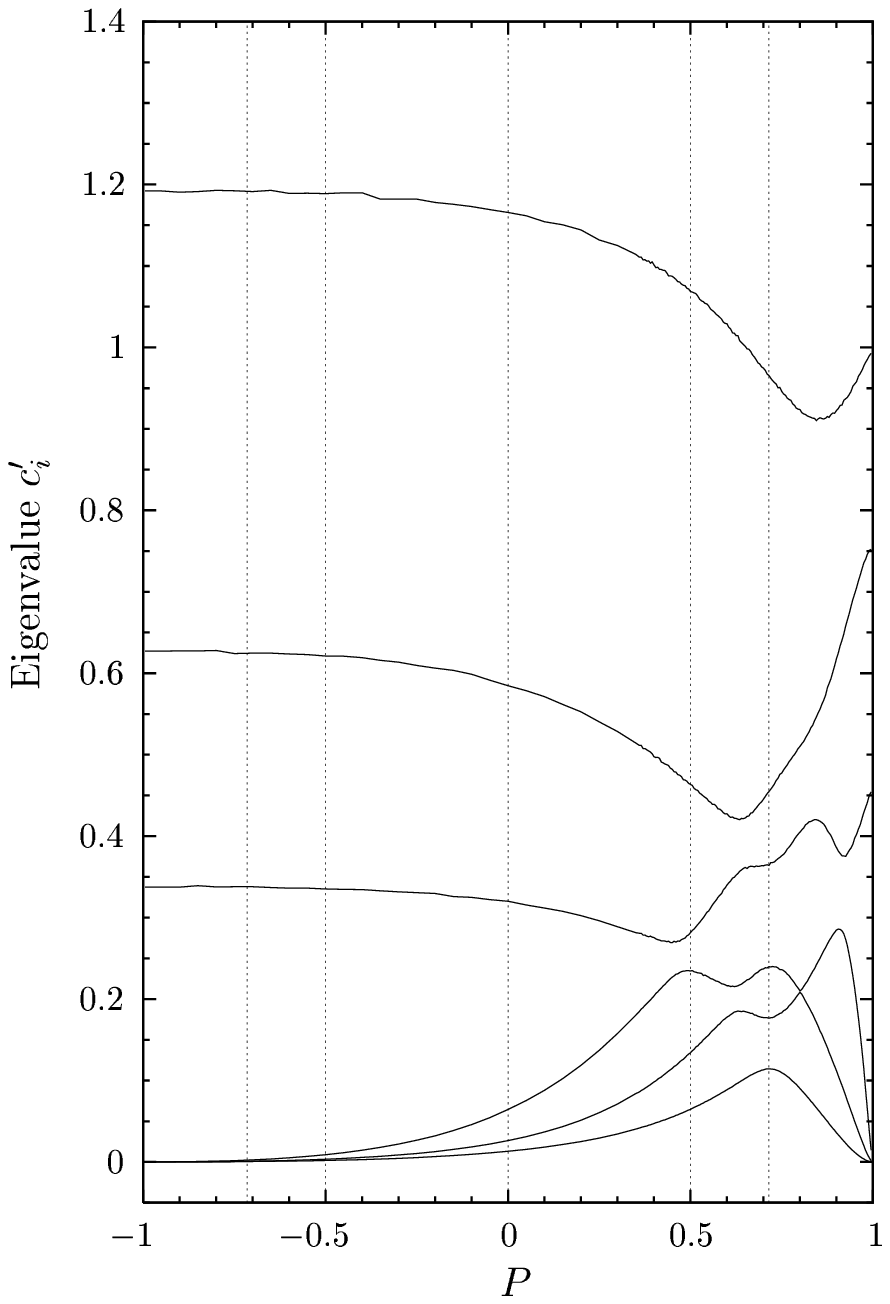}
\end{center}
\caption{\label{fig:eigval500c} Same as Fig.\
\protect\ref{fig:eigval500a} for symmetry class (c).}
\end{figure}

\begin{figure}
\begin{center}
\leavevmode
\includegraphics[totalheight=15cm]{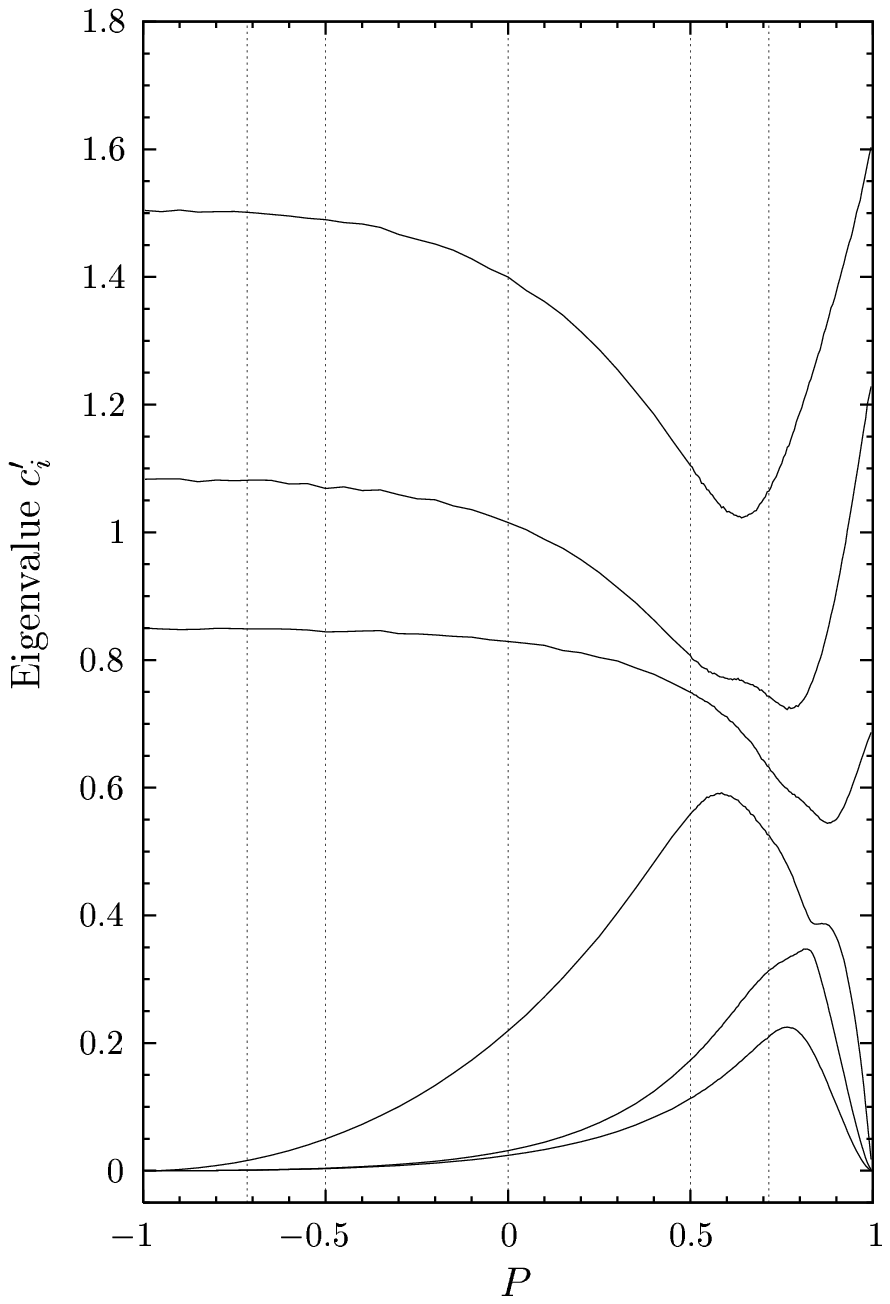}
\end{center}
\caption{\label{fig:eigval500d} Same as Fig.\
\protect\ref{fig:eigval500a} for symmetry class (d).}
\end{figure}
\begin{figure}
\begin{center}
\leavevmode
\includegraphics[totalheight=15cm]{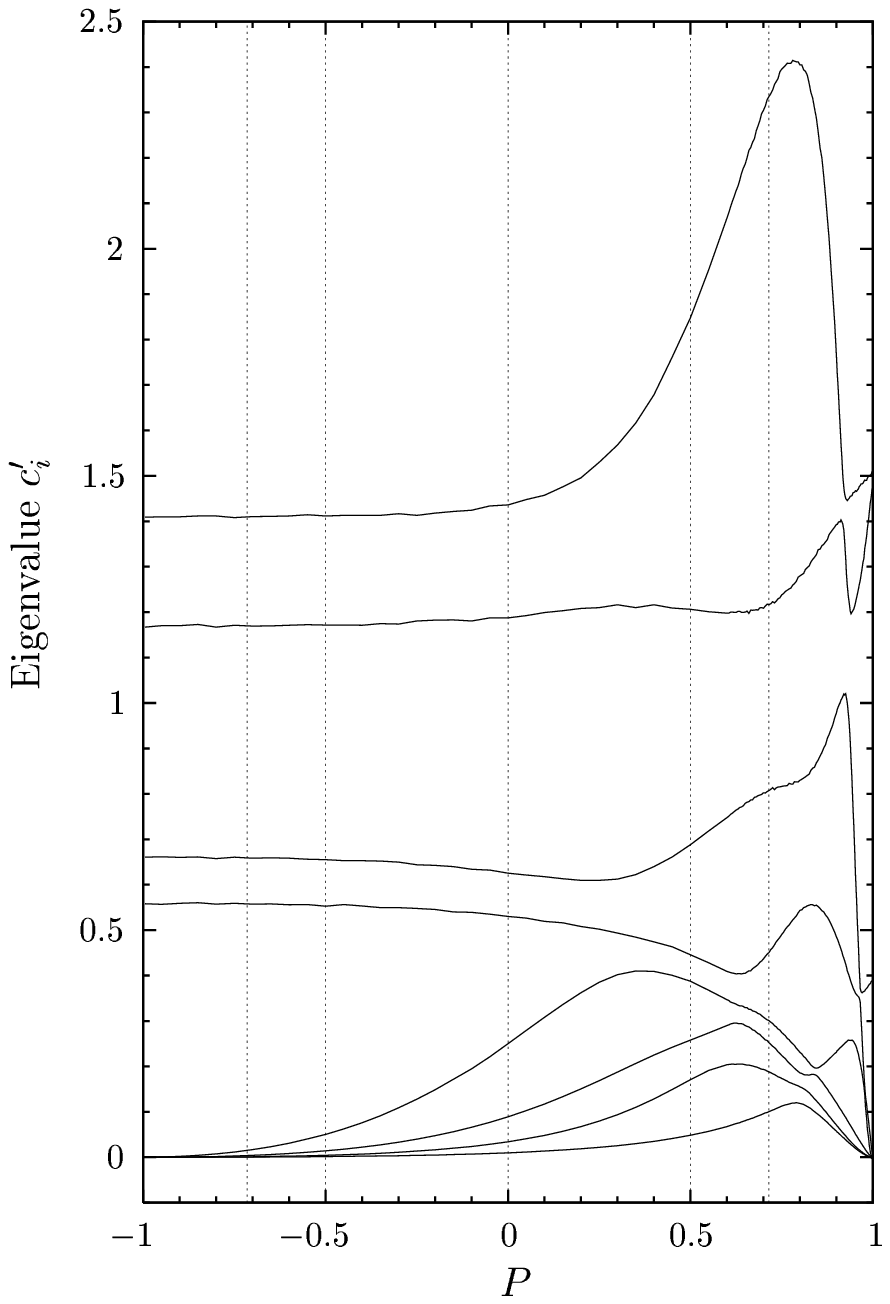}
\end{center}
\caption{\label{fig:eigval800a} Same as Fig.\
\protect\ref{fig:eigval500a} for \mbox{\( \sqrt{s} = 800 {\rm \ GeV}
\)} (symmetry class (a)).}
\end{figure}

\begin{figure}
\begin{center}
\leavevmode
\includegraphics[totalheight=15cm]{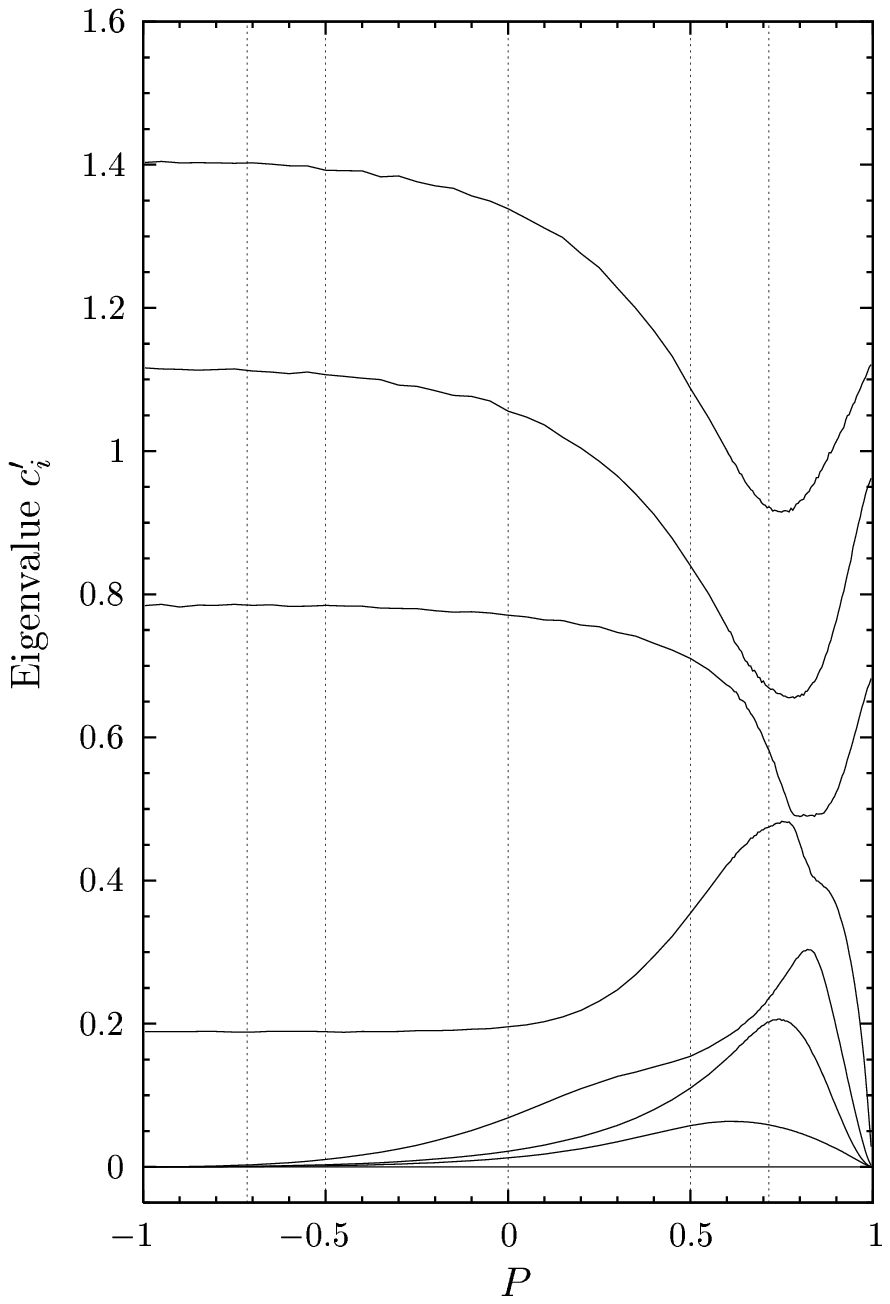}
\end{center}
\caption{\label{fig:eigval800b} Same as Fig.\
\protect\ref{fig:eigval500a} for \mbox{\( \sqrt{s} = 800 {\rm \ GeV}
\)} and symmetry class (b).}
\end{figure}

\begin{figure}
\begin{center}
\leavevmode
\includegraphics[totalheight=15cm]{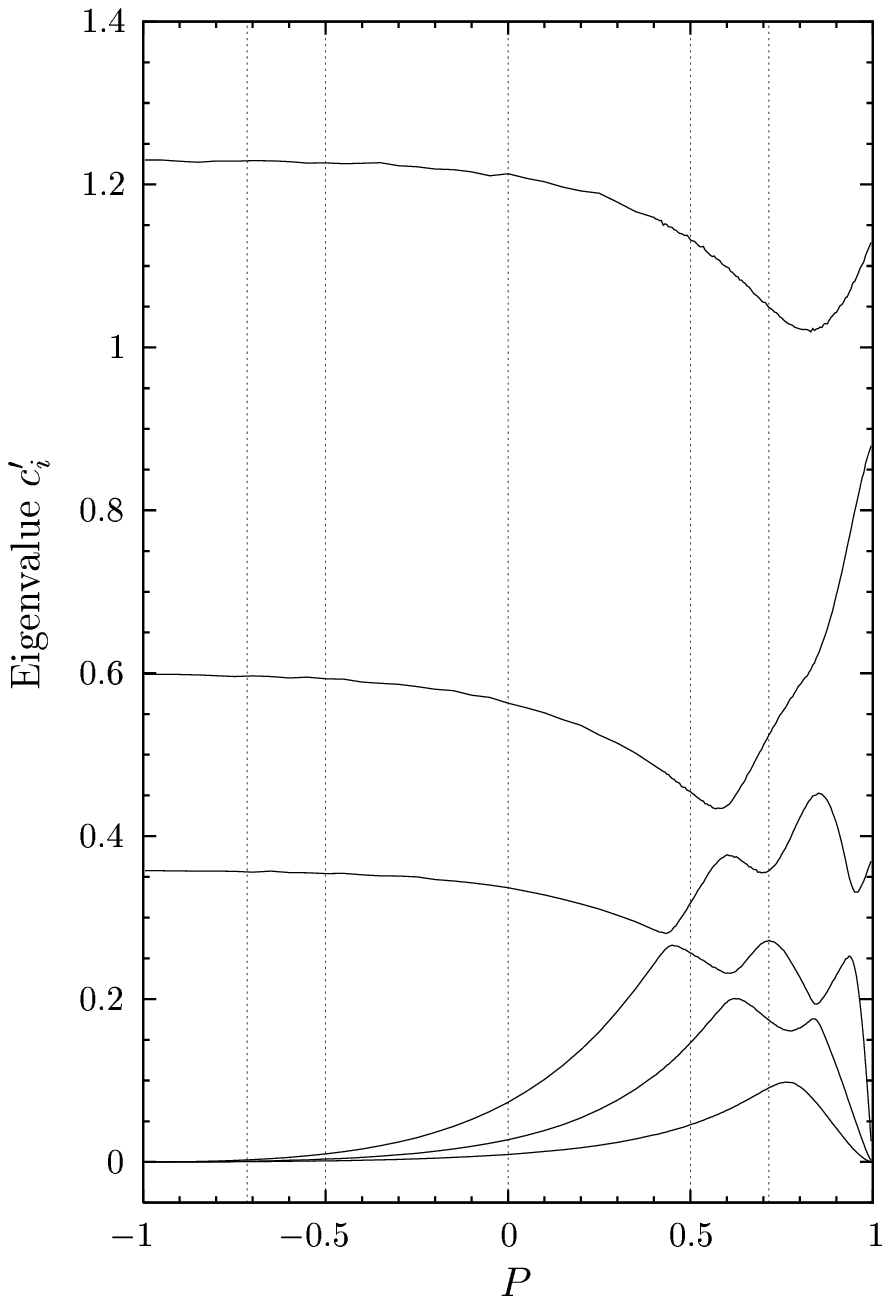}
\end{center}
\caption{\label{fig:eigval800c} Same as Fig.\
\protect\ref{fig:eigval500a} for \mbox{\( \sqrt{s} = 800 {\rm \ GeV}
\)} and symmetry class (c).}
\end{figure}

\begin{figure}
\begin{center}
\leavevmode
\includegraphics[totalheight=15cm]{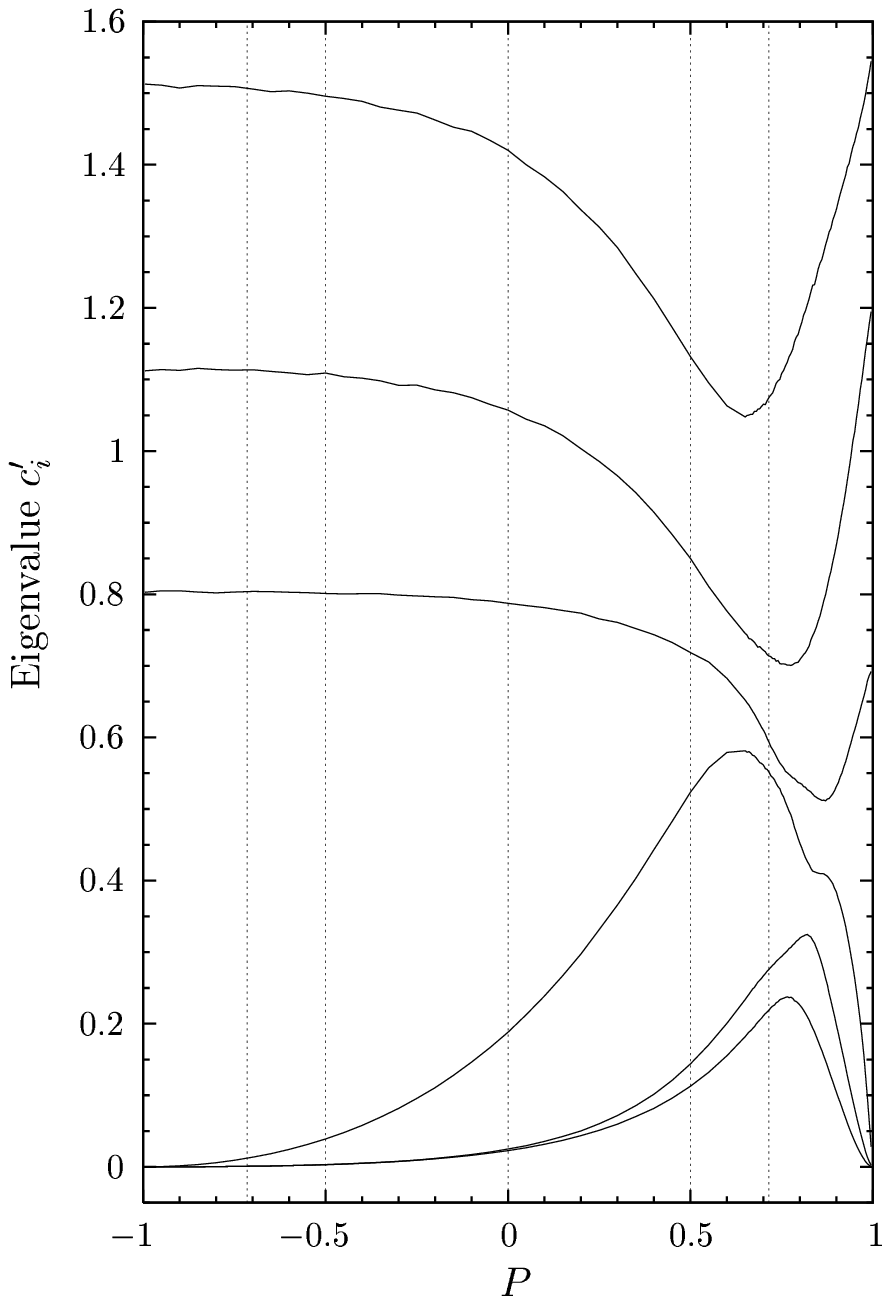}
\end{center}
\caption{\label{fig:eigval800d} Same as Fig.\
\protect\ref{fig:eigval500a} for \mbox{\( \sqrt{s} = 800 {\rm \ GeV}
\)} and symmetry class (d).}
\end{figure}

\begin{figure}
\begin{center}
\leavevmode
\includegraphics[totalheight=15cm]{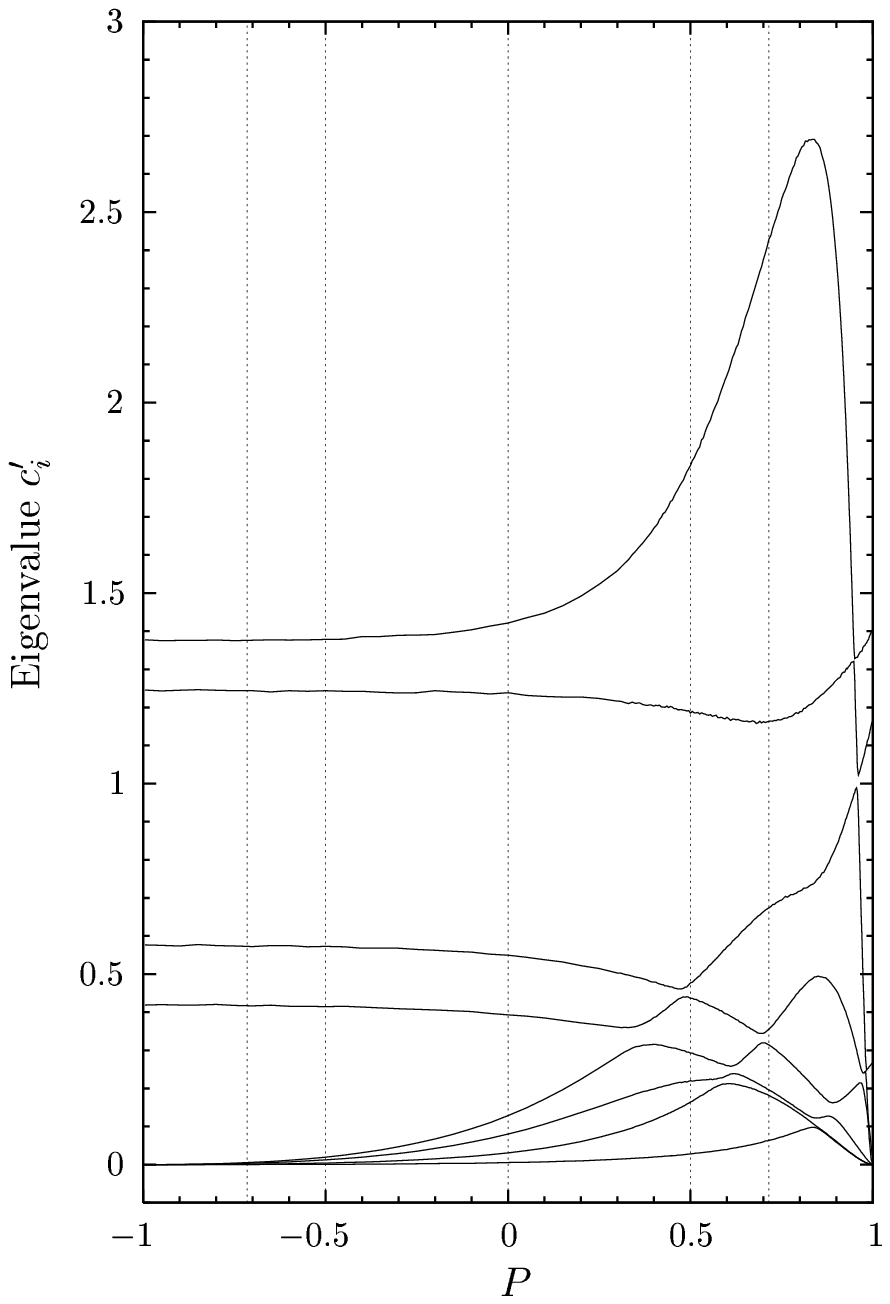}
\end{center}
\caption{\label{fig:eigval3a} Same as Fig.\
\protect\ref{fig:eigval500a} for \mbox{\( \sqrt{s} = 3 {\rm \ TeV}
\)} (symmetry class (a)).}
\end{figure}

\begin{figure}
\begin{center}
\leavevmode
\includegraphics[totalheight=15cm]{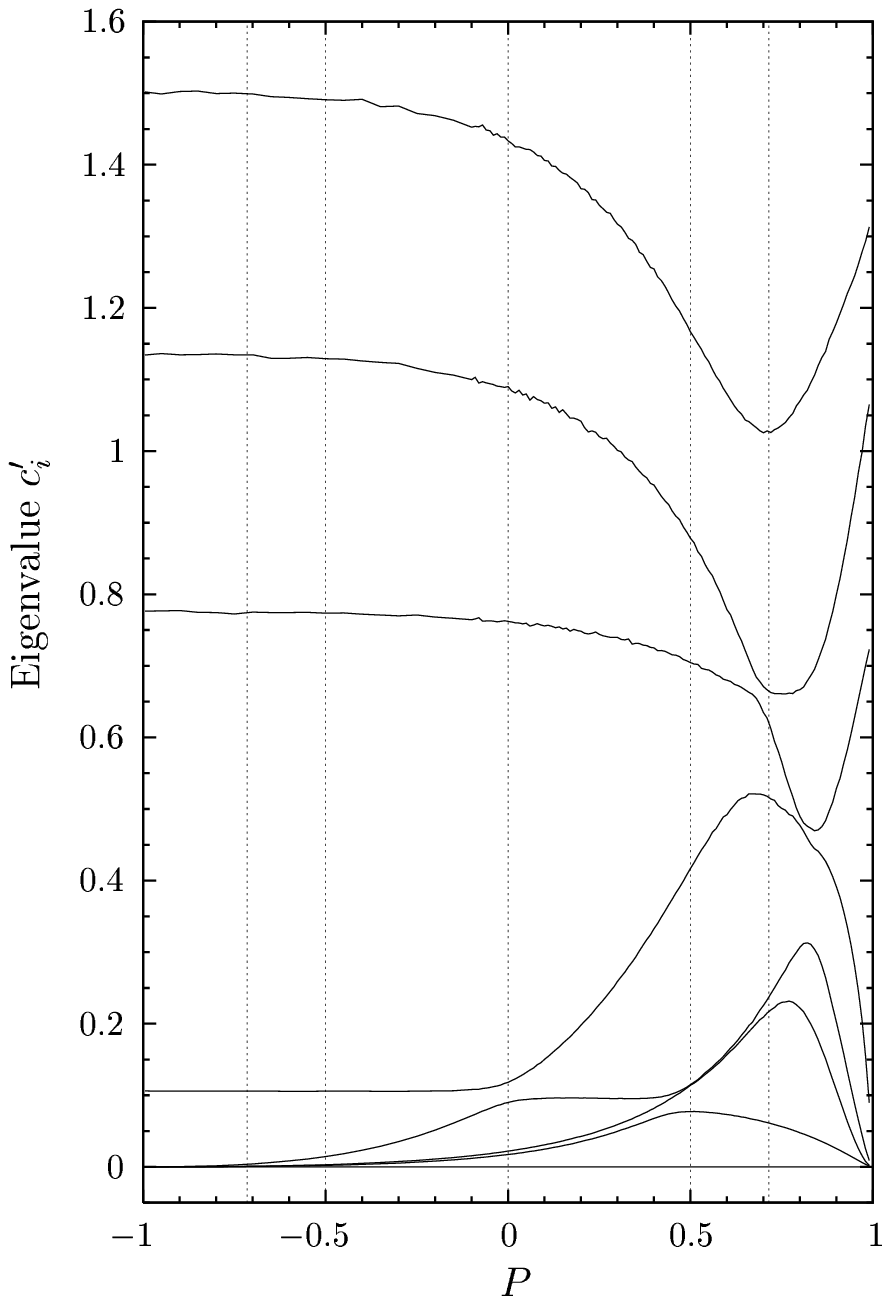}
\end{center}
\caption{\label{fig:eigval3b} Same as Fig.\
\protect\ref{fig:eigval500a} for \mbox{\( \sqrt{s} = 3 {\rm \ TeV} \)}
and symmetry class (b).}
\end{figure}

\begin{figure}
\begin{center}
\leavevmode
\includegraphics[totalheight=15cm]{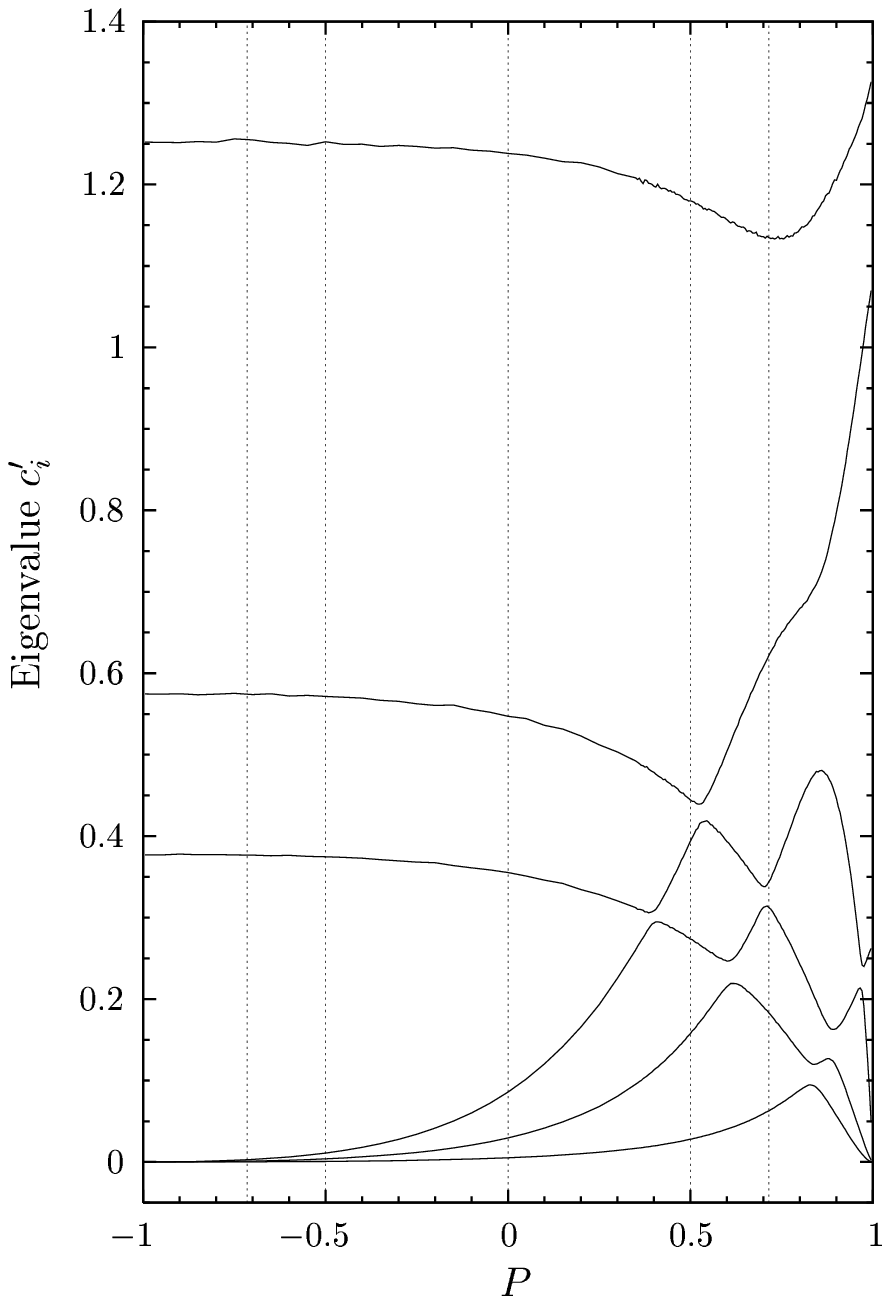}
\end{center}
\caption{\label{fig:eigval3c} Same as Fig.\
\protect\ref{fig:eigval500a} for \mbox{\( \sqrt{s} = 3 {\rm \ TeV} \)}
and symmetry class (c).}
\end{figure}

\begin{figure}
\begin{center}
\leavevmode
\includegraphics[totalheight=15cm]{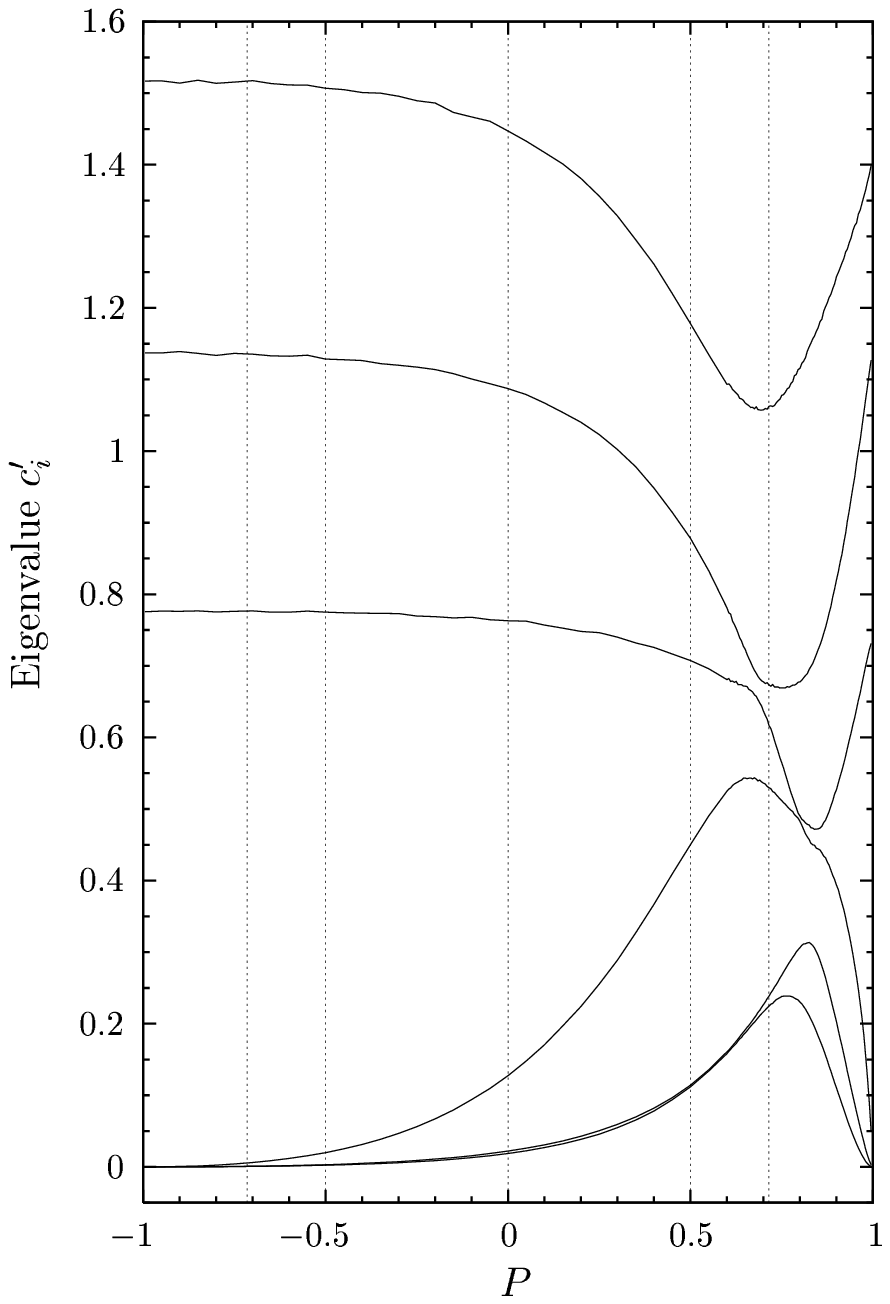}
\end{center}
\caption{\label{fig:eigval3d} Same as Fig.\
\protect\ref{fig:eigval500a} for \mbox{\( \sqrt{s} = 3 {\rm \ TeV} \)}
and symmetry class (d).}
\end{figure}


\clearpage


\appendix

\section{Appendix: Conventions}
\label{app-conv}

Momenta and helicities of incoming and outgoing particles are denoted
as in Fig.\ \ref{fig:pro}.  We evaluate the production amplitude in
the frame obtained from the one in Fig.\ \ref{fig:pro} by a rotation
of $\Theta$ around the $y$-axis, so that the new $z'$-axis points along
the $W^-$ momentum.  For the respective polarisation vectors
$\epsilon_{\lambda}^{}$ and $\epsilon_{\overline{\lambda}}$ of $W^-$
and $W^+$ we choose in this frame
\begin{eqnarray*}
\epsilon_{\pm} & = & \frac{1}{\sqrt{2}}\,(0, \mp 1, -i, 0), \\
\epsilon_0     & = & \frac{1}{m_W}\,(q^3, 0, 0, q^0), \\
\overline{\epsilon}_{\pm} & = & \frac{1}{\sqrt{2}}\,(0, \mp 1, i, 0), \\
\overline{\epsilon}_0     & = & \frac{1}{m_W}\,(-q^3, 0, 0, q^0).
\end{eqnarray*}
The four-spinors for the initial leptons are expressed through
two-spinors $\chi$ in the usual way \cite{Nachtmann:ta}, with
\begin{equation}
\chi_{\tau=+1} = \left(\cos\frac{\Theta}{2}, 
			-\sin\frac{\Theta}{2}\right) , \qquad
\chi_{\tau=-1} = \left(\sin\frac{\Theta}{2}, 
			\cos\frac{\Theta}{2}\right)
\end{equation}
for the electron and 
\begin{equation}
\chi_{\,\hebar=+1} = \left(\sin\frac{\Theta}{2}, 
			\cos\frac{\Theta}{2}\right) , \qquad
\chi_{\,\hebar=-1} = \left(-\cos\frac{\Theta}{2}, 
			\sin\frac{\Theta}{2}\right)
\end{equation}
for the positron.
The evaluation of the diagrams in Fig.\ \ref{fig:feynm} then leads to
(\ref{eq:pram}), where the $d$-functions are defined in the usual
fashion:
\begin{eqnarray}
d_{\he, 0}^1 & = & - \frac{\he }{\sqrt{2}} \sin \Theta , \\
d_{\he, \pm 1}^1 & = & \frac{1}{2} (1 \pm \he \cos \Theta ), \\
d_{\he, \pm 2}^2 & = & \pm \frac{1}{2} (1 \pm \he \cos \Theta) \sin
\Theta.
\end{eqnarray}
For the $W$ decay tensors (\ref{eq:deca}) one has
\begin{eqnarray}
d{\cal D}_{\lprime \lambda} & = & 24 \pi m_W\;\Gamma (W^- \rightarrow
f_1\overline{f_2})\; l_{\lprime}\, l_{\lambda}^{\ast}\;
 d(\cos\!\vartheta)\,d\varphi \; , 
\nonumber \\
d\overline{{\cal D}}_{\lbarprime \lbar} & = & 24 \pi m_W\;\Gamma (W^+
\rightarrow f_3\overline{f_4})\; \overline{l}_{\lbarprime}\,
		\overline{l}_{\lbar}^{\ast} \;
 d(\cos\!\overline{\vartheta})\,d\overline{\varphi}\;,
\end{eqnarray}
where
\begin{eqnarray}
l_- & = & d_+ (\vartheta) \, e^{-i\varphi}, \qquad
\overline{l}_- \;=\; d_+ (\overline{\vartheta}) \, 
		e^{i\overline{\varphi}}, 
\nonumber \\
l_0 & = & -d_0 (\vartheta), \qquad\hspace{1.5em}
\overline{l}_0 \;=\; -d_0 (\overline{\vartheta}) , 
\nonumber \\
l_+ & = & d_- (\vartheta) \, e^{i\varphi}, \hspace{2.5em}
\overline{l}_+ \;=\; d_- (\overline{\vartheta}) \, 
		e^{-i\overline{\varphi}}
\end{eqnarray}
with \( d_{\pm} (x) = (1 \pm \cos x) / \sqrt{2} \) and \( d_0 (x) =
\sin x \).


\end{document}